\documentclass[11pt, a4paper]{article}
\usepackage{jheparxiv}
\usepackage[utf8]{inputenc}
\usepackage{amsmath}
\usepackage{amsfonts}
\usepackage{amssymb}
\usepackage{latexsym}
\usepackage{mathrsfs}
\usepackage{braket}		%Bra Ket Notation
\usepackage{setspace}
\usepackage{graphicx}
\usepackage{color}
\usepackage{xcolor}
\usepackage{slashed}
\usepackage{twistor}
\usepackage[all]{xy}
\usepackage{tikz-cd}
\usepackage{mathtools}
\usepackage{tikz}
\newcommand*\circled[1]{\tikz[baseline=(char.base)]{
            \node[shape=circle,draw,inner sep=2pt] (char) {#1};}}

\newcommand{\bep}{\begin{picture}}
\newcommand{\eep}{\end{picture}}

\newcounter{YoungHeight}\newcounter{YoungWidth}

\newcounter{Mul1}\newcounter{Mul2}\newcounter{Mul3}\newcounter{Mul4}
\newcounter{A0}\newcounter{A1}\newcounter{A2}
\newcounter{B3}
\newcounter{C3}\newcounter{C4}
\newcounter{D1}\newcounter{D2}\newcounter{D3}
\newcounter{T0}\newcounter{T1}

\newlength{\txtHShift}

\newlength{\txtWidth}

\newcommand{\HalfLength}[2]{\setcounter{Mul1}{#1}\setcounter{Mul2}{#1}\addtocounter{Mul1}{\value{Mul2}}\addtocounter{Mul1}{\value{Mul2}}%
\addtocounter{Mul1}{\value{Mul2}}\addtocounter{Mul1}{\value{Mul2}}\setcounter{#2}{\value{Mul1}}}

\newcommand{\Length}[1]{#10}
\newcommand{\shiftedText}[2]{{\hspace{#1}#2}}
\newcommand{\calcHShift}[1]{\settowidth{\txtWidth}{#1}\setlength{\txtHShift}{-0.5\txtWidth}}

\newcommand{\TextCenter}[3]{{\HalfLength{#2}{T0}%
\HalfLength{#3}{T1}\addtocounter{T1}{-3}\calcHShift{#1}%
\put(\value{T0},\value{T1}){\shiftedText{\txtHShift}{#1}}}}

\newcommand{\RectT}[3]{\bep(\Length{#1},\Length{#2})\put(0,0){\line(1,0){\Length{#1}}}\put(0,0){\line(0,1){\Length{#2}}}%
\put(\Length{#1},\Length{#2}){\line(-1,0){\Length{#1}}}\put(\Length{#1},\Length{#2}){\line(0,-1){\Length{#2}}}#3{#1}{#2}\eep}

\newcommand{\RectBRowUp}[4]{{\bep(\Length{#1},20)\put(0,0){\RectT{#2}{1}{\TextCenter{#4}}}%
\put(0,10){\RectT{#1}{1}{\TextCenter{#3}}}\eep}}

\newcommand{\sB}{\tilde{\mathsf{B}}}
\newcommand{\gf}{\mathtt{gf}}
\newcommand{\kbold}{\boldsymbol{k}}
\newcommand{\mso}{\mathfrak{so}}
\newcommand{\msp}{\mathfrak{sp}}
\newcommand{\Ccurl}{\mathscr{C}}

\newcommand{\cT}{\mathcal{T}}
\newcommand{\Tr}{\text{Tr}}
\newcommand{\cI}{\mathcal{I}}
\newcommand{\ta}{\mathtt{a}}
\newcommand{\tb}{\mathtt{b}}

\newcommand{\tx}{\mathtt{x}}
\newcommand{\ty}{\mathtt{y}}
\newcommand{\pl}{\partial}

\newcommand{\hatphi}{\hat\phi}
\newcommand{\BG}{\mathtt{BG}}
\newcommand{\boldJ}{\boldsymbol J}

\newcommand{\tp}{\mathsf{p}}

\usepackage{graphicx}

\newcommand{\nn}{\nonumber}

 \def\one{\mbox{1 \kern-.59em {\rm l}}}
\newcommand{\ths}{\mathfrak{ths}}

\newcommand{\sa}{\mathsf{a}}
\newcommand{\bs}{\mathsf{b}}
\newcommand{\ff}{\mathsf{f}}

\renewcommand{\d}{\mathrm{d}}

\newcommand{\boldI}{\boldsymbol I}

\newcommand{\eps}{\epsilon}

\newcommand{\msf}[1]{\mathsf{#1}}

\newcommand{\sA}{\tilde{\mathsf{A}}}

\newcommand{\msu}{\mathfrak{su}}
\newcommand{\PP}{\mathbb{P}}
\newcommand{\PPb}{\overline{\mathbb{P}}}

\title{Spinorial higher-spin gauge theory from IKKT model\\ in Euclidean and Minkowski signatures}

\author[a]{Harold Steinacker}
\affiliation[a]{Department of Physics, University of Vienna, \\
Boltzmanngasse 5, A-1090 Vienna, Austria}
\emailAdd{harold.steinacker@univie.ac.at}

\author[b]{\& Tung Tran}
\affiliation[b]{
 Service de Physique de l’Univers, Champs et Gravitation,\\
Université de Mons, 20 place du Parc, 7000 Mons, Belgium
}%

\emailAdd{vuongtung.tran@umons.ac.be}

\abstract{We explore the semi-classical relation between the %covariant 
fuzzy 4-hyperboloid $H_N^4$ and  non-compact quantized twistor space $\P_N^{1,2}$ at large $N$. 
This provides two backgrounds of the IKKT matrix model via two natural stereographic projections, leading to higher-spin gauge theories
with Euclidean and Minkowski signature
denoted by HS-IKKT. The resulting higher-spin gauge theory can be understood as an uplift of $\cN=4$ super Yang-Mills to twistor space. The action of  HS-IKKT is written using a spinor formalism in both Euclidean and Minkowski signature. We then compute the tree-level amplitudes of the massless sector within the Yang-Mills part of the HS-IKKT model in the flat limit in Euclidean signature. All $n$-point tree-level scattering amplitudes for $n\geq4$  of this sector are found to vanish in the flat limit. }

\begin{document}

\maketitle
  
%%%%%%%%%%%%%%%%%%%%%%%%%%%%%%%%%%%%%%%%%%

\section{Introduction}\label{sec:1}

The IKKT-matrix model \cite{Ishibashi:1996xs} can be considered as an alternative and constructive description of type IIB superstring theory. It also provides
a promising approach towards a quantum theory of gravity in $3+1$ dimensions, since both 
 spacetime and physical fields emerge from the same matrix degrees of freedom; see e.g. \cite{Steinacker:2019fcb} for a review. Recently, it was shown that the Einstein-Hilbert action can be obtained from the one-loop effective action  on non-commutative branes in the presence of  fuzzy extra dimensions \cite{Steinacker:2021yxt,Steinacker:2023myp}. From this perspective, the gravitational theory induced by the IKKT-matrix model is closely related to the idea of induced gravity of Sakharov \cite{Sakharov:1967pk,Visser:2002ew}.

Some natural backgrounds of the IKKT model are given by covariant fuzzy spaces such as the fuzzy 4-sphere $S^4_N$ \cite{Sperling:2017dts} or fuzzy 4-hyperboloid $H^4_N$ \cite{Sperling:2018xrm}. These are the total spaces of quantized  sphere $\P^1_N$-bundle over the base manifold $S^4$ or $H^4$, respectively. From this point of view, $S^4_N$ can be viewed as a compact fuzzy twistor space $\P^3_N$ while $H^4_N$ is understood as a non-compact fuzzy twistor space $\P^{1,2}_N$. What is significant about these fuzzy spaces is that their algebra of functions $\Ccurl$ is truncated. As a result, the higher-spin algebra $\ths$, which is a subspace of $\Ccurl$, is also truncated for finite $N$ \cite{Sperling:2017dts,Sperling:2017gmy}. Since this truncated symmetry must be accompanied by higher-spin gauge fields,\footnote{Note that the truncated higher-spin algebra $\ths$ coincides with the usual higher-spin algebra $\hs$ of  4-dimensional target space in the semi-classical (large $N$) limit as discussed in e.g. \cite{Steinacker:2019fcb,Steinacker:2022jjv}.} it is obvious that the IKKT-matrix model can induce a higher-spin gauge theory (HS-IKKT) whose spectrum is finite.

The connection between (HS-)IKKT matrix model and twistor theory naturally leads to a spinorial formulation of the resulting gauge theory, which
was explored %in full non-linearity 
for the case of $S_N^4$ in \cite{Steinacker:2022jjv}. However this only leads to a background with Euclidean signature. To overcome that limitation, we generalize in the present paper the analysis in \cite{Steinacker:2022jjv}  to backgrounds defined using $H_N^4$. Indeed, it is known that $H_N^4$ can be projected to a fuzzy FLRW cosmological spacetime with Lorentzian signature. Then the IKKT-matrix model defined on this background becomes a more `conventional' field theory. 

In this work we will consider two projections of $H_N^4$: (i) a stereographic projection to a 4-hyperboloid (or equivalently Euclidean AdS) as the base manifold, and (ii) a $SO(1,3)$-invariant projection which defines a FLRW type cosmology \cite{Steinacker:2017bhb,Steinacker:2017vqw,Sperling:2019xar}. 
In the first case, we essentially recover the same results of \cite{Steinacker:2022jjv} (up to signs) for the spinorial formulation of the IKKT-matrix model on $H^4$. In the second case, we provide a novel formulation of the IKKT-matrix model in terms of twistor/spinor variables on the FLRW spacetime. 
However, in that case the spinorial description turns out to be rather unconventional,  making it hard to study scattering processes in the flat limit.

As an application of the spinorial formulation,
we compute tree-level on-shell scattering amplitudes of massless fields of the Yang-Mills (YM) sector of HS-IKKT theory in the complexified Euclidean case, and show that all $n$-point tree-level amplitudes of this sector vanish for $n\geq 4$ in the flat limit. %This result is not a surprise. In fact, since the 3-point MHV amplitudes between massless external states in the Yang-Mills sector vanish, no higher-point amplitudes can be formed. 
To obtain an alternative understanding of this result, we also project all vertices of this sector to the light-cone gauge, and observe that the MHV sector can be removed by a local field redefinition. %This implies that the YM massless sector of HS-IKKT theory coincides with self-dual higher-spin gravity with two-derivative interactions \cite{Krasnov:2021nsq} in the flat limit.
Due to the unconventional space-like spinor formalism in the Lorentzian case, we compute the tree-level scattering amplitudes of HS-IKKT theory only in Euclidean signature in the present paper, and set the stage to elaborate the Lorentzian case elsewhere.

The paper is structured as follows. Section \ref{sec:2} provides a brief review of the IKKT-matrix model and almost-commutative 4-hyperboloid. Section \ref{sec:3} elaborates the relation between $H_N^4$ and $\P_N^{1,2}$ using spinor formalism from the stereographic projection point of view. Section \ref{SO(1,3)spinors} studies the spinorial description of the $SO(1,3)$-invariant projection. The gauge-fixing procedure and decompositions of modes are specified in Section \ref{sec:shift-vector}. Section \ref{sec:5} derives spacetime action of HS-IKKT. Section \ref{sec:6} computes the scattering amplitudes of the HS-IKKT. We conclude in Section \ref{sec:7} and collect some technicalities in the Appendices.

\paragraph{Notation.} Throughout the paper, we use $a,b$ as $SO(1,4)$-indices where $a,b=0,1,2,3,4$ while $\hat a,\hat b$ stand for $SO(2,3)$-indices for $\hat a,\hat b=0,1,2,3,5$. The $Sp(4)$ and twistorial indices will be denoted as $A,B$ where $A,B=1,2,3,4$. Note that $\alpha,\beta=0,1$ while $\dot\alpha,\dot\beta=\dot 0,\dot 1$. We will use the strength-one symmetrization convention, e.g. $A_aB_a=\frac{1}{2}(A_{a_1}B_{a_2}+A_{a_2}B_{a_1})$; and write fully symmetric rank-$s$ tensor as $T_{a(s)}=T_{a_1\ldots a_s}$ for short.

%%%%%%%%%%%%%%%%%%%%%%%%%%%%%%%%%%%%%%%%%% 
%%%%%%%%%%%%%%%%%%%%%%%%%%%%%%%%%%%%%%%%%%

\section{Review on the IKKT matrix model and the fuzzy 4-hyperboloid}\label{sec:2}
%%%%%%%%%%%%%%%%%%%%%%%%%%

The field content of the $SO(1,9)$-invariant action functional
\begin{align}\label{IKKTSO(1,9)}
    S=\Tr\Big([Y^{\boldsymbol{I}},Y^{\boldsymbol{J}}][Y_{\boldsymbol{I}},Y_{\boldsymbol{J}}]+\Psi_{\cA}(\tilde\gamma^{\boldsymbol{I}})^{\cA\cB}[Y_{\boldsymbol{I}},\Psi_{\cB}]\Big)\,,\qquad {\boldsymbol{I}}=0,1,\ldots,9\,,
\end{align}
describing the IKKT model comprises $N\times N$ hermitian matrices $Y^{\boldsymbol{I}}$, and the  matrix-valued Majorana-Weyl spinors 
$\Psi_{\cB}$ associated with $SO(1,9)$. Since $Y^{\boldI}$ are matrices, they do not commute. This non-commutativity can be interpreted in terms of a quantized Poisson structure $\theta^{\boldI \boldJ}$  on a brane embedded in target space $\R^{1,9}$ viz.
\begin{align}\label{Poisson1}
    [Y^{\boldI},Y^{\boldJ}]:=\im\,\theta^{\boldI \boldJ}\,.
\end{align}
To extract classical coordinate functions $\{y^{\boldsymbol I}\}$ 
 describing $\R^{1,9}$, it is reasonable to assume that there are  localized quasi-coherent states $|y\rangle \in \cH$ (with $\cH$ being some Hilbert space on which the matrices act) such that the $y^{\boldsymbol I}$ arise as expectation value of $Y^{\boldI}$ viz.
\begin{align}
 y^{\boldsymbol{I}} = \langle y|Y^{\boldsymbol{I}}|y\rangle \ \in  \R^{1,9} \ .
 \label{coherent-expect}
\end{align}
The  resolution of the coordinate functions $y^{\boldI}$ will increase with the size of the matrices $Y^{\boldI}$, which suggests to consider the large $N$ (or semi-classical) limit where matrices are almost-commutative \cite{Steinacker:2020nva,Ishiki:2015saa,Schneiderbauer:2016wub,Berenstein:2012ts}. In this limit, we can replace \eqref{Poisson1} with Poisson brackets
\begin{align}
    \{y^{\boldI},y^{\boldJ}\}:=\theta^{\boldI\boldJ}\,.
\end{align}
Here $y^{\boldsymbol I}$ can be used to define an almost-commutative variety embedded in target space  via $y^{\boldsymbol I}:\cM \hookrightarrow \R^{1,9}$. Then classical functions in terms of $y$ are related to matrices via
\be
    \begin{split}
    \Ccurl(\cM)\quad &\sim \quad {\rm Mat}(\cH)\\
    f(y)=\langle y|F(Y)| y \rangle \quad &\sim \quad F(Y)\,.
    \end{split}
\ee
The matrix algebra 
$({\rm Mat}(\cH),[\,,])$ generated by $Y^{\boldsymbol{I}}$ is interpreted as quantized version of the Poisson algebra defined by the pair $(\Ccurl(\cM),\{\,,\})$. Furthermore, the trace in \eqref{IKKTSO(1,9)} will be replaced by some appropriate integral, as discussed below.

For our purpose of constructing a higher-spin theory from the IKKT-matrix model on $H_N^4$, it is sufficient to consider the semi-classical limit or -regime, where we can work with ordinary functions as familiar from field theory. The non-commutativity  then reduces to the explicit Poisson brackets
\begin{align}\label{Poisson2}
    \{y^{a},y^{b}\}=: \theta^{ab}=-\ell_p^2m^{ab}\,,\qquad a,b=0,1,2,3,4\,,
\end{align}
where $m^{ab}$ are functions on twistor space $\P^{1,2}$ (as explained below) arising from the semi-classical limit of $\mso(1,4)$ generators in certain representations \cite{Sperling:2018xrm}, and $\ell_p$ is a natural length scale. 

It is worth noting that the Poisson bracket \eqref{Poisson2} involves two derivatives in $y^a$. This means that even though the IKKT-matrix model has the structure of a  Yang-Mills gauge theory, it behaves like a gravitational theory. To study the IKKT model perturbatively \cite{Aoki:1999vr}, we will consider fluctuations $y^a=\bar y^a + \sa^a$ of a background $\bar{y}^a$ on $H^4_N$ \cite{Hasebe:2012mz}. This defines an almost-commutative Yang-Mills-type gauge theory on $H_N^4$ \cite{Sperling:2018xrm} that is invariant under the gauge transformations $U^{-1}(\bar y^a + \sa^a)U$ where $U$ is any unitary matrix, replacing commutators by Poisson brackets.

\paragraph{Algebra of functions on semi-classical $H_N^4$.} Endowing the ambient space $\R^{1,4}$ with the metric $\eta_{ab}=(-,+,+,+,+)$ allows us to describe a 4-dimensional hyperboloid $H^4_N$ of radius $R$ in terms of a space  of functions $\Ccurl$ with the following $\mso(1,4)$-covariant  semi-classical relations \cite{Sperling:2018xrm}:\footnote{Note the sign difference with the case of $S_N^4$ \cite{Sperling:2017dts}.}
\begin{subequations}\label{eq:so(4,2)algebra}
\begin{align}
    \{m_{ab},m_{cd}\}&=+(m_{ad}\eta_{bc}-m_{ac}\eta_{bd}-m_{bd}\eta_{ac}+m_{bc}\eta_{ad})\,,\\
    \{m_{ab},y_c\}&=+(y_a\eta_{bc}-y_b\eta_{ac})\,,\\
    \{y_a,y_b\}&=-\ell_p^2\, m_{ab}\,,\\
    y_ay^a&=-y_0^2+y_{\ta}y^{\ta}=-R^2=-\frac{\ell_p^2N^2}{4}\,,\qquad \ta=1,2,3,4\,,\label{ysphere}\\
    \epsilon_{abcde}m^{ab}y^{c}&=-\frac{4N}{\ell_p}m_{de}\label{eq:selfdualityso(42)}\,
\end{align}
\end{subequations}
%\textcolor{blue}{Notation is imprecise. $N$ makes sense only in the fuzzy case. Maybe do give fuzzy first and the its semiclassical limit.}
%\textcolor{blue}{This arises in the semi-classical i.e. large $N$ limit of 
%the as
%\begin{align}
%m_{ab} = \cM_{ab}, \qquad y_a = \cM_{a5}
%\label{generators-H4N-so42}
%\end{align}
%where $\cM_{AB}$ are generators of the  doubleton representation $\cH_N$ of $\mso(4,2)$.
%}
%{\bf I would suggest to be as minimal as possible, since we need only semi-classical limit and flat limit.}
with large $N$. Here, \eqref{eq:selfdualityso(42)} is a self-duality relation, which allows us to restrict the space of functions on $H^4_N$  to
\begin{align}
    \Ccurl(y^a,m^{ab})=\sum_{k,m}f_{c(k)a(m),b(m)}y^{c(k)}m^{ab}\ldots m^{ab}=\bigoplus_{k,m}\ \parbox{85pt}{{\bep(70,30)\unitlength=0.4mm%
\put(0,3){\RectBRowUp{7}{4}{$m+k$}{$m$}}%
\eep}}\,
\end{align}
corresponding to two-row Young tableaux,
where the convention $y^{a(m)}$ means $y^{(a_1}\ldots y^{a_m)}$. Within $\Ccurl$, we can find a subspace 
\begin{align}
    \ths(\mso(1,4))=\sum_m g_{a(m),b(m)}m^{ab}\ldots m^{ab}=\bigoplus_m\ \parbox{60pt}{{\bep(70,30)\unitlength=0.4mm%
\put(0,3){\RectBRowUp{5}{5}{$m$}{$m$}}%\,,
\eep}}\,,
\end{align}
which consists of polynomials purely in terms of $m^{ab}$, thus defining a truncated higher-spin algebra $\ths$. Note that  $\ths$ coincides with the usual higher-spin algebra $\hs$ of Euclidean AdS${}_4$ in $N\rightarrow \infty$ limit. 

The above algebra of function implies that the IKKT-matrix model on a $H^4_N$ background induces a higher-spin gauge theory (HS-IKKT) \cite{Sperling:2018xrm}. Remarkably, the higher-spin fields mitigate the violation of Lorentz invariance through noncommutativity (cf. \eqref{Poisson2}), while
parity invariance is broken due to the presence of $\eps_{abcde}$ tensors \cite{Sperling:2017dts}. This causes the HS-IKKT to exhibit some sort of `chiral' feature \cite{Sperling:2018xrm}, and 
reflects the fact that HS-IKKT can be formulated on twistor space \cite{Steinacker:2022jjv}.

\paragraph{Twistor realization of semi-classical $H_N^4$.} 

It is manifest that \eqref{eq:so(4,2)algebra} is related to the Lie algebra $\mso(2,4)$; more precisely, they arise in the semiclassical limit of a specific ``doubleton" representation $\cH_N$ of $\mso(2,4)$. 

Instead of working with $\mso(2,4)$, it is sometimes more convenient to use the $\msu(2,2)$ formulation. As is well-known, the non-compact twistor space $\P^{1,2}$ can be realized either as $SO(1,4)$-equivariant bundle over $H^4$, or a 6-dimensional co-adjoint orbit of $SU(2,2)$, i.e. $\cO_{\Xi}=\{g\,\Xi\, g^{-1}\,,\, g\in SO(2,4)\}$ where $\Xi=(N,0,0)$ has the stabilizer $U(1)\times SU(2)$ \cite{Sperling:2018xrm}. This allows us to identify $H^4_N$ with $\P^{1,2}$. More explicitly, consider the following identifications: 
\begin{align}\label{mapsotosu}
    y^{AB}=-y^{BA}=\ell_p^{-1}y^a\gamma_a^{AB}\,,\qquad l^{AB}= l^{BA}=\frac{1}{2} m^{ab}\Sigma^{AB}_{ab}\, 
     \,,
\end{align}
where $\gamma_a = -2\Sigma_{a5}$ are the gamma matrices of $\mso(1,4)$ realized by the following  basis adapted to the compact $SU(2)_L \times SU(2)_R\subset SU(2,2)$ subgroup:
\begin{align}\label{gamma-so5-explicit}
 (\gamma_0)^{A}_{\ B}=
\begin{pmatrix}
 \one_2 & 0 \\ 0 &-\one_2
\end{pmatrix}\,,\quad (\gamma_{m})^{A}_{\ B}=
\begin{pmatrix}
 0 & -(\sigma_m)^{\alpha}{}_{\dot\beta} \\ (\sigma_m)^{\dot\alpha}{}_{\beta} & 0
\end{pmatrix}\,,
\quad
(\gamma_4)_{\ B}^{A} =
\im\begin{pmatrix}
 0 & \one_2 \\ \one_2 & 0
\end{pmatrix} \, ,
\end{align}
%\textcolor{blue}{the sign of $\gamma_m$ is different in \cite{Sperling:2018xrm}. I dont think it matters, but can we change it here to be coherent?}
for $m=1,2,3$, and
\begin{align}
    \{\gamma_{a},\gamma_b\}^{A}_{\ B}=-2\eta_{ab}\delta^A_{\ B}\,,\qquad \Sigma_{ab}^{AB}=-\Sigma_{ba}^{AB}=\Sigma_{ab}^{BA}=\frac{1}{4\,\im}[\gamma_a,\gamma_b]^{AB}\,.
\end{align}
This allows us to express the $H_N^4$ using $\msu(2,2)$ representation as:
\begin{subequations}\label{eq:sp(4)relations}
\begin{align}
    \{l^{AB},l^{CD}\}&=+(l^{AC}C^{BD}+l^{AD}C^{BC}+l^{BD}C^{AC}+l^{BC}C^{AD})\,,\\
    \{l^{AB},y^{CD}\}&=+(y^{AC}C^{BD}+y^{BC}C^{AD}-y^{AD}C^{BC}-y^{BD}C^{AC})\,,\\
    \{y^{AB},y^{CD}\}&=-(l^{AC}C^{BD}-l^{AD}C^{BC}-l^{BC}C^{AD}+l^{BD}C^{AC})\,,\\
    y_{AB}y^{AB}&=  l_{AB}l^{AB} = -4R^2\,,\\
    \epsilon_{ABCD}y^{AB} &= y_{CD}\, \label{self-duality-sp4} 
\end{align}
\end{subequations}
where 
\begin{align}
    C^{AB}=-C^{BA}=\diag(\epsilon^{\alpha\beta},\epsilon^{\dot\alpha\dot\beta})
    \label{C-matrix}
\end{align}
is the $\mso(1,4)$-invariant matrix
%\textcolor{blue}{please dont use $\msp(4)$ here, this is $\mso(4,1)$, everywhere}; 
and we use 
$\epsilon^{01}=-\epsilon^{10}=1\,,\epsilon^{\alpha\beta}=\epsilon_{\alpha\beta}$
as our $\msp(2)$-invariant matrix. These matrices allow us to raise and lower $\msu(2,2)$ and/or $\msp(2)$ indices as
\begin{align}
    V_{B}C^{AB}=U^{A}\,,\qquad V^{B}C_{BA}=U_{A}\,,\qquad 
    u^{\alpha}=u_{\beta}\epsilon^{\alpha\beta}\,,\qquad u_{\alpha}=u^{\beta}\epsilon_{\beta\alpha}\,.
\end{align}
The Poisson algebra in $\msu(2,2)$ representation becomes
\begin{align}
    \Ccurl(y^{AB},l^{AB})=\sum_{k,m}f_{A(k)B(2m)|C(k)}y^{AC}\ldots y^{AC}l^{BB}\ldots l^{BB}=\bigoplus_{k,m}\ \parbox{75pt}{{\bep(70,30)\unitlength=0.4mm%
\put(0,3){\RectBRowUp{6}{4}{$k+2m$}{$k$}}%
\eep}}\,.
\end{align}
We can identify the following subspace:
\begin{align}
    \ths(\msp(4))=\sum_{m}g_{B(2m)}l^{BB}\ldots l^{BB}=\bigoplus_m\ \parbox{60pt}{{\bep(\Length{6},6)\unitlength=0.4mm\put(0,-1.5){\RectT{5}{1}{\TextCenter{$2m$}}}\eep}}\ 
\end{align}
as the truncated higher-spin algbera $\ths(\msp(4))$. This allows us to identify semi-classical $H_N^4$ with the non-compact semi-classical twistor space $\P^{1,2}_N$ spanned by the $\msu(2,2)$ vectors $Z^A$ and their complex conjugate $Z_A^{\dagger}$. 
%\textcolor{blue}{We dont need the spinors $\lambda,\mu$ at this point, they will be discussed in section 3. I'd remove this here}

For later convenience, we also note that by lowering the indices of the gamma matrices in \eqref{gamma-so5-explicit}, we obtain
\begin{align}
\label{chiralso(1,5)basis}
    \gamma^0_{AB}=\begin{pmatrix}
     \epsilon_{\alpha\beta} &0 \\
     0 &-\epsilon_{\dot\alpha\dot\beta}
    \end{pmatrix}\,,\quad \gamma^m_{AB}=\begin{pmatrix}
     0 & \tilde\sigma^{m}_{\alpha\dot\beta}\\
     -\tilde\sigma^m_{\dot\beta\alpha} &0
    \end{pmatrix}\,,\quad \gamma^4_{AB}=\begin{pmatrix}
     0 &\epsilon_{\alpha\dot\alpha}\\
     -\epsilon_{\dot \alpha\alpha} &0
    \end{pmatrix}\,.
\end{align}

%\textcolor{blue}{I think in $\gamma^4$ the lower-left epsilon should have reversed indices, dotted first}

%The `hat' operator is simply a quaternionic conjugation, i.e. $\hat \rho \cdot \eps=\eps\cdot\rho$ for any quatermion $\rho\in GL(2,\HH)$.
%In what follows, we will consider $U(1)\times SO(4)\sim U(1)\times SU(2)_L\times SU(2)_R$ as
%the maximal compact subgroup of $SO(2,4)$ whenever we want to have a conformally flat spacetime as the base manifold. 

\medskip

The Hilbert space $\cH_N$ underlying $H_N^4$ is the lowest-weight irreducible representation
\begin{align}
    (0,0,N)_{\msu(2,2)}
    %=(0,0,1)^{\otimes_{\text{sym}} N}_{\msu(2,2)}
    % No this is not true in the noncompact case =\hat{Z}^{A_1}\ldots\hat{Z}^{A_N}|0\rangle\.
\end{align}
which can be realized using an oscillator construction in terms of operators $Z^A, \bar Z_A$ subject to the constraint 
%\begin{align}
%\hat{\cN}:= \hat{Z}_{A} Z^{A}=N =\frac{2R}{\ell_p} \, %.
%\end{align}
%Something is wrong here. What is $\hat Z$? It should be 
\begin{align}\label{eq:numberoperator}
    \hat{\cN}:= \bar{Z}_A Z^A  = Z^\dagger_{A} (\gamma^0)^{A}{}_B Z^B =N =\frac{2R}{\ell_p} \, 
\end{align}
where $\bar Z_A = Z^\dagger_{\bullet} (\gamma^0)^{\bullet}{}_A$ is the Dirac conjugate of $Z^A$; and the ``number operator'' $\hat\cN$ is invariant under $SU(2,2)$ rather than $SU(4)$
(cf. section 3.3 in \cite{Sperling:2018xrm}). 
Note that the $A$ indices are chiral wrt. $SU(2,2)$.

%This leads to 
%\begin{align}\label{numberinspinor}
% \hat{\cN} = \langle \lambda \, \bar\lambda\rangle - [\mu \,\bar\mu]\,.
%\end{align}
%\textcolor{blue}{remove here, is in section 3}

The space of operators
\begin{equation}\label{fuzzyP12}
    \begin{split}
    \Ccurl(\P^{1,2}_N)&=\mathrm{End}(\cH_N)=(N,0,0)_{\msu(2,2)}\otimes (0,0,N)_{\msu(2,2)}=\sum_{n=1}^N(n,0,n)_{\msu(2,2)}\\
    &=\sum_{n=0}^N\,f_{A(n)}{}^{B(n)}Z^A\ldots Z^A\bar{Z}_B\ldots \bar{Z}_B\,
    \end{split}
\end{equation}
is given by the space of polynomials in $Z^A, \bar Z_B$ with equal number of $Z$ and $\bar Z$.
Its semi-classical limit is given by 
%Viewing $\hat{Z}^A$ as `creation operator', the $N$-particle Fock space $\cH_N$ is defined by  %% No doesnt work this way in the noncompact case
the space of commutative functions generated by $Z^A, \bar Z_B$ with equal number of $Z$ and $\bar Z$ subject to the constraint \eqref{eq:numberoperator}
modulo $U(1)$ endowed with the Poisson structure
\begin{align}\label{eq:commZ}
    \{Z^A,\bar{Z}_B\}=\delta^A{}_B\,
\end{align}
%\textcolor{blue}{(note that $\bar Z = Z^\dagger \gamma^0$ is the Dirac conjugate here!)}
This is  nothing but the space of functions on $\Ccurl(\P^{1,2})$. 
%Suppose the $SU(2,2)$-invariant product between $Z$ and $\hat{Z}$ is:
%\begin{align}\label{eq:numberoperator}
%\hat{\cN}:= \hat{Z}_{A} Z^{A}=N=\frac{2R}{\ell_p}\,
%\end{align}
%in the semi-classical limit.
We can  check that
\begin{align}\label{eq:gradingZ}
    \{\hat{\cN},Z^A\}=-Z^A\,,\qquad \{\hat{\cN},\bar{Z}_A\}=+\bar{Z}_A\,.
\end{align}
Thus $\hat \cN$ gives the Poisson algebra defined by the pair $(\Ccurl(\P^{1,2}),\{\,,\})$ a gradation as shown. Moreover, we can use $\hat\cN$ to define an $H^{3,4} \subset \C^4$ as
\begin{align}\label{eq:H43}
    H^{3,4}\simeq H^7:=\{Z^A\in \C^4\,|\,\bar{Z}_AZ^A=N\}\,.
\end{align}
This allows us to understand $\P^{1,2}$ from the point of view of the following Hopf fibration\footnote{note that for $\mso(4,1)$ both indices of the gamma matrices are equivalent since the spinor representation is self-dual due to $C_{AB}$, unlike for $\mso(4,2)$.}:
\begin{align}\label{Hopf1}
    \begin{split}
    \P^1\xhookrightarrow{}\P^{1,2}=H^{3,4}/U(1)&\rightarrow H^4\subset \R^{1,4}\,,\\
    Z^A&\mapsto y^a:=\frac{\ell_p}{2}\bar{Z}_A(\gamma^a)^A{}_BZ^B\,,
    \end{split}
\end{align}
where $\gamma^a = 2\Sigma^{a5}$. 
This means that $\P^{1,2}_N$ can be understood as $\P^1_N$-bundle over $H^4$. Here the subscript $N$ indicates not only the radius constraint  \eqref{eq:numberoperator} but also the origin of $\P^1_N$  as a fuzzy sphere $S^2_N$
in the non-commutative regime, as explained further in Section \ref{sec:3}.
Note that $H^4$ has Euclidean signature even though the ambient space $\R^{1,4}$ is endowed with a $SO(1,4)$ metric \cite{Sperling:2018xrm}.
The $\P^1_N$ will be described in terms of `negative chirality' spinors which transform under the local $SU(2)_L$. 

%%%%%%%%%%%%%%%%%%%%%%%%%%%%%%%%%%%%%%%%%%%%%%
\paragraph{Semi-classical $H^{2,2}_N$.} For later discussion related to a matrix-model cosmology with Lorentzian signature, we recall that $\P^{1,2}$ can be also viewed as $\P^1_N$-bundle over a split signature 4-hyperboloid $H^{2,2}$. This is realized by the following Hopf map:
\begin{align}\label{eq:quantizeHopf2}
    \begin{split}
    \P^1\xhookrightarrow{} \P^{1,2}=H^{3,4}/U(1)&\rightarrow H^{2,2}\subset \R^{2,3}\\
    Z^A&\mapsto t_{\hat a}=\frac{1}{R}\bar{Z}_A(\Sigma_{\hat a 4})^A_{\ B}Z^B=\frac{1}{R}m_{\hat a  4}\,,\qquad \hat{a}=0,1,2,3,5\,,
    \end{split}
\end{align}
%\textcolor{red}{there must be the dirac conjugation}
where $t_{\hat{a}}$ transform as vectors under $SO(2,3)$ whose generators are $m_{\hat a\hat b}$. They satisfy the following relations
\begin{subequations}
\begin{align}
    \{t^{\hat a},t^{\hat b}\}&=\frac{1}{R^2}m^{\hat a \hat b}\,,\qquad \qquad \quad \quad \, \, \hat a,\hat b=0,1,2,3,5\,,\label{eq:Phatcommutator}\\
    \eta_{\hat{a}\hat{b}}t^{\hat{a}}t^{\hat{b}}&=-t_{0}^2+t_{\hat{i}}t^{\hat{i}}-t_{5}^2=\frac{1}{\ell_p^2}\,,\qquad \hat i = 1,2,3\,,\label{Psphere}\\
    y_{\hat a}t^{\hat a}&=0=y_{\mu}t^{\mu}\,,\qquad \qquad \quad \quad \ \, \mu=0,1,2,3\,, \label{eq:orthogonalofPY}
\end{align}
\end{subequations}
where the metric of $\R^{2,3}$ is $\eta_{\hat{a}\hat b}=\diag(-,+,+,+,-)$.
Due to the last relations, the $t^{\mu}$ can be understood as generators of the internal $S^2_N$ underlying the higher spin structure. They are associated with 
the gamma matrices $\Sigma^{\mu 4}=: \frac{1}{2\,\im}\gamma^\mu \gamma^4 = 
  \frac{1}{2}\underline{\gamma}^{\mu}$ given by
\begin{align}\label{gammaSO(1,3)}
    (\underline{\gamma}^0)^{A}_{\ B}=\frac 12 \begin{pmatrix}
     0 &\one_2\\
     -\one_2 & 0
    \end{pmatrix}\,,\quad  (\underline{\gamma}^{\hat i})^A_{\ B}\equiv (\underline{\gamma}^i)^A_{\ B}=\frac 12  \begin{pmatrix}
     \sigma_i &0\\
     0 & -\sigma_i
    \end{pmatrix}\,,\qquad \hat i=1,2,3\,.
\end{align}
%\textcolor{blue}{I fixed a factor $i$ in the above definition, but I think there should be a factor $-1$ in front of the $\sigma$s, as before. (Do we ever need that?)} {\bf only for the stereographic projection, I think}
We also note that the above definitions imply $ y_{4}=-\ell_pR\,t_{ 5}\,$.

\paragraph{Flat limit of semi-classical $H^4_N$.}

In the later sections of this paper, we will focus on the flat limit $R\to \infty$ of the semi-classical $H^4_N$. If $y^\ta, \ \ta = 1,\ldots,4$ are local coordinates near the ``south pole" $y^a = (R,0,0,0,0)$, it is natural to consider a second set of vector generators $t^\ta$ 
(different from but analogous to the generators $t^{\hat a}$ in \eqref{eq:quantizeHopf2}) which arises from the underlying representation of $\mso(2,4)$:
\begin{align}
    y^\ta &= \ell_p m^{\ta 5}  \nn\\
    t^\ta &= \frac 1R m^{\ta 0}
    \label{t-generators-flatH4}
\end{align}
for $\ta = 1,\ldots ,4$, which satisfy 
\begin{subequations}\label{projected-so(4,2)algebra}
\begin{align}
    \{y_{\ta},y_{\tb}\}&=-\ell_p^2m_{\ta\tb}\,,\qquad \qquad \quad \qquad \quad\ \, \ta,\tb=1,2,3,4\,,\\
    y_{\ta}y^{\ta}&=-R^2+y_0^2=-\frac{\ell_p^2N^2}{4}+y_0^2\,,\label{projected-ysphere}\\
    \{t^{\ta},t^{\tb}\}&=\frac{1}{R^2}m^{\ta \tb}\,,\label{eq:Thatcommutator}\\
    \delta_{\ta\tb}t^{\ta}t^{\tb}&=\frac{1}{\ell_p^2}\,,\label{Tsphere}\\
    y_{\ta}t^{\ta}&=0\,. \label{eq:orthogonalofTY}
\end{align}
\end{subequations}
Due to the last relations, $t^{\ta}$ can be considered as generators of the internal $S^2_N$ underlying the higher spin structure. We will use this representation of ``momentum'' on $H^4$ to analyze the degrees of freedom for higher-spin fields in Section \ref{sec:shift-vector}.

\paragraph{Semi-classical $\cM^{1,3}_N$ spacetime.} To get the desired Lorentzian signature from either $H^4$ or $H^{2,2}$, we can consider the following projections to a $SO(1,3)$-covariant spacetime:
\begin{subequations}
\label{eq:M13projections}
\begin{align}
    \pi_y\,&:
    m^{ab}\mapsto y^{\mu}=\ell_p m^{\mu 5}\,,\\
    \pi_t\,&:m^{\hat a\hat b}\mapsto t^{\mu}=\frac{1}{R}m^{\mu 4}\,.
\end{align}
\end{subequations}
The projections \eqref{eq:M13projections} can be realized explicitly as follows:
\begin{subequations}\label{eq:YPonR13}
\begin{align}
    \eta_{\mu\nu}y^{\mu}y^{\nu}&=-R^2-y_4^2\, \, =-R^2\cosh^2(\tau)\,,\\
    \eta_{\mu\nu}t^{\mu}t^{\nu}&=\frac{1}{\ell_p^2}+\frac{y_4^2}{\ell_p^2R^2}=+\ell_p^{-2}\cosh^2(\tau)\,, \label{S2sphereM13}
\end{align}
\end{subequations}
where $\eta_{\mu\nu}=\diag(-,+,+,+)$ and $\tau$ is a time parameter that defines a space-like foliation for a cosmological FLRW spacetime with $k=-1$. In particular, $\tau$ is defined by 
\begin{align}
    y_4=R\sinh (\tau)\,,
\end{align}
which features a big-bounce at $\tau=0$. Here, $t^{\mu}$ defines internal space-like sphere $S^2$ with radius $\ell_p^{-2}\cosh^2(\tau)$ (cf., \eqref{eq:orthogonalofPY} and \eqref{S2sphereM13}) whose local stabilizer is $SO(3)\simeq SU(2)$. For this reason, one may use $t^{\mu}$ as auxiliary vectors to describe higher-spin modes. Further relations include \cite{Sperling:2018xrm,Sperling:2019xar}:
\begin{subequations}
\begin{align}
    \{t^{\mu},y^{\nu}\}&=+\frac{\eta^{\mu\nu}}{R}y^4=\eta^{\mu\nu}\sinh(\tau)\,,\label{eq:PoissonM13}\\
    \{t^{\mu},y^4\}&=-\frac{y^{\mu}}{R}\,,\\
    m^{\mu\nu}&= R^2 \{t^\mu,t^\nu\}   =-\frac{1}{\cosh^2(\tau)}\Big(\sinh(\tau)(y^{\mu}t^{\nu}-y^{\nu}t^{\mu})+\eps^{\mu\nu\sigma\rho}y_{\sigma}t_{\rho}\Big)\label{mgenerator}\,.
\end{align}
\end{subequations}
The above suggests that we can identify $y^{\mu}$ as coordinates on $\cM^{1,3}_N$, and $t^{\mu}$ as momentum generators. In fact, by virtue of \eqref{eq:PoissonM13}, we have
\begin{align}\label{tderivative}
    \{t_{\mu},\phi(y)\}=\sinh(\tau)\p_{\mu}\phi(y)\,.
\end{align}
There is an important global time-like vector field:
\begin{align}\label{timelikeT}
    \cT=y^{\mu}\p_{\mu}
\end{align}
which describes the time evolution of the FLRW cosmological background with $k=-1$ and is compatible with $SO(1,3)$ isometry. The coordinates of the FLRW patch read \cite{Steinacker:2017bhb}:
\begin{align}
    \begin{pmatrix}
     x^0\\
     x^1\\
     x^2\\
     x^3
    \end{pmatrix}=R\cosh(\tau)\begin{pmatrix}
     \cosh(\chi)\\
     \sinh(\chi)\sin(\theta)\cos(\varphi)\\
     \sinh(\chi)\sin(\theta)\sin(\varphi)\\
     \sinh(\chi)\cos(\theta)
    \end{pmatrix}\,,
\end{align}
for which the metric can be computed as
\be
    \begin{split}
    \d s^2&=-R^2\sinh^3(\tau)\d\tau^2+R^2\sinh(\tau)\cosh^2(\tau)\d\Sigma^2\\
    &=-\d t^2+a^2(t)\d\Sigma^2\,.
    \end{split}
\ee
Here, $\d\Sigma^2=\d\chi^2+\sinh^2(\chi)\d\Omega^2$ is the metric on the unit hyperboloid $H^3$; and the scale parameter $a(t)$ is determined by
\begin{subequations}
\begin{align}
    a(t)^2&=R^2\sinh(\tau)\cosh^2(\tau)\,,\\
    \d t&=R\sinh(\tau)^{\frac{3}{2}}\d\tau\,.
\end{align}
\end{subequations}
One can show that around
\begin{subequations}
\begin{align}
    &\text{early time} &: && &a(t)\sim t^{1/5}\,,\\
    &\text{late time} &: && &a(t)\sim t\,.
\end{align}
\end{subequations}
The above features a FLRW cosmology that is asymptotically coasting at late times, and has a Big Bounce at the initial time $t=0$ since the timeline changes its direction as it `jumps' from one sheet of $\cM^{1,3}$ to the other \cite{Karczmarek:2022ejn,Battista:2022hqn}. 
%%%%%%%%%%%%%%%%%%%%%%%%%%%%%%%%%%%%

%%%%%%%%%%%%%%%%%%%%%%%%%%%%%%%%%%%%%%%%%%%%%%%%%%%%%%%%%%%%%%%%%%%%%%%%%%%%%%%%%%%%
 \section{Fuzzy twistor geometry and 4-hyperboloid}\label{sec:3}

We now study the spinorial versions of $H_N^4$ in  a stereographic projection which naturally admits a smooth flat limit. The spinorial effective vielbein and metric are also derived. We begin with a brief review of twistor geometry following \cite{Tran:2022mlu}.
%%%%%%%%%%%%%%%%%%%%%%%%%%%%%%%%%%%%%%%%%%
\subsection{Twistor space}
We define the commutative twistor space $\PT$ to be an open subset of $\P^{1,2}$ (for a review, see e.g. \cite{Adamo:2017qyl} and \cite{Penrose:1985bww,Penrose:1986ca,Ward:1990vs,Krasnov:2020lku})
\begin{align}
\label{twistor-def}
    \PT=\{Z^A=(Z^1,Z^2,Z^3,Z^4)=(\lambda^{\alpha},\mu^{\dot\alpha})\in \P^{1,2}\,|\, \hat \cN\neq 0\}\, .
\end{align}
Here $Z^A\sim r\,Z^A\,,\ \forall r\in \C^*$ are homogeneous coordinates of $\P^{1,2}$, and $\lambda^{\alpha},\mu^{\dot\alpha}$ 
transform in the fundamental of the compact subgroups $SU(2)_L \times SU(2)_R$ of $SU(2,2)$. Note that these are {\em not} Weyl spinors, as any non-compact transformation 
(such as boosts in $SO(1,3)$) will mix these two spinors.
%\textcolor{blue}{this explains some strange formulas later}
The complex conjugation of the twistor $Z^{A}$ denoted as $Z_{A}^{\dagger}$:
\begin{align}
Z_{A}^\dagger&= \begin{pmatrix}
    {\bar\lambda}_{\alpha} \\
{\bar\mu}_{\dot\alpha}
\end{pmatrix}\, 
\end{align}
transforms in the anti-fundamental representation of $\msu(2,2)$ so that the Dirac conjugation of $Z^A$ reads
\begin{align}
    \bar{Z}_A=Z^{\dagger}_B(\gamma_0)^B{}_A=\binom{\bar{\lambda}_{\alpha}}{-\bar{\mu}_{\dot\alpha}}\,.
\end{align}
Restricting ourselves to $\mso(1,4)\subset \msu(2,2)$, we 
can also use the anti-symmetric matrix $C^{AB}$
to define the `quaternionic' conjugate twistor $\hat{Z}^{A}$ of $Z^{A}$ as
\begin{align}
    \hat{Z}^{A}&=(\hat{\lambda}^{\alpha},\hat{\mu}^{\dot\alpha})
    =Z_{B}^{\dagger}C^{A B}\,,
\end{align}
which is manifestly compatible with $SO(1,4)$. Note that the hat conjugation acts on spinors with the following rules
\begin{align}
    \lambda^{\alpha} &=(\lambda^0,\lambda^1)\mapsto \hat{\lambda}^{\alpha}
    = \overline{\lambda}_\beta \varepsilon^{\alpha\beta}=(\overline{\lambda_1},-\overline{\lambda_0})\,,\nonumber \\ 
    \mu^{\dot\alpha} &=(\mu^{\dot 0},\mu^{\dot 1})\mapsto \hat{\mu}^{\dot\alpha}=\overline{\mu}_{\dot\beta} \varepsilon^{\dot\alpha\dot\beta}=(\overline{\mu_{\dot 1}},-\overline{\mu_{\dot 0}})\,.
\end{align}
It is useful to check that
\begin{subequations}
    \begin{align}
    \langle \lambda\,\bar{\lambda}\rangle &= \lambda_{\beta}\eps^{\alpha\beta}\bar\lambda_{\alpha}=-\lambda_{\beta}\hat\lambda^{\beta}=\langle \lambda\,\hat\lambda\rangle\,,\\
    \hat{\hat\lambda}^{\alpha}&=-\lambda^{\alpha}\,,\\
    \hat{\hat\mu}^{\dot\alpha}&=-\mu^{\dot\alpha}\,.
\end{align}
\end{subequations}
This means at the level of spinors, we can interchangeably use $(\hat\lambda,\hat\mu)$ for $(\bar{\lambda},\bar{\mu}$). In terms of the hatted spinors, the number operator $\hat\cN=\bar{Z}_AZ^A$ becomes
\begin{align}\label{eq:Ninspinors}
     \hat\cN=\langle \lambda\,\hat\lambda\rangle -[\mu\,\hat\mu]=N\,,
\end{align}
where the angle and square brackets are defined by $\langle u \,v\rangle = u^{\alpha}v_{\alpha}\,,\ [u\,v]=u^{\dot\alpha}v_{\dot\alpha}$. 

The correspondence between $\PT$ and $H^4$ is given by the incidence relations 
\begin{align}\label{eq:incident}
    \mu^{\dot\alpha}= \tx^{\alpha\dot\alpha}\lambda_{\alpha}\qquad \Leftrightarrow \qquad \tx^{\alpha\dot\alpha}=\frac{\lambda^{\alpha}\hat\mu^{\dot\alpha}-\hat\lambda^{\alpha}\mu^{\dot\alpha}}{\langle \lambda\,\hat\lambda\rangle}\,,
\end{align}
 which state that a point $\tx\in H^4$ corresponds to a holomorphic embedded Riemann sphere. We emphasize that all spinors and the $2\times 2$ matrix $\tx^{\alpha\dot\alpha}$ are dimensionless \cite{Steinacker:2022jjv}. Furthermore, we have the following reality condition:
\begin{align}\label{reality-x}
    \hat \tx^{\alpha\dot\alpha}=\tx^{\alpha\dot\alpha}\,.
\end{align}
%\textcolor{blue}{good. But the hat is tiny ... maybe widehat? but is ok, no problem}
The Poisson structure \eqref{eq:commZ} in terms of spinors read:
\begin{align}\label{eq:Poissonspinors}
    \{\lambda^{\alpha},\bar{\lambda}_{\beta}\}=\delta^{\alpha}{}_{\beta}\,,\qquad \qquad \{\mu^{\dot\alpha},\bar{\mu}_{\dot\beta}\}=\delta^{\dot\alpha}{}_{\dot\beta}\,,
\end{align}
which leads us to the following gradations:
\begin{subequations}
\begin{align}
    \{\hat{\cN},\lambda^{\alpha}\}&=-\lambda^{\alpha}\,,\qquad \qquad  \{\hat{\cN},\mu^{\dot\alpha}\}=-\mu^{\dot\alpha}\,,\\
    \{\hat{\cN},\bar\lambda_{\alpha}\}&=+\bar\lambda_{\alpha}\,,\qquad \qquad  \{\hat{\cN},\bar\mu_{\dot\alpha}\}=+\bar\mu_{\dot\alpha}\,,\\
    \{\hat\cN,\tx^{\alpha\dot\alpha}\}&=0\,.
\end{align}
\end{subequations}
The algebra of functions \eqref{fuzzyP12} on $\P^{1,2}$ in this language becomes:
\begin{align}
    \Ccurl(\P^{1,2})=\sum_{n+m=p+q}f_{\alpha(n)\dot\alpha(m)}{}^{\beta(p)\dot\beta(q)}\lambda^{\alpha(n)}\mu^{\dot\alpha(m)}\bar{\lambda}_{\beta(p)}\bar{\mu}_{\dot\beta(q)}\,,
\end{align}
which can be reduced further to polynomials in terms of $\lambda,\hat\lambda$:
\begin{align}
    \Ccurl(\P^{1,2})=\sum_nf_{\alpha(n)}{}^{\beta(n)}(\tx)\lambda^{\alpha(n)}\bar{\lambda}_{\beta(n)}\,
\end{align}
by using the incident relation \eqref{eq:incident}. Due to the constraint $\{\hat\cN,f\}=0$ where $f\in \Ccurl(\P^{1,2})$, the number of $\lambda$ and $\bar\lambda$ generators  must be the same at this point.
Therefore, all higher-spin modes can be viewed as functions on the internal (fuzzy) Riemann sphere $\P^1_N$.
%%%%%%%%%%%%%%%%%%%%%%%%%%%%%%%%%

%%%%%%%%%%%%%%%%%%%%%%%%%%%%%%%
\subsection{The stereographic projection to \texorpdfstring{$H^4$}{H4}} 

Following the lines of \cite{Steinacker:2022jjv}, we can recover the above twistorial construction and the incidence relations from the Hopf map \eqref{Hopf1}
$\P^{1,2} \to H^4$ (or Euclidean AdS${}_4$) followed by a stereographic projection $H^4 \to \R^4$.
 Using \eqref{Hopf1}, we get
\begin{subequations}\label{Yhopfmap}
\begin{align}
    y^{\ta}&=\frac{\ell_p}{2}\bar{Z}_A(\gamma^{\ta})^{A}{}_{B}Z^B=-\frac{\ell_p}{2}\langle \lambda\, \bar\lambda\rangle (\hat\sigma^{\ta})_{\alpha\dot\alpha}\tx^{\alpha\dot\alpha}\,,\qquad \quad \ta=1,2,3,4\,,\label{eq:yahopf}\\
    y^0&=\frac{\ell_p}{2}\bar{Z}_A(\gamma^0)^{A}{}_{B}Z^B=+\frac{\ell_p}{2}([\mu\,\bar\mu]+\langle \lambda\,\bar\lambda\rangle)
    % =-R+\ell_p\langle \lambda\,\hat\lambda\rangle\,  \nn\\
    =R+\ell_p [\mu\,\bar\mu] \geq  R 
    \label{eq:y0hopf}
\end{align}
\end{subequations}
where $\hat{\sigma}^{\ta}_{\alpha\dot\alpha}=(\tilde\sigma^m_{\alpha\dot\alpha},\epsilon_{\alpha\dot\alpha})$ for $m=1,2,3$. Note that when the spinors $\mu^{\dot\alpha} = (0,0)$, we will be at the center $y^0 = R$ of the projection, reflecting the fact that the underlying doubleton irrep $\cH_N$ of $\mso(4,2)$ is a lowest-weight representation \cite{Sperling:2018xrm}. Furthermore, $y^0$ can be also written as 
\begin{align}
    y^0=-R+\ell_p\langle \lambda\,\bar\lambda\rangle=-R+\ell_p\langle \lambda\,\hat\lambda\rangle\,.
\end{align}
It is then convenient to define
\begin{align}
    x^{\ta}=\frac{R}{\ell_p\langle \lambda\,\bar\lambda\rangle}y^{\ta}\,,\qquad y^{\ta}=\Big(1+\frac{y^0}{R}\Big)x^{\ta}\,
\end{align}
for $\ta=1,2,3,4$.
We obtain
\begin{align}
\label{eq:Yparametrization}
    y^0=\frac{R(x^2+R^2)}{(R^2-x^2)}\,,\qquad y^{\ta}=\frac{2R^2x^{\ta}}{(R^2-x^2)}\,,\qquad  x^2=x_{\ta}x^{\ta}<R^2\, .
\end{align}
It is a simple matter to show that
\begin{align}
    -y_0^2+y_{\ta}y^{\ta}=-R^2\,,\qquad \ta=1,2,3,4\,,
\end{align}
so that  the  above $x^{\ta}$ define a stereographic projection $H^4 \to \R^4$.
The conformally flat metric corresponding to $H^4$ can then be obtained by the pullback:
\begin{align}\label{eq:EAdSmetric}
    \d s^2=\Big(\frac{\pl y^a}{\pl x^{\ta}}\frac{\pl y^b}{\pl x^{\tb}}\eta_{ab}\Big)\d x^{\ta}\d x^{\tb}:=g_{\ta\tb}\d x^{\ta}\d x^{\tb}=\frac{4R^4\eta_{\ta\tb}\d x^{\ta}\d x^{\tb}}{(R^2-x^2)^2}=\Omega^2\eta_{\ta\tb}\d x^{\ta}\d x^{\tb} \,
\end{align}
thus defining the conformal factor $\Omega^2$,
where $\eta_{\ta\tb}=\diag(+,+,+,+)$.
 Although this coordinate system does not give us the desired Lorentzian signature, the metric \eqref{eq:EAdSmetric} can have a smooth flat limit where $R\rightarrow \infty$. Lastly, it is worth noting that \eqref{eq:y0hopf} implies
\begin{subequations}
     \begin{align}
    \langle \lambda\,\hat\lambda\rangle &= \frac{N}{2}\Big(1+\frac{y^0}{R}\Big)=\frac{NR^2}{(R^2-x^2)}=\frac{N}{2}\Omega\,,\\
    [\mu\,\hat{\mu}]&=\frac{Nx^2}{2(R^2-x^2)}\,,
\end{align}
\end{subequations}
%\textcolor{blue}{there is an issue with signs, cf. \eqref{eq:y0hopf}, }
which allows us to parametrize $\lambda^{\alpha},\hat{\lambda}^{\alpha}$ fiber coordinates as
\begin{align}\label{lambdaparametrization}
    \lambda^{\alpha}:=\frac{R}{\sqrt{R^2-x^2}}\binom{z}{-1}\,,\qquad \hat{\lambda}^{\alpha}:=\frac{ R}{\sqrt{R^2-x^2}}\binom{+1}{\bar{z}}\,,
\end{align}
where $1+|z|^2=N$ for $z\in \C^*$.

%%%%%%%%%%%%%%%%
\paragraph{Effective vielbein and metric.}

Using the explicit form of the gamma matrices we recover the 
incidence relation \eqref{eq:incident}, which takes a simpler form in terms of 
\begin{align}
    \ty^{\alpha\dot\alpha}:=\langle \lambda\,\hat\lambda\rangle \tx^{\alpha\dot\alpha}=\lambda^{\alpha}\hat{\mu}^{\dot\alpha}-\hat{\lambda}^{\alpha}\mu^{\dot\alpha}\,,\qquad \ty^0:=\langle \lambda\,\hat\lambda\rangle\, .
\end{align}
The reality condition of $\ty^{\alpha\dot\alpha}$, which is $\hat \ty^{\alpha\dot\alpha}=\ty^{\alpha\dot\alpha}$, follows directly from \eqref{reality-x} since $\hat{\hat\lambda}=-\lambda$ and $\hat{\hat\mu}=-\mu$.

It is convenient to work with the projective spinor bundle $\P^{1,2}\simeq \PS=\P^1\times \R^4$ where $\PS$ has coordinates $(\lambda^{\alpha},\ty^{\alpha\dot\alpha})$ and is a trivial bundle \cite{Steinacker:2022jjv}. Note that we can make a conformal transformation to  recover $H^4$ whenever it is appropriate. From this consideration, the algebra of functions $\Ccurl(\P^{1,2})$ reduces to:
\begin{align}\label{algebraPS}
    \Ccurl(\PS)=\Ccurl(\P^1)\times C^{\infty}(\R^4)\,.
\end{align}

As in \cite{Steinacker:2022jjv}, we will consider $H^4_N$ as a background in the IKKT model via the matrix configuration given by the (fuzzy version of) $y^a$. Such a background defines an effective frame or vielbein via the Hamiltonian vector field $\{y^a,-\}$ \cite{Sperling:2018xrm}.
Using \eqref{eq:Poissonspinors}, we obtain the spinorial form of the effective vielbein in analogous to \cite{Steinacker:2022jjv}:
\begin{subequations}\label{effectivevielbeins1}
\begin{align}
    \cE^{\alpha\dot\alpha|\beta\dot\beta}&:=\{\ty^{\alpha\dot\alpha},\ty^{\beta\dot\beta}\}= 2(\lambda^{(\alpha}\hat{\lambda}^{\beta)}\epsilon^{\dot\alpha\dot\beta}+\mu^{(\dot\alpha}\hat{\mu}^{\dot\beta)}\epsilon^{\alpha\beta})\,,\label{Evielbein1}\\
    \cE^{0|\alpha\dot\alpha}&:=\{\ty^0,\ty^{\alpha\dot\alpha}\}=\hat{\lambda}^{\alpha}\mu^{\dot\alpha}+\lambda^{\alpha}\hat{\mu}^{\dot\alpha}\,.\label{Evielbein2}
\end{align}
\end{subequations}
We also find
\begin{subequations}\label{eq:Poissononcoordinates2}
\begin{align}
     \{\ty^{\alpha\dot\alpha},\lambda^{\beta}\}=-\epsilon^{\alpha\beta}\mu^{\dot\alpha}\,,\qquad \qquad 
    \{\ty^{\alpha\dot\alpha},\hat{\lambda}^{\beta}\}=+\epsilon^{\alpha\beta}\hat{\mu}^{\dot\alpha}\,.
\end{align}
\end{subequations}
If $\varphi(\ty)$ is a $\hs$-valued smooth section of $\Ccurl(\PS)$, then
\begin{align}\label{nabla-def-1}
    \begin{split}
    \{\ty^{\alpha\dot\alpha},\varphi(\ty|\lambda,\hat\lambda)\}:&=\big(\{\ty^{\alpha\dot\alpha},\ty^{\beta\dot\beta}\}\pl_{\beta\dot\beta}+\{\ty^{\alpha\dot\alpha},\lambda^{\beta}\}\pl_{\beta}+\{\ty^{\alpha\dot\alpha},\hat{\lambda}^{\beta}\}\hat{\pl}_{\beta}\big)\varphi\,\\
    &=\big(\cE^{\alpha\dot\alpha|\beta\dot\beta}\pl_{\beta\dot\beta}+\cE^{\alpha\dot\alpha|\beta}\pl_{\beta}+\hat{\cE}^{\alpha\dot\alpha|\beta}\hat{\pl}_{\beta}\big)\varphi\,,
    \end{split}
\end{align}
where $\partial_{\alpha\dot\alpha}:= \p/\p \ty^{\alpha\dot\alpha}$, $\p_{\alpha}:=\p/\p \lambda^{\alpha}$ and $\hat\p_{\alpha}:=\p/\p \hat{\lambda}^{\alpha}$. Similarly, we define
\begin{align}\label{eq:0thderivation}
    \begin{split}
    \{\ty^0,\varphi(\ty|\lambda,\hat\lambda)\}:&=\big(\{\ty^0,\ty^{\beta\dot\beta}\}\pl_{\beta\dot\beta}+\{\ty^0,\lambda^{\alpha}\}\pl_{\alpha}+\{\ty^0,\hat{\lambda}^{\alpha}\}\hat\pl_{\alpha}\big)\varphi\,\\
    &=\big(\cE^{0|\beta\dot\beta}\pl_{\beta\dot\beta}+\cE^{0|\beta}\pl_{\beta}+\hat{\cE}^{0|\beta}\hat\pl_{\beta}\big)\varphi\,.
    \end{split}
\end{align}
Note that all contributions from $\cE^{\alpha\dot\alpha|\beta}\,,\,\hat{\cE}^{\alpha\dot\alpha|\beta}$ and $\cE^{0|\beta}\,,\,\hat{\cE}^{0|\beta}$ are subleading in the flat limit as in the case of $S^4_N$ (cf., \cite{Steinacker:2022jjv}). Therefore, whenever we take a flat limit, they can be neglected. Finally, to compute effective metric, say $G^{\alpha\dot\alpha\beta\dot\beta}$, it is sufficient to consider a scalar field $\vartheta(\ty)$ whose kinetic Lagrangian reads
\begin{align}\label{eq:preeffectivemetric}
    \begin{split}
    -\cE^{0|\alpha\dot\alpha}\pl_{\alpha\dot\alpha}\vartheta\,\cE_{0|\beta\dot\beta}\pl^{\beta\dot\beta}\vartheta+\cE^{\zeta\dot\zeta|\alpha\dot\alpha}\pl_{\alpha\dot\alpha}\vartheta\, \cE_{\zeta\dot\zeta|\beta\dot\beta}\pl^{\beta\dot\beta}\vartheta:=G^{\alpha\dot\alpha\beta\dot\beta}\pl_{\alpha\dot\alpha}\vartheta \,\pl_{\beta\dot\beta}\vartheta\,.
    \end{split}
\end{align}
Proceeding as in \cite{Steinacker:2022jjv}, we get
\begin{subequations}
\begin{align}
    G^{\alpha\dot\alpha\beta\dot\beta}(\ty)&=N^2\epsilon^{\alpha\beta}\epsilon^{\dot\alpha\dot\beta}-\ty^{\alpha\dot\alpha}\ty^{\beta\dot\beta}\,,\\
    G^{\alpha\dot\alpha\beta\dot\beta}(\tx)&=\langle \lambda\,\hat\lambda\rangle^2\Big(\frac{N^2}{\langle \lambda\,\hat \lambda \rangle^2}\epsilon^{\alpha\beta}\epsilon^{\dot\alpha\dot\beta}+\tx^{\alpha\dot\alpha}\tx^{\beta\dot\beta}\Big)\,,
\end{align}
\end{subequations}
which matches with the result of \cite{Steinacker:2016vgf}. 
Note that in the flat limit where we send the dimensionless ratio
$\frac{x^2}{R^2} \to 0$,
%\textcolor{blue}{I think this is a better characterization of the flat limit}
%$R\rightarrow \infty$ or equivalently $\langle \lambda\,\hat\lambda\rangle \gg N$, 
the effective metric $G^{\alpha\dot\alpha\beta\dot\beta}\mapsto N\,\eps^{\alpha\beta}\eps^{\dot\alpha\dot\beta}$, which is simply the standard metric of flat space in spinor form. 

%\textcolor{blue}{Btw it would be interesting to undertstand how translations are then implemented on the spinors, this would similarly perhaps allow to understand how conformal transformations (?) are recovered in the FLRW setting. Different project :)}
%{\bf Sure! I am happy to learn group theory from you.}

%%%%%%%%%%%%%%%%%%%%%%%%%%%%%%%%%%%%%
%%%%%%%%%%%%%%%%%%%%%%%%%%%%%%%%%%%%%%

%%%%%%%%%%%%%%%%%%%%%%%%%%%%%%%%%%%%%
%%%%%%%%%%%%%%%%%%%%%%%%%%%%%%%%%%%%%%
\section{Spinor description in the \texorpdfstring{$SO(1,3)$}{SO13}-invariant projection}\label{SO(1,3)spinors}

So far we have developed a natural spinor formalism for the fuzzy 4-hyperboloid along with a suitable stereographic projection to $H^4$ and then $\R^4$ (when we consider the flat limit), generalizing the framework in \cite{Steinacker:2022jjv}. However, the corresponding fields of the gauge theory resulting from the IKKT matrix model do not propagate, since the base manifold $H^4$ has all-plus (Euclidean) signature. This  can be circumvented by considering a different $SO(1,3)$-invariant projection \eqref{eq:M13projections} as in \cite{Sperling:2019xar}, which leads to a FLRW type  spacetime with Lorentzian signature.

Recall that the projection \eqref{eq:M13projections} is realized in the IKKT model by choosing the background to be (the fuzzy versions of) $t^{\mu}$ defined as
\begin{align}
    t^{\mu}=\frac{1}{R}\bar{Z}_A(\Sigma^{\mu 4})^A_{\ B}Z^B
    = \frac{1}{2R}\bar{Z}_A(\underline{\gamma}^{\mu})^A{}_B Z^B\,,\qquad \mu=0,1,2,3\,.
    %=\frac{1}{2R}Z^{\dagger}_A(\gamma_0)^A{}_B(\underline{\gamma}^{\mu})^B{}_CZ^C\,,\qquad \mu=0,1,2,3\,.
\end{align}
Furthermore, as discussed in Section \ref{sec:3}, the spinors $\lambda^\alpha,\mu^{\dot\alpha}$  are not Lorentz spinors here, but 
spinors of the compact subgroups $SU(2)_L$ and $SU(2)_R$ of $SU(2,2)$ which are mixed by the $SO(1,3)$ generators $\Sigma^{\mu\nu}$.
Using the gamma matrices \eqref{gammaSO(1,3)}, we obtain explicit spinorial expressions for $t^\mu$:
\begin{subequations}\label{t-description-spinor}
    \begin{align}
 %   \begin{split}
    t^0&=\frac{1}{4R}\big( [\mu\,\bar\lambda]  + \langle \lambda\,\bar\mu\rangle   
      \big) \equiv \frac{1}{4R}\big( \mu^\dagger\lambda  + \lambda^\dagger \mu 
      \big)   \\
    t^{i}&=\frac{1}{4R}\Big(\bar\lambda_{\alpha}(\sigma^i)^{\alpha}_{\ \beta}\lambda^{\beta}+\bar{\mu}_{\dot\alpha}(\sigma^i)^{\dot\alpha}{}_{ \dot\beta}\mu^{\dot\beta}\Big)\,,\qquad i=1,2,3\,.
   % \end{split}
\end{align}
\end{subequations}
%\textcolor{blue}{I prefer the explicit form for $t^0$ since then the argument below is more transaprent)}
This leads to the following Poisson brackets  
\begin{subequations}\label{PoissonFLRW}
\begin{align}
    \{t^i,\lambda^{\alpha}\}&=+\frac{\im}{2R}(\sigma^i)^{\alpha\beta}\lambda_{\beta}\,,\qquad &\{t^i,\hat\lambda^{\alpha}\}&=-\frac{\im}{2R}\hat\lambda_{\beta}(\sigma^i)^{\beta\alpha}\,,\\
    \{t^i,\mu^{\dot\alpha}\}&=-\frac{\im}{2R}(\sigma^i)^{\dot\alpha\dot\beta}\mu_{\dot\beta}\,,\qquad &\{t^i,\hat\mu^{\dot\alpha}\}&=+\frac{\im}{2R}\hat\mu_{\dot\beta}(\sigma^i)^{\dot\beta\dot\alpha}\,.
    \end{align}
\end{subequations}
We also provide the explicit spinorial expressions for the $y^\mu$ (cf. \eqref{Yhopfmap}):
\begin{align}
\label{y-i-spinor-FLRW}
    y^i &= -\frac{1}{4R}\big(\bar{\lambda}_{\alpha} (\sigma^i)^{\alpha}{}_{\dot\beta} \mu^{\dot\beta}  + \bar{\mu}_{\dot\alpha}(\sigma^i)^{\dot\alpha}{}_{\beta}\lambda^{\beta}
      \big) 
       \equiv -\frac{1}{4R}\big(\lambda^\dagger \sigma^i \mu + \mu^\dagger \sigma^i\lambda  \big) \\
%      \big) 
  y^0 &= R+\ell_p [\mu\,\bar\mu] 
\end{align}
%Note that $y^a=0$ implies $\lambda \simeq \bar\lambda$ cf \eqref{eq:yahopf}.
%\textcolor{blue}{I think this is only at the south pole, which is too special in the FLRW case}
%{\bf Yes, this can't work globally. But for our purpose it is enough. }
%This means that around $y^i=0$, $\mu\simeq \bar{\mu}$ so that $\hat\cN=\langle \lambda \bar\lambda\rangle -[\mu\bar\mu]=0$ and $Z^{\dagger}\simeq Z$. So $\bar{Z}_A= (\lambda_{\alpha},-\mu_{\dot\alpha})^T$. It also seems like we are working with 3d spinors at this point so there is no different between dotted and undotted indices? So we may write
%\begin{align}
%    t^0=\frac{1}{R}Z_A(\underline{\gamma}^0)^A{}_BZ^B=\frac{1}{4R}(\bar{c}_1\lambda \mu -\bar{c}_2\mu \lambda)  \,.
%\end{align}
%Thus, $t^0$ implies $\lambda\simeq \mu$. Is this roughly what you want to say?
%In principle, it is possible to express also the 
% space-time generators $y^\mu$ 
%and any vector fields in terms of spinors
% using \eqref{y-i-spinor-FLRW}.

It is clear from these formulas that the spinorial 
representation of coordinates and derivatives on the FLRW background
is quite distinct -- and perhaps less appealing --
than the familiar form on Minkowski space, because our spinors transform under the space-like isometry group $SU(2)\times SU(2)\subset SO(1,4)$ rather than the local $SL(2,\C)$. This reflects the lack of manifest local Lorentz invariance in the model, which is expected to be recovered only effectively for the physical fields.  We delegate 
a full treatment this problem to future work, and avoid this step in the following. However, we shall illustrate how
the local relativistic propagation is recovered properly in this spinorial setting.

\paragraph{Effective vielbein and metric.}

To derive the effective frame and metric on the FLRW background\footnote{Note that the present background $t^\mu$
play the role of momentum generators, while the $y^a$ in the case of $H^4$ are position generators. Nevertheless, both backgrounds define a higher-spin gauge theory on the respective base manifold.}
$\cM^{1,3}$ defined by $t^\mu$ in the spinorial framework, we
need an explicit realization of general $\hs$-valued fields on $\cM^{1,3}$ in terms of spinors. 
We have just seen that in contrast to the previous case of $H^4$, it is no longer possible to define the $t^{\mu}$ globally in terms of a single pair of spinors. 
However since we are mainly interested in local scattering amplitudes,  we can choose (using the $SO(1,3)$ isometry of the FLRW background) some reference point $\mathsf{p}= (p^0,0,0,0)\in \cM^{1,3}$. Then $t^0=0$ and $t^{\mu}|_{\mathsf{p}}\rightsquigarrow t^i$ spans the local $\P^1$, which can be described
effectively  in terms of spinors $\lambda^\alpha$. 
To see this, the following observation is useful:
combining the above expression for $t^0$
with  \eqref{y-i-spinor-FLRW}
and recalling that $t^0 = 0$ at the reference point $y^\mu = (y^0,0,0,0)$, it follows that 
\begin{align}
    \lambda^\dagger \sigma^\mu \mu = -\mu^\dagger \sigma^\mu\lambda 
\end{align}
%or
%\begin{align}
%   \bar{\lambda}_{\alpha} (\sigma^i)^{\alpha}{}_{\dot\beta} \mu^{\dot\beta} = \bar{\mu}_{\dot\alpha}(\sigma^i)^{\dot\alpha}{}_{\beta}\lambda^{\beta}
%\end{align}
which implies $\lambda^\dagger\otimes \mu = -\mu^\dagger \otimes \lambda$
and therefore $\lambda \propto \mu$.
As a consequence, the generator of the local $\P^1$
\begin{align}
\label{ti-lambda-relation-M31}
    t^i = \lambda^\dagger \sigma^i \lambda + \mu^\dagger \sigma^i \mu \propto\lambda^\dagger \sigma^i \lambda    
\end{align}
can be expressed in terms of $\lambda$ only. 
Hence we can choose an open subset $U_{\mathsf{p}}\subset \cM^{1,3}$ around the reference point $\tp$, with 
local trivialization $\P^{1,2}|_{U_{\mathsf{p}}}=\P^1\times U_{\mathsf{p}}$, and 
write the most general function as (cf. \eqref{mode-expansion-t-lambda})
\begin{equation}
    \begin{split}
     \label{varphi-local-M31}
\varphi&= \varphi(y|t) =
\sum_{s=0}^{\infty}t^{i(s)}\varphi_{i(s)}\simeq\sum_{s=0}^{\infty}\lambda^{\beta(s)}\hat\lambda^{\beta(s)}\varphi_{\beta(2s)}\, .
    \end{split}
\end{equation}
This provides a spinorial representation for the $\hs$ modes on the FLRW background.
Now consider \begin{align}
    \begin{split}
    \{t^{\mu},\varphi(y|\lambda,\hat\lambda)|_{U_{\tp}}\}&=\Big(\{t^{\mu},y^{\nu}\}\frac{\p}{\p y^{\nu}}+\{t^{\mu},\lambda^{\alpha}\}\frac{\partial}{\p \lambda^{\alpha}}+\{t^{\mu},\hat\lambda^{\alpha}\}\frac{\partial}{\p \hat\lambda^{\alpha}}\Big)\varphi(y|\lambda,\hat\lambda)|_{U_{\tp}}\\
    &=:\Big(E^{\mu\nu}\frac{\p}{\p y^{\nu}}+E^{\mu|\alpha}\frac{\p}{\p \lambda^{\alpha}}+\hat{E}^{\mu|\alpha}\frac{\p}{\p\hat\lambda^{\alpha}}\Big)\varphi(y|\lambda,\hat\lambda)|_{U_{\tp}}\,,
    \end{split}
\end{align}
where $\varphi(y|\lambda,\hat\lambda)|_{U_{\tp}}\in \Ccurl(\P^{1,2})|_{U_{\tp}}$ and $E^{\mu|\bullet}$ is the effective vielbein in this coordinate. To compute effective metric of FLRW patch, it is sufficient to consider a scalar field $\varphi(y)$ and the kinetic term $\{t^{\mu},\varphi(y)\}\{t_{\mu},\varphi(y)\}$. We obtain\footnote{Strictly speaking the effective metric involves an extra conformal factor given by the dilaton, which we can assume to be constant on the local patch.}
\begin{align}\label{effectivemetricFLRW}
    \{t^{\mu},\varphi(y)\}\{t_{\mu},\varphi(y)\}:=\gamma^{\mu\nu}\p_{\mu}\varphi(y)\p_{\nu}\varphi(y)\,,\qquad \gamma^{\mu\nu}=\eta^{\mu\nu}\sinh^2(\tau)\,.
\end{align}

%transform $\eta^{\mu\nu}$ to $\eps^{\alpha\beta}\eps^{\dot\alpha\dot\beta}$. 

%\textcolor{blue}{why call this $\vartheta$? I'd call it $\varphi$}

\paragraph{Flat limit.} In the limit where $R\rightarrow \infty$, it is obvious that 
\begin{align}
    E^{\mu\nu}=\eta^{\mu\nu}\sinh(\tau)
\end{align}
 will be the only effective vielbein that survives. Since contributions resulting from  the Poisson bracket \eqref{eq:Poissonspinors} on fiber coordinates $\lambda,\bar\lambda$ are subleading in the flat limit (except in the extreme IR regime), they can be neglected to a good approximation; this is denoted as ``asymptotic regime" in \cite{Steinacker:2020xph}. Hence, in the flat limit, we simply factorize all fiber coordinates $\lambda,\bar\lambda$ outside the Poisson brackets, which implies that the Poisson bracket only acts on functions in the flat limit. This is also the limit where we can effectively replace $t^i$ for a pair of spinors $\lambda^{\alpha},\bar\lambda^{\beta}$ as discussed above.

%%%%%%%%%%%%%%%%%%%%%%%%%%%%%%%%%%%%%%

%%%%%%%%%%%%%%%%%%
%%%%%%%%%%%%%%%%%%%%%%%%%%%%%

%%%%%%%%%%%%%%%%%%%%%%%%%%%%%%
\section{Vector description of higher-spin modes in the flat limit}\label{sec:shift-vector}

Before analyzing higher-spin modes on semi-classical $\P^{1,2}_N$ in the spinor formalism, it is worthwhile to work out the vector description using $t^{\ta}$ and $t^{\mu}$ as  generators of the internal $S^2_N$ to describe higher-spin modes in the flat limit. The goal of this section is to simplify some of the results obtained using the group theory approach in \cite{Sperling:2018xrm}. 

If we consider $H^4$ and $\cM^{1,3}$ in the flat limit, then higher-spin valued functions can be parametrized by the internal generators $t^{\ta}$ and $t^{\mu}$
as: 
\begin{equation}
\begin{split}
 \phi &= \phi_{\ta(s)}(y) t^{\ta(s)}\,,  \qquad \, \phi\in \Ccurl(\R^4\times S^2)\,, \nn\\
       \phi &= \phi_{\mu(s)}(y) t^{\mu(s)}\,, \qquad \phi\in \Ccurl(\R^{1,3}\times S^2)\,,
       \end{split}
\end{equation}
and we recall that $t^{\ta}$ is defined in 
\eqref{t-generators-flatH4} while $t^{\mu}$ is defined in \eqref{eq:M13projections}. The $\hs$-valued gauge potential $\cA_{\ta}$ and $\cA_{\mu}$ can be defined analogously as
\begin{align}
    \cA_{\ta}=\cA_{\tb(s)|\ta}t^{\tb(s)}\,,\qquad \cA_{\mu}=\cA_{\nu(s)|\mu}t^{\nu(s)}\,.
\end{align}
Since $t^{\ta}$ and $t^{\mu}$ are generators of $SO(3)$, we can use the Littlewood-Richardson rule to count independent components in $\cA_{\ta}$ or $\cA_{\mu}$, respectively. It is worth noting that even though HS-IKKT theory does not possess a mass parameter, its higher-spin modes should be viewed as ``would-be massive''. In the above representation, the tensors $\cA_{\nu(s)|\mu}$
are space-like (due to \eqref{eq:orthogonalofTY}) but not divergence-free\footnote{It is also possible to represent the modes in terms of divergence-free but not space-like tensors.}. For instance, $\cA_{\mu}$ has a total of $4(2s+1)$ off-shell degrees of freedom as previously shown in \cite{Sperling:2019xar}.

\paragraph{Gauge fixing and propagating dof.} To count the physical degrees of freedom in the higher-spin gauge potential $\cA$, we must impose a gauge fixing condition.\footnote{We shall not discuss the corresponding ghost sector explicitly here but refer the readers to \cite{Blaschke:2011qu}.}  
As usual in Yang-Mills matrix models (see e.g. \cite{Blaschke:2011qu}), a suitable gauge fixing function on a fluctuating background $Y^{\boldsymbol{I}} + \cA^{\boldsymbol{I}}$ is given by 
\begin{align}
\cG(\cA) = \{y_{\boldsymbol{I}},\cA^{\boldsymbol{I}}\} \ .
 \label{gaugefix-intertwiner}
 \end{align}
This is a good choice because
 there is always a gauge  such that
$\cG(\cA) = 0$, provided  $\Box = \{y_{\boldsymbol{I}},\{y^{\boldsymbol{I}},\cdot\}\}$ is surjective. 
Note that an admissible/integrable fluctuation mode $\cA$
satisfies the gauge-fixing condition 
$\{y_{\boldsymbol{I}},\cA^{\boldsymbol{I}}\} = 0$ if and only if it is orthogonal to all pure gauge modes $\cA^{(g)}$, i.e.
\begin{align}
 \langle \cA^{(g)},\cA\rangle = 0 \ , 
 \qquad  \cA^{(g)}[\xi] = \{y^{\boldsymbol{I}},\xi\} \ .
 \label{gaugefix-inner}
\end{align}
for $\xi \in \Ccurl(\cM)$. 
Since the gauge fixing condition removes $2s+1$ modes, the gauge-fixed potential $\cA_{\tb(s)|\ta}$ has $3(2s+1)$ degrees of freedom.
Moreover, we can use \eqref{gaugefix-inner} to remove further $(2s+1)$ on-shell pure gauge components of $\cA_{\nu(s)|\mu}$ on the Lorentzian FLRW background. As a result, 
%the Euclidean higher-spin valued gauge potential $\cA_{\tb(s)|\ta}$ has $3(2s+1)$ dof. while 
the Lorentzian $\cA_{\nu(s)|\mu}$ has $2(2s+1)$ physical dof. Note that in contrast to more conventional gauge theories, these physical modes always include the extra dof of the would-be massive higher-spin fields.

\paragraph{Mode ansatz and decomposition.} In the flat limit,
we can further decompose the above into the following irreducible modes:
%\begin{equation}\label{mode-separation-tangential-vector}
%    \begin{split}
%    \cA_\ta &= \Big[A_{(\tb(s)\ta)}
%    + g_{\ta\tb} \sA_{\tb(s-1)} +\p_{\ta}\xi_{\tb(s)}+\p_{\tb}\xi_{\ta\tb(s-1)}\Big]t^{\tb(s)}\,, \\
%    \cA_\mu &= \Big[A_{(\nu(s)\mu)}
%    + g_{\mu\nu} \sA_{\nu(s-1)}
%    + \del_{\mu}\xi_{\nu(s)}+\del_{\nu}\xi_{\mu\nu(s-1)}\Big]t^{\nu(s)}\,,
%    \end{split}
%\end{equation}
%or rather 
\begin{equation}\label{mode-separation-tangential-vector}
    \begin{split}
    \cA_\ta &= \Big[A_{(\tb(s)\ta)}
    +  \delta_{\ta\tb}\sA_{\tb(s-1)}
    +y_{\ta}\xi_{\tb(s)}
    +\p_{\ta}\tilde\xi_{\tb(s)}
    \Big]t^{\tb(s)}\,, \\
    \cA_\mu &= \Big[A_{(\nu(s)\mu)}
   + \eta_{\mu\nu} \sA_{\nu(s-1)}
    + y_{\mu}\xi_{\nu(s)}
    + \del_{\mu}\tilde\xi_{\nu(s)}
    \Big] t^{\nu(s)}\, .
    \end{split}
\end{equation}
Note that the last mode is the pure gauge mode in the Minkowski case, and
the second mode can be written as 
\begin{align}
 \eta_{\mu\nu} \sA_{\nu(s-1)} t^{\nu(s)}
 \propto  t_{\mu} \sA_{\nu(s-1)} t^{\nu(s-1)} 
\end{align}
and similarly in the Euclidean case,
up to normalization.
This is simpler and more coherent than the organization of modes in the curved case \cite{Sperling:2018xrm,Steinacker:2019awe}, which should be useful to study the resulting physics. 
Since the $A_{\tb(s)\ta}$ coefficient has $2s+3$ components while $\sA_{\tb(s-1)}$ has $2s-1$ components, we recover precisely 
all 
$$(2s+3) + (2s-1) + 2 (2s+1) = 4(2s+1)$$ 
components of the $\hs$-valued gauge potential $\cA_\ta$ (or $\cA_{\mu}$, repsectively).\footnote{Recall that a massive spin-$s$ field in $4d$ has $2s+1$ degrees of freedom.} Furthermore, gauge-fixing removes %$2s+1$ degrees of freedom in the Euclidean case, and 
$2\times(2s+1)$ on-shell modes in Minkowski signature;
%\textcolor{blue}{I think is nonsense, always removes two dof. Let me think}. 
essentially, the higher-spin modes associated with the coefficients $\xi$ and $\tilde\xi$  will be removed by gauge fixing. To show that the two modes $A$ and $\sA$ are linearly independent, it suffices to show that their inner product matrix is non-degenerate, which we will verify for the spin 1 modes below.

%\textcolor{blue}{In fact we need to keep the  $Y(s,1)$ (hook) term. This arises via 
%$\theta^{\mu\nu} \sim \frac{\ell_{Pl}^2}{\cosh\eta}(y^\mu t^\nu - y^\nu t^\mu)$ in the Minkowski case, and probably similar in the Euclidean case (need to think). Then that mode is}
%$$
%\cA^\mu = \theta^{\mu\nu}\tilde A_\nu
%$$
%or better
%$$
%\cA^\mu = \theta^{\mu\nu}\p_\nu \phi^{(s)}
%\sim (t^\mu \cT - y^\mu D)\phi^{(s)}
%$$
%\textcolor{blue}{
%where  $D = t^\mu \del_\mu$.
%These provide independent modes, which are the $\cA^{\pm}$ modes in my previous papers. Together with one of the modes above this gives two spin $s$ modes. But this is even more tricky because they are in superpositions of $\cC^{s\pm 1}$, and we are faced with the complications in my previous paper with Marcus.
%The operators 
%\begin{align}
%    D^+ = t^\mu \del_\mu 
%\end{align}
%and $D^- = ...$  were useful.
%Maybe we could use the mode
%\begin{align}
%   B^\mu &\sim t^\mu \phi^{(s)} 
%\end{align}
%and/or $y^\mu \phi^{(s)}$ 
%which should provide an independent spin $s$ mode. This is nicer than the above. Gauge fixing is not clear though. }

\paragraph{Kinetic action in the flat limit.} Now, let us focus on the Lorentzian case and consider the kinetic actions of the $\cA_{\mu}$ mode \cite{Sperling:2017gmy}
\begin{align}
    S=\int \{y^{\mu},\cA^{\nu}\}\{y_{\mu},\cA_{\nu}\}+ 2\{y^{\mu},y^{\nu}\}\{\cA_{\mu},\cA_{\nu}\}-\{y_{\mu},\cA^{\mu}\}^2\,.
\end{align}
Using \eqref{projected-so(4,2)algebra}, we obtain for instance
%    \begin{align}
%    \{y^{\ta},\cA^{\tb}\}\{y_{\ta},\cA_{\tb}\}&=\theta^{\ta\tc}\p_{\tc}\cA^{\tb}\theta_{\ta\td}\p^{\td}\cA_{\tb}\simeq \p_{\tc}\cA^{\ta}\p^{\tc}\cA_{\ta}\,,
%\end{align}
%where we recall that $\theta^{\ta\tb}=-\ell_p m^{\ta\tb}\,,\ \theta^{\ta\tc}\theta_{\tb\tc}\simeq \delta^{\ta}{}_{\tb}$. 
%\textcolor{blue}{where else do you use that? this is the definition of $\theta_{\tb\tc}$? I use opposite sign conventions. BETTER:}
 \begin{align}
    \{y^{\mu},\cA^{\nu}\}\{y_{\mu},\cA_{\nu}\}&=
    E^{ \mu i} E_{\mu}^{\ j}\p_i \cA^{\nu}\p_{j}\cA_{\nu}\
    =:  \p_{i}\cA^{\mu}\p^{i}\cA_{\mu}\,,
\end{align}
Here, we rise and lower coordinate indices with 
\begin{align}
    \gamma^{ij} = \eta_{\mu\nu} E^{\mu i} E^{\nu j}\,,\qquad i,j=0,1,2,3\,, 
\end{align}
which we can consider as (locally constant) effective  metric in the flat limit. 
On the other hand, the term $\{y^{\mu},y^{\nu}\}\{\cA_{\mu},\cA_{\nu}\}$ is sub-leading for local scales where
$\{y^{\mu},y^{\nu}\}\sim \theta^{\mu\nu}$ can be treated as constant, i.e. for modes with wavelength much shorter than the curvature scale. This is certainly true in the flat limit, where  $\cA_{\mu} \{  \{y^{\mu},y^{\nu}\},\cA_{\nu}\}$ (after partial integration) is a one-derivative term with scale set by the geometric curvature, and therefore can be
neglected. This should be sufficient for the study of scattering amplitudes. Then the kinetic action without the gauge fixing term is simply the Box operator, i.e.
\begin{align}\label{Seff}
    S_{\text{eff}}=-\int  \cA_{\mu}\Box\cA^{\mu}\,,\qquad \Box:=\p_{i}\p^{i}
\end{align}
as it should, since $\{y_{\mu},\cA^{\mu}\}^2$ is removed upon gauge fixing.
%The second term of the above reads [This is not true in general]
%\begin{align}
%    -2\{y^{\ta},\cA^{\tb}\}\{y_{\tb},\cA_{\ta}\}=-2\theta^{\ta\tc}\p_{\tc}\cA^{\tb}\theta^{\tb\td}\p_{\td}\cA_{\ta}\simeq -\frac{2}{3}\p_{\tc}\cA_{\tb}\p^{\tc}\cA^{\tb}+\frac{1}{3}(\p_{\tc}\cA^{\tc})^2
%\end{align}
%while the first term can be combined with the gauge fixing term $\{y_{\ta},\cA^{\ta}\}^2$. Note that
%\begin{align}
%    \{y_{\ta},\cA^{\ta}\}^2=\frac{1}{3}\p_{\tc}\cA^{\ta}\p^{\tc}\cA_{\ta}-\frac{1}{3}(\p_{\tc}\cA^{\tc})^2\,.
%\end{align}
%\textcolor{blue}{careful - this is no longer true if the $\cA$ are $\hs$ valued! I dont think we should do that. We may consider some  regime where the $\{y^{\ta},y^{\tb}\}\{\cA_{\ta},\cA_{\tb}\}$ is negligible to simplify the expression. This first-derivative term is only important in the IR regime, may me negligible for local considerations.}
%Here, the averaging over $\theta^{\ta\tb}$ denoted as $[\theta\theta]_0$ is given by the following rules in the flat limit:
%\begin{align}
%    [\theta^{\ta\tb}\theta^{\tc\td}]_0\simeq \frac{1}{3}\Big(\eta^{\ta\tc}\eta^{\tb\td}-\eta^{\ta\td}\eta^{\tb\tc}\Big)
%\end{align}
%Intriguingly, this can be combined with the standard kinetic term to form the kinetic effective action with no Poisson structure remaining. 
%Observe that we have transformed the gauge-fixing term $\{y_{\ta},\cA^{\ta}\}^2$ to $(\p_{\tc}\cA^{\tc})^2$. Using the mode expansion \eqref{mode-separation-tangential-vector}, the term $(\p_{\tc}\cA^{\tc})^2$ will allow us to remove the higher-spin modes $\xi_{\tb(s)}$ from $\cA_{\tb(s)|\ta}$. As such, 
It is then sufficient to analyze the physical modes using the expansion
\begin{align}\label{modeexpansioninter1}
\sa_{\mu}=\cA_{\nu(s)|\mu}t^{\nu(s)}=\Big(A_{(\nu(s)\mu)}+\eta_{\mu\nu}\sA_{\nu(s-1)}\Big)t^{\nu(s)}
\end{align}
where
$A_{(\nu(s)\mu)}$ and $\sA_{\nu(s-1)}$ are traceless,
after gauge fixing and removing pure gauge modes. This encodes the $2(2s+1)$ physical components of the $\hs$-valued gauge potential $\cA_{\mu}$. 
Since $t^\mu$ is space-like \eqref{eq:orthogonalofTY}, we can demand the tensors  $A_{(\nu(s)\mu)}$ and $\sA_{\nu(s-1)}$ to be not only traceless but also space-like\footnote{The time-like component of $\cA$ is contained in the third term in \eqref{mode-separation-tangential-vector} which is unphysical.}
\begin{align}\label{spacelikeconstraints}
    A_{\nu}\cT^{\nu} = 0 = \sA_{\nu}\cT^{\nu}\, ,
\end{align}
but not divergence-free\footnote{Note that since the fields are not divergence-free, one cannot introduce a shift-symmetry of type $\eta_{\mu\nu}\vartheta_{\nu(s-1)}$ to gauge away the second $\sA$ modes as in standard approach in higher-spin literature, see e.g. Section 3 of \cite{Ponomarev:2022vjb}.}.
This also shows that the resulting theory is  ghost-free (cf., \cite{Steinacker:2019awe}), even though (local) Lorentz invariance is not manifest, in accordance with the fact that the matrix model defines a preferred frame.

\paragraph{Averaging over the fiber.} Following the procedure in \cite{Steinacker:2019awe}, we can average over the fiber coordinates $t^{\nu}$ to obtain the spacetime kinetic action of higher-spin modes in vectorial description. We illustrate this for the  spin 1 fields, whose modes expansion \eqref{modeexpansioninter1} gives
    \begin{align}\label{effS1}
        S_{\text{eff}}
        &= -\int A_{\mu}\Box A^{\mu}
        +(A_{\mu}\Box\sA_{\sigma} + \sA_{\sigma}\Box A_{\mu})
        t^{\mu}t^{\sigma}
        + t_{\nu}t^{\nu} \sA_{\rho}\Box \sA_{\sigma}t^{\sigma}t^{\rho}\,  \nn\\
         &=  -\int A_{\mu}\Box A^{\mu}
        +  \frac{1}{3\ell_p^2}(A_{\mu}\Box\sA_{\sigma} +\sA_{\sigma}\Box A_{\mu})
      \kappa^{\mu\sigma}
        +\frac{\cosh^2(\tau)}{3\ell_p^4} \sA_{\rho}\Box \sA_{\sigma}\kappa^{\sigma\rho} \ 
    \end{align}
    using \eqref{S2sphereM13}, and
 averaging over fiber coordinates in the flat limit using 
\begin{align}
    [t^{\mu}t^{\nu}]_0 
    &= \frac{1}{3\ell_p^2}\kappa^{\mu\nu}\ .
\end{align}
 Here
\begin{align}\label{kappa}
  \kappa^{\mu\nu} &= \cosh^2(\tau)\eta^{\mu\nu} +\frac{{y^\mu}{y^{\nu}}}{R^2}
\end{align}
is the space-like projector orthogonal to the time-like FLRW vector field $\cT$ cf. \eqref{timelikeT}.
By virtue of the space-like constraints \eqref{spacelikeconstraints}, the second term in \eqref{kappa} will not contribute
upon choosing space-like tensors as discussed above\footnote{The restriction to time-like tensors might be avoided by absorbing the third mode  in \eqref{mode-separation-tangential-vector} into the first. This issue will be avoided in the spinorial approach below.}, and
the kinetic action takes the form
\begin{align}\label{matchpattern}
    S_{\text{eff}}=-\int A_{\mu}\Box A^{\mu}+\alpha A_{\mu}\Box \sA^{\mu}+\alpha\sA_{\mu}\Box A^{\mu}+\beta\sA^{\mu}\Box\sA_{\mu}\,,
\end{align}
where $\alpha=\cosh^2(\tau)/3\ell_p^2$ and $\beta=\cosh^4(\tau)/3\ell_p^4$. 
The inner product of the modes $(A,\tilde A)$ is obtained similarly, dropping the $\Box$.
This boils down to the matrix 
\begin{align}
   M= \begin{pmatrix}
        1 & \alpha \\
        \alpha & \beta
    \end{pmatrix}
\end{align}
with non-vanishing determinant
  $\det M=\frac{2\cosh^4(\tau)}{9\ell_p^4} \neq 0$.
 Therefore these two modes are independent, but not orthogonal\footnote{For a similar analysis of the full fluctuation spectrum on the curved FLRW background see \cite{Steinacker:2019awe}.}.

%We conclude even though the space-like tensors $A$ and $\sA$ constitute two independent set of modes, they are not orthogonal from each others. 

While the analysis can be continued also at non-linear level, we find it cumbersome to integrate out fiber coordinates using vectorial description. This problem can be simplified by using spinorial formalism (cf., Sections \ref{sec:3} and \ref{SO(1,3)spinors} and Appendix \ref{app:2}). In particular, when the two higher-spin modes $A$ and $\sA$ written in spinoral language will completely disentangle with each others in the flat limit. As a result, we can make a field redefinition to bring them into a ``helicity'' base and perform scattering amplitude calculation.

%\textcolor{blue}{Now section 5 is good. Note that the tensors are not divergence-free, which explains the extra dof}

%%%%%%%%%%%%%%%%%%%%%%%%%%%%%%%%%%
\section{Spinorial reduction to 4 dimensions}\label{sec:5}

In  this section we rewrite the IKKT-matrix model in spinorial language and obtain its spacetime actions for the massless sector from two aforementioned projections of $H^4_N$. Note that unlike \cite{Steinacker:2022jjv}, we will study the IKKT-matrix model in second-order formalism as in 
\cite{Sperling:2018xrm}. Moreover, while the stereographic projection allows us to obtain the spacetime action of HS-IKKT in a global manner, reducing the twistor action to spacetime one in $SO(1,3)$-invariant projection can only be achieved locally, i.e. around certain reference point $\tp\in \cM$ where we can set the temporal component $t^0$ of the momentum $t^{\mu}$ to zero. 
%%%%%%%%%%%%%%%%%%%%%%%%%%%%%%%%%%%%%%%%

%%%%%%%%%%%%%%%%%%%%%%%%%%%%%%%%%%
\subsection{Reduced Euclidean action on \texorpdfstring{$H^4$}{H4}}

We start with the $SO(1,4)$-invariant action on the Euclidean hyperboloid $H^4_N$, which is embedded along the first $1+4$ coordinates of the target space.
The chiral basis of $\mso(1,4)$ $\gamma$-matrices in \eqref{gamma-so5-explicit} allows us to cast $y^a$ into spinorial form:
\begin{align}\label{eq:YABdecomposition}
    y^{AB}=q^{AB}+p^{AB}=\begin{pmatrix}
     q^{\alpha\beta} & 0\\
     0& q^{\dot\alpha\dot\beta}
    \end{pmatrix}+\begin{pmatrix}
     0 & p^{\alpha\dot\beta}\\
     -p^{\dot\beta\alpha}&0
    \end{pmatrix}\,
\end{align}
where $q^{AB}$ represents the zeroth direction associated to $\gamma_0$ matrix. Since the external symmetry group $SO(1,4)$ acts on $q^0$, supersymmetry will be broken by the background,
%\textcolor{blue}{I dont understand the reasoning - here we still preserve the full isometry! }, 
while the underlying matrix model is of course still supersymmetric. We also note that the reality condition for $y^{AB}$ is 
\begin{align}
    (y^{AB})^*=-(C^{-1}YC)_{BA}\,   = (C^{-1}YC)_{AB} \ .
\end{align}
%\textcolor{blue}{I would write $(\gamma^{AB})^*$, since we're working with components. Same below}
%\textcolor{blue}{why not latter? %is easier, using antisymmetry}
In terms of components, we have
\begin{align}
    (q^{\alpha\beta})^{*}=-(\eps^{-1}q \eps)_{\beta\alpha}=(\eps^{-1}q\eps)_{\alpha\beta}\,,\qquad (p^{\alpha\dot\alpha})^{*}=-(\eps^{-1}p\eps)_{\dot\alpha\alpha}\,
    = (\eps^{-1}p\eps)_{\alpha\dot\alpha}
    \label{reality-cond-background-spinorial}
\end{align}
due to anti-symmetry. Next, let the remaining 5 coordinates of $SO(1,9)$ in \eqref{IKKTSO(1,9)}, which $SO(1,4)$ does not act on, be the scalar fields $\widetilde{y}_i$ on $H^4$ associated to the internal symmetry group $SO(5)$ \cite{Steinacker:2010rh}, where $i=5,6,7,8,9$.
%\textcolor{blue}{better  use $SO(5)$ here, $SU(4)$ would require 6 fields}
The local isomorphism $\mso(5)\simeq \msp(4)$ allows us to write:
\begin{align}\label{eq:extrascalars}
    \widetilde{y}^i\gamma_i^{\cI \cJ}\mapsto \phi^{\cI \cJ}\,,\qquad \qquad  \cI,\cJ=1,2,3,4\,,
\end{align}
where $\phi^{\cI\cJ}=-\phi^{\cJ\cI}$ can be written in terms of $2\times 2$ block matrices $\phi^{IJ}$. 

Next, consider the fermions 
$\Psi := \Psi^{A\cI}$ as 4-spinors of $SO(1,4)$. Let us clarify on the notation of $\Psi$. Here, $A$ is a 4-spinor index of $SO(1,4)$ and $\cI$ is a spinor index of $SO(5)$.
Upon imposing the 9+1-dimensional Majorana-Weyl condition, this amounts to 4 Weyl spinors on space-time. Using the decomposition \eqref{eq:YABdecomposition} of bosonic coordinates, we can write the Yukawa term as
\begin{align}
    \circled{Y}=\im \,\bar{\Psi}_{A\cI}\{p^{AB},\Psi_{B}{}^{\cI}\}+\im \,\bar{\Psi}_{A\cI}\{q^{AB},\Psi_{B}{}^{\cI}\}+\im \,\bar{\Psi}_{A\cI}\{\phi^{\cI\cJ},\Psi^{A}{}_{\cJ}\}
\end{align}
In terms of 2-spinors, $\Psi^{A\cI}$ and $\bar{\Psi}_{A\cI}:=\Psi_{B\cI}^{\dagger}(\gamma_0)^B{}_A$ can be decomposed as
\begin{align}
    \Psi^{A\cI}=(\chi^{\alpha\cI},\tilde{\chi}^{\dot\alpha\cI})\,,\qquad \bar{\Psi}_{A\cI}=(\bar{\chi}_{\alpha\cI},-\bar{\tilde\chi}_{\dot\alpha\cI})\,
\end{align}
where the two Weyl spinors for each $\cI$ are related by the Majorana condition $\Psi = C \bar\Psi^t$ through the charge conjugation matrix $C$ in $9+1$ -dimensions, which means that 
\begin{align}
    \tilde{\chi}^{\dot\alpha\cI} = 
 (C^{3+1})^{\dot\alpha}{}_{\beta} (C^6)^\cI{}_\cJ \bar\chi^{\beta\cJ}\,.
\end{align}
%\textcolor{blue}{more-or-less}.
Then, the Yukawa term can be written in the following form:
\begin{align}
    \circled{Y}=&+\im \,\bar{\chi}_{\alpha \cI}\{p^{\alpha\dot\beta},\tilde\chi_{\dot\beta}{}^{\cI}\}-\im \chi_{\alpha \cI}\{p^{\alpha\dot\beta},\bar{\tilde\chi}_{\dot\beta}{}^{\cI}\}+\im \bar{\chi}_{\alpha\cI}\{q^{\alpha\beta},\chi_{\beta}{}^{\cI}\}-\im\bar{\tilde\chi}_{\dot\alpha\cI}\{q^{\dot\alpha\dot\beta},\tilde\chi_{\dot\beta}{}^{\cI}\} \nonumber\\
    &+\im \bar{\chi}_{\alpha \cI}\{\phi^{\cI\cJ},\chi^{\alpha}{}_{\cJ}\}-\im \bar{\tilde\chi}_{\dot\alpha\cI}\{\phi^{\cI\cJ},\tilde\chi^{\dot\alpha}{}_{\cJ}\}
\end{align}
where half of the terms are redundant due to the Majorana condition.

Let the $H^4 \subset \R^{1,4} $ background be parametrized by $\ty^{\alpha\dot\alpha}$ (the tangential direction) and $\ty_0$ (the transversal direction). Then $p^{\alpha\dot\alpha}$ and $q^{\alpha\beta}$ can be decomposed as:
\begin{align}
    \binom{p^{\alpha\dot\alpha}}{q^{\alpha\beta}}=\binom{\ty^{\alpha\dot\alpha}}{\ty_0\epsilon^{\alpha\beta}}+\binom{\sa^{\alpha\dot\alpha}}{\hat\phi\epsilon^{\alpha\beta}}\,,
\end{align}
where $(\sa^{\alpha\dot\alpha},\hatphi)$ stand for $\hs$-valued fluctuations, which are subject to reality conditions analogous to \eqref{reality-cond-background-spinorial}. Together, $(\sa^{\alpha\dot\alpha},\hat\phi,\phi^{IJ})$ constitute the set of bosonic dynamical fields, while $(\chi,\widetilde\chi)$ are fermionic dynamical fields. Note that we do not have a fermionic `background' in our setup.
Using the above organization, we arrive at the following spinorial action for the IKKT-matrix model in the semi-classical limit on the projective spinor bundle $\PS$ :
\begin{equation}\label{affineaction}
    \begin{split}
    S=&\int_{\PS}\mho\,\Big(\frac{1}{2}\ff_{\alpha\alpha}\ff^{\alpha\alpha}+\frac{1}{2}\{p^{\alpha\dot\alpha},\hatphi\}\{p_{\alpha\dot\alpha},\hatphi\}+\frac{1}{2}\{p^{\alpha\dot\alpha},\phi^{IJ}\}\{p_{\alpha\dot\alpha},\phi_{IJ}\}\\
    &-\frac{\im}{2}\bar{\chi}^{\alpha}{}_{\cI}\{p_{\alpha\dot\beta},\tilde{\chi}^{\dot\beta\cI}\}+\frac{\im}{2} \chi^{\alpha}{}_{\cI}\{p_{\alpha\dot\beta},\bar{\tilde\chi}_{\dot\beta}{}^{\cI}\}+\frac{\im}{2}\bar{\tilde\chi}_{\dot\alpha\cI}\{\ty_0,\tilde{\chi}^{\dot\alpha\cI}\}-\frac{\im}{2}\bar{\chi}_{\alpha\cI}\{\ty_0,\chi^{\alpha\cI}\}\\
    &+\frac{1}{2}\{\ty_0,p^{\alpha\dot\alpha}\}\{\ty_0,p_{\alpha\dot\alpha}\}+\frac 12 \{\ty_0,\hat{\phi}\}\{\ty_0,\hatphi\}+\frac{1}{4}\{\ty_0,\phi^{IJ}\}\{\ty_0,\phi_{IJ}\}\\
    &+\frac{\im}{2}\bar{\tilde\chi}_{\dot\alpha\cI}\{\hatphi,\tilde{\chi}^{\dot\alpha\cI}\}-\frac{\im}{2}\bar{\chi}_{\alpha\cI}\{\hatphi,\chi^{\alpha\cI}\}-\frac{\im}{2} \bar{\chi}_{\alpha}{}^{\cI}\{\phi_{\cI\cJ},\chi^{\alpha\cJ}\}+\frac{\im}{2} \bar{\tilde\chi}_{\dot\alpha}{}^{\cI}\{\phi_{\cI\cJ},\tilde{\chi}^{\dot\alpha\cJ}\}\\
    &+\frac{1}{2}\{\hatphi,\hatphi\}\{\hatphi,\hatphi\}+\frac{1}{2}\{\hatphi,\phi^{IJ}\}\{\hatphi,\phi_{IJ}\}+\frac{1}{2}\{\phi^{IJ},\phi^{MN}\}\{\phi_{IJ},\phi_{MN}\}\Big)\,,
    \end{split}
\end{equation}
%\begin{equation}\label{affineaction}
%    \begin{split}
%    S=&\int_{\PS}\mho\,\Big(\frac{1}{2}\ff_{\alpha\alpha}\ff^{\alpha\alpha}+\frac{1}{2}\{p^{\alpha\dot\alpha},\hatphi\}\{p_{\alpha\dot\alpha},\hatphi\}+\frac{1}{2}\{p^{\alpha\dot\alpha},\phi^{IJ}\}\{p_{\alpha\dot\alpha},\phi_{IJ}\}-\im\bar{\chi}^{\alpha}\{p_{\alpha\dot\beta},\widetilde{\chi}^{\dot\beta}\}\\
%    &+\frac{1}{2}\{\ty_0,p^{\alpha\dot\alpha}\}\{\ty_0,p_{\alpha\dot\alpha}\}+\frac 12 \{\ty_0,\hat{\phi}\}\{\ty_0,\hatphi\}+\frac{1}{4}\{\ty_0,\phi^{IJ}\}\{\ty_0,\phi_{IJ}\}+\frac{\im}{2}\widetilde{\bar{\chi}}_{\dot\alpha}\{\ty_0,\widetilde{\chi}^{\dot\alpha}\}\\
%    &-\frac{\im}{2}\bar{\chi}_{\alpha}\{\ty_0,\chi^{\alpha}\}+\frac{\im}{2}\widetilde{\bar{\chi}}_{\dot\alpha}\{\hatphi,\widetilde{\chi}^{\dot\alpha}\}-\frac{\im}{2}\bar{\chi}_{\alpha}\{\hatphi,\chi^{\alpha}\}-\frac{\im}{2} \bar{\chi}^I\{\phi_{IJ},\chi^J\}-\frac{\im}{2} \widetilde{\bar{\chi}}^{\dot I}\{\phi_{\dot{I}\dot{J}},\widetilde{\chi}^{\dot{J}}\}\\
%    &+\frac{1}{2}\{\hatphi,\hatphi\}\{\hatphi,\hatphi\}+\frac{1}{2}\{\hatphi,\phi^{IJ}\}\{\hatphi,\phi_{IJ}\}+\frac{1}{2}\{\phi^{IJ},\phi^{MN}\}\{\phi_{IJ},\phi_{MN}\}\Big)\,,
%    \end{split}
%\end{equation}
where 
\begin{align}
    \ff^{\alpha\alpha}=\{p^{\alpha}{}_{\dot\gamma},p^{\alpha\dot\gamma}\}\,,\qquad \hat f^{\alpha\alpha} = f^{\alpha\alpha}\,,
\end{align}
%\textcolor{blue}{what is $$\ff_{\alpha\alpha}$?
%Also, the $\{\ty_0,p^{\alpha\dot\alpha}\}{\ty_0,p_{\alpha\dot\alpha}\}$ term is strange; you can drop the fixed background and replace it by $$\{\ty_0,a^{\alpha\dot\alpha}\}\{\ty_0,a_{\alpha\dot\alpha}\}$
%}
and we have rescaled fields/coordinates appropriately. Here, the commutators $[\,,]$ have been replaced by the Poisson brackets $\im\{\,,\}$, and the `trace' in \eqref{IKKTSO(1,9)} is substituted by an appropriate integral on $\PS$ with the measure
\begin{align}\label{affinemeasure}
    \mho:=\D^3Z\wedge \D^3\bar{Z}=\frac{R^8}{(R^2-x^2)^4}\,\d^4\tx \frac{\langle \lambda \,\d\lambda\rangle\wedge \langle \hat\lambda \,\d\hat\lambda\rangle}{\langle \lambda\,\hat\lambda\rangle^2}\,
\end{align}
which comes from wedging the $SU(4)$-invariant holomorphic measure $\D^3Z$ \cite{Witten:2003nn}:
\begin{align}\label{eq:DZmeasure}
    \D^3Z=\epsilon_{ABCD}Z^{A}\d Z^{B}\d Z^{C}\d Z^{D}=\frac{R^4\lambda_{\alpha}\lambda_{\beta}\langle \lambda \,\d \lambda \rangle \wedge \d\tx^{\alpha\dot\alpha} \wedge \d\tx^{\beta}{}_{\dot\alpha}}{(R^2-x^2)^2}\,,
\end{align}
and the anti-holomorphic measure  $\D^3\bar{Z}=\eps_{ABCD}\hat{Z}^A\d \hat Z^{B}\d \hat Z^{C}\d \hat Z^{D}$. It can be easily checked that $\{\hat\cN,\mho\}=0$ where $\hat\cN$ is the number operator defined in \eqref{eq:numberoperator}.

%by virtue of \eqref{eq:incident} and the parametrization \eqref{lambdaparametrization}. The anti-holomorphic measure  %$\D^3\bar{Z}$ is defined analogously. 

It is worth noting that the IKKT-matrix model on $H^4$  admits a smooth flat limit where $R\rightarrow \infty$, and all contributions associated with $\ty_0$ can be dropped as shown in 
\cite{Steinacker:2022jjv}. We will assume this limit from now on to study scattering amplitudes of the IKKT matrix model. %Finally, it is easy to notice that the 4-dimensional Euclidean action of (HS-)IKKT theory
%is already recognized in \eqref{affineaction}
%\textcolor{blue}{? what do you mean by "is already recognized"?}, and it naturally admits a higher-spin extension by virtue of \eqref{algebraPS}.

%%%%%%%%%%%%%%%%%%%%%%%%%%%%%%%%%%%%%%%%%%%%%
\paragraph{A sketch on self-dual sector.} Let us look at the Yang-Mills term in the action \eqref{affineaction}. In terms of `background' coordinates $\ty^{\alpha\dot\alpha}$ and fluctuations $\sa$, the Yang-Mills term reads:
\begin{align}
    \begin{split}
    L^{\text{YM}}=\frac 12 \ff_{\alpha\alpha}\ff^{\alpha\alpha}=\bar{L
    }^{\ff}_{\BG}&+2\{\ty^{\alpha}{}_{ \dot\gamma},\sa^{\alpha\dot\gamma}\}\{\ty_{\alpha\dot\sigma},\sa_{\alpha}{}^{\dot\sigma}\}+\{\ty^{\alpha}{}_{\dot\gamma},\ty^{\alpha\dot\gamma}\}\{\sa_{\alpha\dot\sigma},\sa_{\alpha}{}^{\dot\sigma}\}\,\\
    &+2\{\ty^{\alpha}{}_{\dot\gamma},\sa^{\alpha\dot\gamma}\}\{\sa_{\alpha\dot\sigma},\sa_{\alpha}{}^{\dot\sigma}\}+\frac 12 \{\sa^{\alpha}{}_{\dot\gamma},\sa^{\alpha\dot\gamma}\}\{\sa_{\alpha\dot\sigma},\sa_{\alpha}{}^{\dot\sigma}\}\,,
    \end{split}
    \label{YM-action-L}
\end{align}
which is analogous with the field strength term in non-commutative gauge theory, cf. \cite{Sperling:2017gmy,Blaschke:2011qu}. Here, the `background' action $\bar{
L}^{\ff}_{\BG}$ consists of terms that are 0th order or 1st order in fields. The term $\{\ty^{\alpha}{}_{\dot\gamma},\ty^{\alpha\dot\gamma}\}\{\sa_{\alpha\dot\sigma},\sa_{\alpha}{}^{\dot\sigma}\}$, which always occurs in non-commutative field theory, can be absored in the background by introducing an auxiliary field $\bs_{\alpha\alpha}$, and writing the Yang-Mills term in the first-order form as
\begin{align}\label{BF-BB}
    S^{\text{YM}}=\int_{\PS}\mho\Big(\bs_{\alpha\alpha}\ff^{\alpha\alpha}-\frac{1}{2}\bs_{\alpha\alpha}\bs^{\alpha\alpha}\Big)\,.
\end{align}
Note that we can also obtain the self-dual sector of \eqref{BF-BB} by dropping the $\bs_{\alpha\alpha}\bs^{\alpha\alpha}$  and maintaining gauge-invariance. Intriguingly, in the flat limit, all higher-spin fields in the self-dual sector of HS-IKKT theory are effectively massless since there is an `emergent' shift symmetry which can be used to gauge away the higher-spin modes $\sA$ in $\cA$ \cite{Steinacker:2022jjv}. We discuss below that this shift-symmetry, however, is \emph{not} a symmetry of the full HS-IKKT model.

%which was not seen previously in the second-order formalism of HS-IKKT \cite{Sperling:2017dts,Sperling:2017gmy,Sperling:2018xrm,Sperling:2019xar} (see also discussion in Section \ref{sec:shift-vector}). 

%%%%%%%%%%%%%%%%%%%%%%%%%%%%%%%%%%%%%%%%%%
\paragraph{Higher-spin modes in spinor formalism.} For concreteness, let us analyze the  pure Yang-Mills sector of the IKKT-matrix model in the flat limit. 

\medskip

If $\sa_{\alpha\dot\alpha}$ is a section with values in the algebra $\Ccurl(\PS)$, it has the following 
expansion:\footnote{Recall that the  ``time-like" fluctuations $\hat\phi$ drop 
out in the flat limit.}
\begin{align}
    \label{a-vectorfield-expansion}  \sa_{\alpha\dot\alpha}=\sum_{s=0}^{\infty}\lambda^{\beta(s)}\hat{\lambda}^{\beta(s)}\cA_{\beta(2s)|\alpha\dot\alpha}\,,\qquad \lambda^{\beta(s)}:=\lambda^{(\beta_1}\ldots \lambda^{\beta_s)}\quad \text{etc.}
\end{align}
Observe that 
\begin{align}
    \hat\sa_{\alpha\dot\alpha}&=\sum_{s=0}^{\infty}(-)^s \lambda^{\beta(s)}\hat\lambda^{\beta(s)}\hat \cA_{\beta(2s)|\alpha\dot\alpha}\,,\\
    \sa_{\alpha\dot\alpha}^{\dagger}&=\sum_{s=0}^{\infty}(-)^s \lambda^{\beta(s)}\hat\lambda^{\beta(s)}\ (\cA_{\beta(2s)|\alpha\dot\alpha})^{\dagger}\,.
\end{align}
Therefore, if we want $\sa_{\alpha\dot\alpha}$ to be real $\hs$-valued field, $\cA_{\beta(2s)|\alpha\dot\alpha}$ need to be complex, subject to the reality conditions
\begin{align}\label{reality-cA}
    \hat \cA_{\beta(2s)|\alpha\dot\alpha}=\im^{2s} \cA_{\beta(2s)|\alpha\dot\alpha}\,,\qquad (\cA_{\beta(2s)|\alpha\dot\alpha})^{\dagger}=\im^{2s}\cA_{\beta(2s)|\alpha\dot\alpha}
\end{align}
As a consequence, if we expand the $\hs$-valued field strength $\ff^{\alpha\alpha}$ in terms of $\cA$ will become complex. %Thus, for the action \eqref{BF-BB} to be real, we require $\bs_{\alpha\alpha}$ in \eqref{BF-BB} to be $\ff^{\alpha\alpha}$ complex conjugate when expanding $\bs_{\alpha\alpha}$ in terms of the following higher-spin modes
%\begin{align}
%    \bs_{\alpha\alpha}=\sum_s\cB_{\beta(2s)\alpha\alpha}\lambda^{\beta(s)}\hat\lambda^{\beta(s)}=\sum_s\Big(B_{\beta(2s)\alpha\alpha}+\eps_{\alpha\beta}\tilde B_{\alpha(2s-1)\alpha}\Big)\lambda^{\beta(s)}\hat\lambda^{\beta(s)}\,.
%\end{align}
%\textcolor{blue}{??what do you mean by that?}. It is implicit that $\bs^{\alpha\alpha}=(\bs_{\alpha\alpha})^{\dagger}$. 
This fact will be exploited in Appendix \ref{app:2} when we obtained the interacting vertices from the self-dual sector.

For now, we will decompose the coefficients $\cA_{\beta(2s)|\alpha\dot\alpha}$ into the following irreps:
%\textcolor{blue}{since our spinors transform under $SU(2)_L \times SU(2)_R$, we should not use $SL(2,\C)$ here. Just drop $SL(2,\C)$}
\begin{align}
\label{vector-modes-spinor-EAdS}
\cA_{\beta(2s)|\alpha\dot\alpha}=A_{(\beta(2s)\alpha)\,\dot\alpha}
+\eps_{\alpha\beta}\sA_{\beta(2s-1)\,\dot\alpha}\,.
\end{align}
We therefore have $\sum_s 2(2s+2) + 2(2s) = \sum_s 4(2s+1)$ off-shell degrees of freedom, in complete agreement with the organization using the  vector formalism in Section \ref{sec:shift-vector}. %In addition, the spin of $A_{\alpha(2s-1)\,\dot\alpha}$ always increase by an integer whenever we insert a pair of fiber coordinates $(\lambda,\hat\lambda)$. 
Notice that when we integrate out all fiber coordinates, the coefficients $A_{\alpha(2s-1)\,\dot\alpha}$ and $\sA_{\alpha(2s-3)\,\dot\alpha}$ will become tensorial fields in the maximally unbalanced/chiral representation \cite{Krasnov:2021nsq,Adamo:2022lah}.\footnote{\label{footnotechiral} A representation features fields with more un-dotted and dotted indices and allows us to work with lowest possible number of derivatives in the interactions.} This is in perfect agreement with the results in vector description \cite{Sperling:2018xrm} where the $\eps_{abcde}$ tensor breaks parity-invariance (see the discussion in Section \ref{sec:2}).

\paragraph{Gauge fixing and dof.} In spinorial language and in the flat limit, the gauge-fixing condition takes the form 
 \begin{equation}\label{gaugefixing}
 \begin{split}
 0 = 
 \{\ty_{\alpha\dot\alpha},\cA^{\beta(2s)|\alpha\dot\alpha}\} 
 &= \{\ty_{\alpha\dot\alpha},A^{(\beta(2s)\alpha)\dot\alpha}\}
  + \varepsilon^{\alpha\beta}\{\ty_{\alpha\dot\alpha},\sA^{\beta(2s-1)\dot\alpha}\}\\
  &=\{\ty_{\alpha\dot\alpha},A^{(\beta(2s)\alpha)\dot\alpha}\}
  - \{\ty^{\beta}{}_{\dot\alpha},\sA^{\beta(2s-1)\dot\alpha}\}\,.
  \end{split}
\end{equation}
The linearized gauge transformation for higher-spin gauge potentials $\cA_{\beta(2s)|\alpha\dot\alpha}$ reads
\begin{align}\label{gaugetransform}
    \delta_{\xi} \cA_{\beta(2s)|\alpha\dot\alpha}=\{
\ty_{\alpha\dot\alpha},\xi_{\beta(2s)}\}\,
 \equiv \cA^{(g)}[\xi_{\beta(2s)}] \ .
\end{align}
We can then use \eqref{gaugefixing} and \eqref{gaugetransform} to remove $(2s+1)$ components of $\cA^{\beta(2s)|\alpha\dot\alpha}$ which leaves us with $3(2s+1)$ off-shell degrees of freedom of $\cA^{\beta(2s)|\alpha\dot\alpha}$ on $H^4$ (without the ghosts). To understand this, we recall from Section \ref{sec:shift-vector} and \cite{Sperling:2018xrm} that even though the IKKT-matrix model does not possess a mass parameter, it leads to three off-shell towers of  ``would-be massive'' higher-spin degrees of freedom on $H^4$. These extra dof. arise from divergence-like components, which are physical on the curved background \cite{Steinacker:2019awe}, but expected to decouple in the flat limit\footnote{It would be interesting to understand whether they can be described by a more complicated net of gauge symmetry as in Zinoviev's system \cite{Zinoviev:2001dt}. }.

%%%%%%%%%%%%%%%%%%%%%%%%%%%%%%%%%%%%%

\paragraph{On shift symmetry.} Note that in other constructions of $3d$ or $4d$ higher-spin gauge theories using spinor formalism (see e.g. \cite{Krasnov:2021nsq,Adamo:2022lah,Zinoviev:2021cmi}), one often imposes by hand an extra ``shift" gauge symmetry of the form
\begin{align}\label{gaugetransform-theta}
 \delta_{\vartheta} \cA_{\beta(2s)|\alpha\dot\alpha}= \epsilon_{\alpha\beta}\,\vartheta_{\beta(2s-1)\,\dot\alpha}\,
 \end{align}
to remove the second mode $\sA_{\beta(2s-1)\,\dot\alpha}$ from $\cA_{\beta(2s)|\alpha\dot\alpha}$ given that $\sa_{\alpha\dot\alpha}$ is divergence-free. This symmetry was shown to be a symmetry of the self-dual Yang-Mills sector in HS-IKKT theory in the flat limit \cite{Steinacker:2022jjv}. However, in the full matrix model, there is no such gauge symmetry (see Appendix \ref{app:1}). As a result, we cannot use $\vartheta$ to remove the second higher-spin modes $\sA_{\alpha(2s-3)\,\dot\alpha}$. This is consistent with the fact that all higher-spin modes are not divergence-free a priori.

%Nevertheless, we show that in the flat limit, all the would-be massive dof. become effectively massless degrees of freedom since we do not need to distinguish the first higher-spin modes $A^{\alpha(2s-1)\,\dot\alpha}$ and the second higher-spin modes $\sA^{\alpha(2s-1)\,\dot\alpha}$ (see Appendix \ref{app:2}).

%As such, we find more physical modes, as discussed above. Nevertheless, this symmetry is an approximate symmetry of the matrix model in the flat limit \textcolor{blue}{really?} where the two physical modes $A_{\alpha(2s-1)\,\dot\alpha}$ and $\sA_{\alpha(2s-3)\,\dot\alpha}$ are totally decoupled from each others. \textcolor{blue}{However it is not a gauge symmetry and therefore does not reduce the number of degrees of freedom.}

\paragraph{On kinetic action.} Let us first look at the kinetic part of the action \eqref{YM-action-L} in the flat limit for the $\hs$-valued gauge potential $\sa_{\alpha\dot\alpha}$:
\begin{align}
    S_2=\int \mho\,\Big(2\{\ty^{\alpha}{}_{ \dot\gamma},\sa^{\alpha\dot\gamma}\}\{\ty_{\alpha\dot\sigma},\sa_{\alpha}{}^{\dot\sigma}\}+\{\ty^{\alpha}{}_{\dot\gamma},\ty^{\alpha\dot\gamma}\}\{\sa_{\alpha\dot\sigma},\sa_{\alpha}{}^{\dot\sigma}\}\Big)\,.
    \label{S2-action-quadratic-nogaugefix}
\end{align}
%\textcolor{blue}{I dont understand what you do below. Do you check gauge invariance?}
Let us first check gauge invariance explicitly:
Under a local gauge transformation, we get
\begin{align}\label{step1}
    \delta_{\xi}S_2=\int \mho\,\Big(2\{\ty^{\alpha}{}_{ \dot\gamma},\sa^{\alpha\dot\gamma}\}\{\ty_{\alpha\dot\sigma},\{\ty_{\alpha}{}^{\dot\sigma},\xi\}\}+2\sa_{\alpha\dot\sigma}\{\{\ty^{\alpha}{}_{\dot\gamma},\ty^{\alpha\dot\gamma}\},\{\ty_{\alpha}{}^{\dot\sigma},\xi\}\}\Big)\,.
 \end{align}
where we have made an integration by part. The second term can be rewritten  as 
\begin{align}\label{step1S2}
&\int \{\ty^{\alpha}{}_{\dot\gamma},\ty^{\alpha\dot\gamma}\}
    \{\sa_{\alpha\dot\sigma},\{\ty_{\alpha}{}^{\dot\sigma},\xi\}\}
    = -\int \{\ty^{\alpha}{}_{\dot\gamma},\ty^{\alpha\dot\gamma}\}
   \big(\{\ty_{\alpha}{}^{\dot\sigma},\{\xi,\sa_{\alpha\dot\sigma}\}\}
      + \{\xi,\{\sa_{\alpha\dot\sigma},\ty_{\alpha}{}^{\dot\sigma}\}\} \big)
   \nn\\
    &= \int 
   \{\ty_{\alpha}{}^{\dot\sigma},\{\ty^{\alpha}{}_{\dot\gamma},\ty^{\alpha\dot\gamma}\}\}\{\xi,\sa_{\alpha\dot\sigma}\}
      + \{\xi,\{\ty^{\alpha}{}_{\dot\gamma},\ty^{\alpha\dot\gamma}\}\} \{\sa_{\alpha\dot\sigma},\ty_{\alpha}{}^{\dot\sigma}\}
\end{align}
using Jacobi identity. The first term in \eqref{step1S2} vanishes due to background eom
\begin{align}
    \{\{\ty^{\alpha}{}_{\dot\gamma},\ty^{\alpha\dot\gamma}\},\ty_{\alpha}{}^{\dot\sigma}\}=0\,.
\end{align}
Furthermore, the second term in \eqref{step1S2} cancels with the first tem in \eqref{step1} if we use Jacobi identity to write
\begin{align}
\int \{\xi,\{\ty^{\alpha}{}_{\dot\gamma},\ty^{\alpha\dot\gamma}\}\} \{\sa_{\alpha\dot\sigma},\ty_{\alpha}{}^{\dot\sigma}\}
= -\int \{\ty^{\alpha}{}_{\dot\gamma},\{\ty^{\alpha\dot\gamma},\xi,\}\} \{\sa_{\alpha\dot\sigma},\ty_{\alpha}{}^{\dot\sigma}\}\,.
\end{align}
noting that $\int\{\xi,-\} = 0$. This establishes the gauge invariance of \eqref{S2-action-quadratic-nogaugefix}.

%{\bf In the past I wanted to check gauge invariance first. Then I go to the gauge-fixed action (6.24) and observe there is a simplification. Maybe we should rename this paragraph.}

%ok. Just make the logic clear, is fine. Maybe write the reduced form for $S_2$ explicitly.

%\textcolor{blue}{ ...}

\paragraph{Gauge-fixed kinetic action.} Note that the quadratic action \eqref{S2-action-quadratic-nogaugefix} can be simplified further in the flat limit by dropping the term term $\{\ty^{\alpha}{}_{\dot\gamma},\ty^{\alpha\dot\gamma}\}\{\sa_{\alpha\dot\sigma},\sa_{\alpha}{}^{\dot\sigma}\}$ in the flat limit (see Appendix \ref{app:2}). Therefore, we can consider the following gauge-fixed quadratic action
\begin{align}
    S_2^{\gf}=S_2+2\int \{\ty_{\alpha\dot\alpha},\sa^{\alpha\dot\alpha}\}^2\,.
\end{align}
Here, the contributions from the gauge-fixing term removes certain contributions
%\textcolor{blue}{This is cryptic, maybe just write it explicitly}
from $S_2$ in the flat limit leaving us with the standard kinetic $\Box$ terms in the quadratic action as shown in Appendix \ref{app:2}. In
particular, we obtain
\begin{align}
\label{S2-action-gf-1}
    S_2^{\gf}= \int \d^4x \,\A_-^{\alpha(2s-1)\,\dot\alpha}\Box \A^+_{\alpha(2s-1)\,\dot\alpha}\,, \qquad \A^{\pm}_{\alpha(2s-1)\,\dot\alpha}=A_{\alpha(2s-1)\,\dot\alpha}\pm \im \sA_{\alpha(2s-1)\,\dot\alpha}\,.
\end{align}
This fact is convenient for us to compute the propagator of HS-IKKT theory using spinor formalism. We detail this fact further below.

%%%%%%%%%%%%%%%%%%%%%%%%%%%%%%%%%%%%%%%%%%%%%

%%%%%%%%%%%%%%%%%%%%
\paragraph{Euclidean action of massless sector.} A complete action of the (HS-) IKKT matrix model on $H_N^4$ can be obtained using the following integral over $\P^1$ \cite{Woodhouse:1985id,Boels:2006ir,Jiang:2008xw}:
    \begin{align}\label{eq:bridge}
    \int_{\P^1}\tK\,\frac{\hat{\lambda}_{\alpha(m)}\,\lambda^{\beta(m)}}{\langle \lambda \,\hat{\lambda}\rangle^{m}}:=\int_{\P^1}\frac{\langle \lambda\, \d\lambda\rangle\wedge\langle \hat\lambda \,\d\hat\lambda\rangle}{\langle \lambda\,\hat\lambda\rangle^2}\, \frac{\hat{\lambda}_{\alpha(m)}\,\lambda^{\beta(m)}}{\langle \lambda \,\hat{\lambda}\rangle^{m}}=-\frac{2\pi i}{(m+1)}\epsilon^{\ \beta}_{ \alpha}\ldots\epsilon^{\ \beta}_{ \alpha}\,.
\end{align}
As discussed in \cite{Steinacker:2022jjv,Tran:2022mlu}, in the present almost-commutative twistor construction, there are more contributions (after integrating out fiber coordinates) than the usual twistor construction using only holomorphic data on twistor space, see e.g. \cite{Bittleston:2020hfv,Bittleston:2022nfr}. The reason is that
 the Poisson bracket (cf., \eqref{eq:Poissonspinors}) acts both on spacetime fields and fiber coordinates. However in flat limit, the effective vielbeins $\cE^{\alpha\dot\alpha|\beta},\,\hat{\cE}^{\alpha\dot\alpha|\beta}$ and $\cE^{0|\beta},\hat{\cE}^{0|\beta}$ are sub-leading, and all contributions associated to $\mu,\hat\mu$ spinors will be suppressed since they scale as $1/\sqrt{R}$ \cite{Steinacker:2022jjv}. For this reason, we will only need to consider terms where all fiber coordinates factorized outside the Poisson brackets. For example,
\begin{align}
    \lambda^{\alpha(s-1)}\hat{\lambda}^{\alpha(s-1)}\{\ty_{\alpha\dot\alpha},\A_{\alpha(2s-1)}{}^{\dot\alpha}\}=\lambda^{\alpha(s-1)}\hat{\lambda}^{\alpha(s-1)}\cE_{\alpha\dot\alpha,\beta\dot\beta}\p^{\beta\dot\beta}\A_{\alpha(2s-1)}{}^{\dot\alpha}
\end{align}
will be the leading contribution in the higher-spin extension for $\{\ty_{\alpha}{}^{\dot\alpha},\sa_{\alpha\dot\alpha}\}$ in the flat limit. Below we only present the final result of the spacetime action of HS-IKKT on $H^4$ for massless sector, delegating the detail of the computation  to Appendix \ref{app:2}. 

The Euclidean gauge-fixed kinetic action for the $\hs$-valued gauge potential $\sa_{\alpha\dot\alpha}$ reads
\be\label{kineticYM}
    \begin{split}
    S_2&=\int_{\PS}\mho\,\Big(2\{\ty^{\alpha}{}_{ \dot\gamma},\sa^{\alpha\dot\gamma}\}\{\ty_{\alpha\dot\sigma},\sa_{\alpha}{}^{\dot\sigma}\}+\{\ty^{\alpha}{}_{\dot\gamma},\ty^{\alpha\dot\gamma}\}\{\sa_{\alpha\dot\sigma},\sa_{\alpha}{}^{\dot\sigma}\}+2\int \{\ty_{\alpha\dot\alpha},\sa^{\alpha\dot\alpha}\}^2\Big)\,\\
    &\simeq 2\sum_s\int \d^4\tx\, \A^{\alpha(2s-1)\,\dot\alpha}_-\Box\A^+_{\alpha(2s-1)\,\dot\alpha}\,.
    \end{split}
\ee
where the composite field $\A$, see \eqref{S2-action-gf-1}. Here the $\Box=\p^{\alpha\dot\alpha}\p_{\alpha\dot\alpha}$ in the flat  limit, cf. \eqref{eq:preeffectivemetric}, and we have integrated out all fiber coordinates in the second line of \eqref{kineticYM}. We also note that the $\sA^{\alpha(2s-1)\,\dot\alpha}$ component originate from the second higher-spin mode of the spin-$(s+1)$ gauge potential $\cA_{\alpha(2s+1)|\beta\dot\beta}$. This combination of higher-spin fields allows us to diagonalize the $\Box$ operator in the flat limit in a trivial way using spinorial formalism. Intriguingly, one may think of $\A_+$ as a positive helicity field while $\A_-$ as a negative helicity field in the flat limit. It is worth emphasizing once again that even though $\A_{\pm}$ may look massless, they have more degrees of freedom than usual massless higher-spin fields due to the fact that their original twistor fields are not divergence-free a priori.

Next, we find at cubic order that
\be\label{cubicaction}
    S_3\simeq 4\sum_{s_2+s_3=s_1+2}\int \d^4\tx \,\p_{\alpha\dot\alpha} \A^{\pm}_{\alpha(2s_1-1)}{}^{\dot\alpha}\p_{\alpha\dot\gamma}\A_{\pm}^{\alpha(2s_2-1)\,\dot\sigma}\p_{\alpha}{}^{\dot\gamma}\A_{\pm}^{\alpha(2s_3-1)}{}_{\dot\sigma}  +\widetilde{S}_3   \,,
\ee
where $\tilde S_3$ are terms that are irrelevant for scattering process of massless modes as they vanish upon plugging in the plane-wave solutions with higher-spin polarization tensors in \eqref{YMhel}. Unlike the usual story of higher-spin Yang-Mills \cite{Adamo:2022lah}, here all helicity configurations are allowed. However, most of the 3-pt amplitudes resulting from gluing the cubic vertices \eqref{cubicaction} with external states vanish on-shell (see Section \ref{sec:6}). One can check that the above cubic vertex has two transverse derivatives $\p^{0\dot 1}=\bar{\p}$ in the light-cone gauge by following the procedure in \cite{Krasnov:2021nsq,Tran:2021ukl}. For this reason, the HS-IKKT matrix model can be referred to as a two-derivative higher-spin theory. Lastly, the quartic term reads
\small
\begin{align}
    S_4\simeq     2\int \d^4x  \Big(\p_{\alpha\dot\gamma_1}\A_{\pm}^{\alpha(2s_1-1)\,\dot\sigma}\p_{\alpha}{}^{\dot\gamma_1}\A_{\pm}^{\alpha(2s_2-1)}{}_{\dot\sigma}\Big)\Big(\p^{\alpha}{}_{\dot\gamma_2}\A^{\pm}_{\alpha(2s_3-1)}{}^{\dot\tau}\p^{\alpha\dot\gamma_2}\A^{\pm}_{\alpha(2s_4-1)\,\dot\tau}\Big)+\widetilde{S}_4\,.
\end{align}
\normalsize
Similar to $\tilde S_3$, $\tilde S_4$ contains irrelevant terms that do not contribute in scattering amplitudes upon plugging plane-wave solutions \eqref{YMhel}; and thus can be ignored in the flat limit. We use the above action to compute scattering amplitudes of the HS-IKKT in Section \ref{sec:6}. Note that while the above action is defined on a background with Euclidean signature, it is nevertheless possible to use some of the standard techniques in quantum field theory to compute the scattering amplitudes if we work with complexified kinematics. In fact, it is natural to do so since our `spacetime' fields are generically chiral, i.e. they have more un-dotted indices than dotted ones.

It is worth emphasizing that although there are more structures after integrating out fiber coordinates in almost-commutative twistor approach compared to the conventional twistor construction (see e.g. \cite{Bittleston:2022nfr}), the non-vanishing contributions of the massless sector in spacetime action on flat space $S$-matrix in both approaches turn out to be the same.\footnote{We leave the study of the scattering of extra divergence modes for future study.} The main advantage of the present non-commutative twistor approach in \cite{Sperling:2017dts,Sperling:2017gmy,Sperling:2018xrm,Sperling:2019xar,Steinacker:2022jjv} is that it allows us to study quantization of twistor space, and to consider models which appear to define a well-defined quantum theory.

%%%%%%%%%%%%%%%%%%%%%%%%%%%%%%%%%%%%%%%%%%%%%%%%
\subsection{Reduced action on the FLRW spacetime}\label{sec:5.2}
%%%%%%%%%%%%%%%%%%%%%%%%%%%%%%%%%

Now we consider similarly the 
HS-IKKT theory on 
the FLRW-like spacetime after doing an $SO(1,3)$-invariant projection as described in Section \ref{SO(1,3)spinors}. Using the spinorial description of $t^{\mu}$ in \eqref{t-description-spinor} where we set $t^0=0$ at the referent point $\tp\in \cM^{1,3}$, we can effectively replace the ``momentum'' generators $t^i$ by a pair of fiber coordinates $(\lambda^{\alpha},\hat\lambda^{\alpha})$ in the flat limit. Recall that by flat limit, we simply mean the limit where we can 
factorize all other fiber coordinates $(\lambda,\hat\lambda)$ outside the Poisson brackets, and drop the explicit
$\mu,\hat\mu$ spinors for the $x$-dependent fields.

Consider a local trivialization at  $\tp\in \cM^{1,3}$ such that $\P^{1,2}|_{U_{\tp}}=\P^1\times U_{\tp}$, where $U_{\tp}$ is an open subset around $\tp$. Then
the measure on our twistor space is given by \cite{Sperling:2019xar}
\begin{align}
   \mho= \tK\, \rho_{\cM}\,d^4y=\tK\, \frac{1}{R|\sinh(\tau)|}\,d^4y
\end{align}
where $\tK$ is again the top form on $\P^1$ fiber (cf., \eqref{eq:bridge}). This measure is invariant under symplectomorphism on $\PT$,
and it is in fact globally well defined.
% It is important to stress that while $\Theta$ is not globally well-defined, the measure $\rho_{\cM}d^4y$ is. 
The local split allows us to  average over fiber coordinates of $\P^1$ on an open subset $U_{\tp}$ around $\tp\in \cM$, and to obtain a spacetime action for the HS-IKKT matrix model on our background defining a FLRW cosmology.

Recalling that the background $\cM^{1,3}$ is defined in terms of  `momentum' generators $t^{\mu}$, the action for the fluctuations $\sa^{\mu}$ of the (HS-)IKKT model in the semi-classical limit reads
\be\label{FLRWaction}
    \begin{split}
    S=\int_{\P^{1,2}} &\mho\,\Big(\frac 12 \{t^{\mu},\sa^{\nu}\}\{t_{\mu},\sa_{\nu}\}+\frac 12 \{t_{\mu},t_{\nu}\}\{\sa^{\mu},\sa^{\nu}\}+\{t^{\mu},\sa^{\nu}\}\{\sa_{\mu},\sa_{\nu}\}+\frac{1}{4}\{\sa^{\mu},\sa^{\nu}\}\{\sa_{\mu},\sa_{\nu}\}\\
    &-\frac{\im}{4}\bar{\Psi}\gamma^{\mu}\{t_{\mu},\Psi\}-\frac{\im}{4}\bar{\Psi}\gamma^{\mu}\{\sa_{\mu},\Psi\}+\frac{1}{4}\{\phi^{\hat\mu},\phi^{\hat\nu}\}\{\phi_{\hat\mu},\phi_{\hat\nu}\}-\frac{\im}{4}\bar{\Psi}\gamma^{\hat\mu}\{\phi_{\hat\mu},\Psi\}\Big)+S_{\BG}
    \end{split}
\ee
%\textcolor{blue}{replace $\Theta$ everywhere}
where $\hat\mu=4,5,\ldots,9$ indicates the 6 extra dimensions and $S_{\BG}$ are the `background' action which consists of zeroth or first order in fluctuations. Note that we have identified $t^{\hat\mu}\equiv \phi^{\hat\mu}$ as scalar fields in non-commutative $\cN=4$ SYM.

%%%%%%%%%%%%%%%%%%%%%%%%%%%%
\paragraph{Higher-spin modes and dof.} As pointed out above, we can realize $t^i$ in terms of the pair of $(\lambda^{\alpha}\,,\hat\lambda^{\alpha})$ fiber coordinates. This allows to parametrize the higher-spin modes of the Yang-Mills gauge potential $\sa_{\mu}$
around a reference point $\tp\in \cM$ (cf., \eqref{t-description-spinor}) as follows 
\begin{equation}
    \begin{split}
     \label{mode-expansion-t-lambda}
\sa_{\mu}&=\sum_{s=0}^{\infty}t^{i(s)}\cA_{i(s)|\mu}\simeq\sum_{s=0}^{\infty}\lambda^{\beta(s)}\hat\lambda^{\beta(s)}\cA_{\beta(2s)|\mu}\,.
    \end{split}
\end{equation}
cf. \eqref{modeexpansioninter1}, which encodes $\sum_s 4(2s+1)$ off-shell degrees of freedom of the higher-spin valued gauge potential $\sa_{\mu}$.
The degeneracy of the kinetic term is removed as usual by imposing the gauge-fixing condition
\begin{equation}
 \begin{split}
 0 =  \{t^\mu,\sa_{\mu}\} 
  \end{split}
\end{equation}
and factoring out the pure gauge modes,  defining the physical Hilbert space as 
\begin{align}
 \cH_{\rm phys} = \{\mbox{gauge-fixed on-shell modes}\}/_{\{\mbox{pure gauge modes}\}  }
 \label{H-phys-0}
\end{align}
(at ghost number zero). This removes two towers of higher-spin modes, leaving us with $\sum_s 2(2s+1)$
physical degrees of freedom\footnote{recall that these spinorial modes are not divergence-free here.}.
A more detailed analysis in vectorial form is given in \cite{Steinacker:2019awe},
where $\cH_{\rm phys}$ was shown to be free of negative modes, i.e. ghosts. The basic reason is that $t^0=0$ 
around the reference point $\tp$, so that there are no time-like higher-spin 
components in \eqref{mode-expansion-t-lambda}. 

In the following, we will focus on certain specific degrees of freedom among these modes.

\paragraph{Spacetime action in the flat limit.} 
Let us focus on the Yang-Mills sector of the action \eqref{FLRWaction} in quadratic, cubic and quatic orders as:
\be\label{FLRWactionflat1}
    \begin{split}
    S=\int_{\P^{1,2}} \mho\,\Big(\frac 12 \{t^{\mu},\sa^{\nu}\}\{t_{\mu},\sa_{\nu}\}&+\frac 12\{t_{\mu},t_{\nu}\}\{\sa^{\mu},\sa^{\nu}\}+
    \{t^\mu,a_\mu\}\{t^\nu,a_\nu\}
    \\
    &
    +\{t^{\mu},\sa^{\nu}\}\{\sa_{\mu},\sa_{\nu}\}+\frac{1}{4}\{\sa^{\mu},\sa^{\nu}\}\{\sa_{\mu},\sa_{\nu}\}+\ldots\Big)\,.
    \end{split}
\ee
where un-hatted indices are contracted with $\eta^{\mu\nu}$. The last term in the first line drops out upon gauge fixing. Using the effective metric \eqref{effectivemetricFLRW} and \eqref{tderivative}, we can write the gauge-fixed kinetic term explicitly as
\begin{align}
    S_2=\int_{\P^{1,2}}d^4y\,\tK\,\Big[\frac{1}{2} \frac{\sinh(\tau)}{R} \Big(\gamma^{\mu\nu}\p_{\mu}\sa^{\rho}\p_{\nu}\sa_{\rho}+(\p_{\mu}\sa^{\mu})^2\Big)+\frac{m^{\mu\nu}}{2R^3|\sinh(\tau)|}\{\sa_{\mu},\sa_{\nu}\}\Big]\,.
\end{align}
where 
$m^{\mu\nu}= R^2 \{t^\mu,t^\nu\}$ defined in \eqref{mgenerator}, and the metric $\gamma^{\mu\nu}$ is given in \eqref{effectivemetricFLRW}, 
which locally reduces to $\eta^{\mu\nu}$. As explained above, the term $\frac 12\{t_{\mu},t_{\nu}\}\{\sa^{\mu},\sa^{\nu}\}$ is suppressed in the flat limit of the FLRW matrix model-like spacetime, where $R\rightarrow 
\infty$ and $\sinh(\tau)$  can be treated as (large) constant at late time $\tau$. Therefore, in the flat limit
\begin{align}
    S_2\simeq -\frac{1}{2}\int_{\PS} \d^4y\,\tK\Big(\,\sa^{\mu}\Box\sa_{\mu}+(\p_{\mu}\sa^{\mu})^2\Big)\,,\qquad \mu=0,1,2,3\,
\end{align}
where the last term drops out upon gauge fixing.
The fluctuations $\sa^\mu$ are real functions on $\PT$, since the matrices of the IKKT model are hermitian. 
In particular, the d'Alembertian is given by
\begin{align}
    \Box = \gamma^{\mu\nu}\del_\mu\del_\nu 
\end{align}
where $\mu,\nu=0,1,2,3\,$ dropping a conformal factor from the FLRW background, which can be considered as locally constant in the present context.

The appropriate spinorial formulation of this action is not evident, since the spinors 
on the present background transforms under the  $SU(2)_L\times SU(2)_R$ space-like isometry group rather than $SL(2,\C)$. 
%We can now use the same organization of the higher-spin modes in terms of complexified modes $\A^{\pm}_{\alpha(2s-1)\,\dot\alpha}$ in \eqref{Apm-def-euclid} as in the Euclidean case,  even though the explicit gauge-fixing and the gauge transformations are different.
%yes. We dont extend anything, this is just rewriting using \eqref{ti-lambda-relation-M31
%It is now tempting to exchange the Lorentz index $\nu$ by a pair of spinor indices $(\alpha,\dot\alpha)$
%using \eqref{ti-lambda-relation-M31}.
%However, this is quite distinct from the usual procedure on Minkowski space, because our spinors transform under $SU(2)\times SU(2)\subset SO(1,4)$ rather than $SL(2,\C)$; this reflects the lack of manifest local Lorentz invariance in the model, which is expected to be recovered only effectively. 
%The problem is that we cannot even write $x^{\alpha\dot\alpha}$ in the Minkowski setting (put this in earlier section). Therefore
%\textcolor{blue}{The appropriate ogranization of functions in terms of spinors is not clear, because in the Minkowski setting  our spinors are $SU(2)\times SU(2)$ rather than Lorentz spinors. Probably can relate divergence-free vector field, and generalize this to the irreducible spinors, future project }.
Then the time-like components take a non-standard form, and local Lorentz-invariance is not manifest.
Nevertheless, a close relation with the Euclidean case on $H^4$ is expected, since the underlying space of functions on $\PT$ is the same,  given by (principal series) unitary irreps of $\mso(2,4)$. Therefore the interactions arising from the matrix model are the same in both signatures, while the kinetic terms should be related by some sort of Wick rotation.
We expect that this relation should work most naturally for the physical fields, because then the time-like components of $\sa_\mu$ 
are unphysical, while for the space-like components
we do recover the ususal spinorial representation 
 due to \eqref{y-i-spinor-FLRW}.  We could thus declare that the spinors transforms as $SL(2,\C)$ spinors under the local Lorentz group, thereby extending the local $SO(3)$ to $SO(1,3)$. 
Since the averaging over the local $S^2$ fiber is uniquely defined by $SO(3)$ invariance for irreducible fields, it should respect the local Lorentz invariance automatically. 

Another possible strategy would be to perform an analytic continuation in the $y^0 y^4$ plane, so that  the actions on two different coordinates can be analytically continued into each others in the flat limit.
Moreover for irreducible (divergence-free) tensor fields, there is a natural map from the FLRW background to the $H^4$ background. This could also provide some sort of Wick rotation, relating these two backgrounds with different signature via the embedding in $\R^{1,4}$.
However, a thorough treatment of this issue is left for future work.

%To this end, we note that although matrix-model type FLRW cosmology provides us the desired Lorentzian signature, higher-spin fields are \emph{not} Lorentzian-real since they have more un-dotted indices than dotted ones
%\textcolor{blue}{?? I dont agree here. If you expand them, they have the same numbers, and they are real essentially}. 

%This is, in fact, not a surprise since we have broken parity invariance when introducing higher-spin fields to the spectrum to retain Lorentz invariance. As a result, we also have complex-valued higher-spin fields when working with spinorial formalism in $\cM^{1,3}$ spacetime. Note that this result is also reflected in vector description of HS-IKKT theory by the presence of $\eps_{abcde}$ tensors \cite{Sperling:2018xrm}. 

%%%%%%%%%%%%%%%%%%%%%%%%%%%%%%%%%
%%%%%%%%%%%%%%%%%%%%%%%%%%%%%%%%%
\section{Amplitudes of the Yang-Mills massless sector in Euclidean signature}\label{sec:6}

In this section, we start with the reduced action of the HS-IKKT matrix model 
on $H^4$ in  the flat limit elaborated in Section \ref{sec:5} (see also Appendix \ref{app:2}), and study 
tree-level scattering amplitudes of the higher-spin modes 
\begin{align}\label{Apm-def-euclid}
    \A^{\pm}_{\alpha(2s-1)\,\dot\alpha}=A_{\alpha(2s-1)\,\dot\alpha}\pm \im\,\sA_{\alpha(2s-1)\,\dot\alpha}\, .
\end{align}
Of course, fields do not propagate on a manifold with Euclidean signature, instead they ``decay'' with the distance. Nevertheless,  it is possible to study scattering amplitudes
  by analytic continuation of the real kinematics into the complex domain. This  will allow us to compute scattering amplitudes using well-known recursion techniques as in \cite{Britto:2005fq}. %As a side remark, if we work with HS-IKKT theory in FLRW cosmological-type spacetime the step of analytic continuation will be redundant. However, it is quite challenging to describe HS-IKKT theory using spinorial description since local Lorentz is not manifest. 
Note that all higher-spin fields with $s>1$ thereby become complex-valued.  
%\textcolor{blue}{what does this have to do with parity violation? We introduced the complexified fields by hand just as a handy tool!}

We recall from section  \ref{sec:shift-vector} that the bosonic higher-spin gauge fields carry  $3(2s+1)$ degrees of freedom in Euclidean signature, and $2(2s+1)$ propagating dof on the physical FLRW space-time. 
Since these are more degrees of freedom compared to the usual cases of massless with 2 and massive with $2s+1$ degrees of freedom in 4 dimension, there will be more higher-spin modes to consider when computing scattering amplitudes. For simplicity, we will only consider the massless sector in this section which contains the fields that satisfy the Lorenz gauge condition $\p^{\alpha\dot\alpha}\A_{\beta(2s-2)\alpha\,\dot\alpha}=0$. 
Our notation for an $n$-point scattering amplitude is then $\cM_n(1_{s_1}^{h_1},\ldots,n_{s_n}^{h_n})$, where $h_i=\pm$ indicates whether the $i$th particle of spin-$s_i$ has positive or negative helicity.

%\textcolor{red}{?? The original fields are real! maybe the spinor fields are complex, does it matter? Why is parity important for this?} {\bf yes, the spinor fields are complex. I think we break parity the moment higher-spin modes are introduced and restricted to two row Young diagrams by $\eps_{abcde}$ tensor.}

To determine helicity of the external states, we note that the kinetic action can be written in the form \cite{Adamo:2022lah}:
\be\label{FreeGT}
S_2=\,\int_{\M}\d^4\tx\,\A^{\alpha(2s-1)\,\dot\alpha}_-\Box \A^+_{\alpha(2s-1)\,\dot\alpha}\,,\qquad s\geq 1\,,
\ee
for the appropriately chosen integration domain $\M$, in terms of the complex fields \eqref{Apm-def-euclid}, 
where $\pm$ denotes positive/negative helicity fields. %The above quadratic action is invariant under the gauge transformation $\delta \A_{\alpha(2s-1)\,\dot\alpha}=\partial_{\alpha\dot\alpha}\xi_{\alpha(2s-2)}$ on the base manifold.
%\textcolor{red}{yes, but the gauge invariance of the 
%IKKT model on $H^4$ the interactions looks very different !?!}
%\textcolor{blue}{this is correct for the FLRW case, not for the Euclidean case. We need to keep things separate?! What is the setting in this section?} {\bf Euclidean}
%This gauge transformation can be obtained directly from the local gauge transformation associated to $\hs$-valued $\xi$ parameter in \eqref{gaugetransform} on almost-commutative twistor space by integrating out fiber coordinates on $\P^1$.
Then by imposing the Lorenz gauge 
\be\label{LorG}
\partial^{\gamma\dot\alpha}\A_{\alpha(2s-2)\gamma\,\dot\alpha}=0\,.
\ee
on the base manifold $\M$, we can select \emph{only} massless modes out of $3(2s+1)$ components of the $\A_{\pm}^{\alpha(2s-1)\,\dot\alpha}$ fields. Suppose $k^{\alpha\dot\alpha}=\kappa^{\alpha}\tilde{\kappa}^{\dot\alpha}$ is an on-shell complex 4-momentum. We define positive and negative helicity polarization tensors associated to external higher-spin states as \cite{Adamo:2022lah}:
\begin{align}\label{YMhel}
    \eps^{+}_{\alpha(2s-1)\,\dot\alpha}=\frac{\zeta_{\alpha(2s-1)}\tilde\kappa_{\dot\alpha}}{\langle\kappa\,\zeta\rangle^{2s-1}}\,,\qquad  \eps^{-}_{\alpha(2s-1)\,\dot\alpha}&=\frac{\kappa_{\alpha(2s-1)}\,\tilde{\zeta}_{\dot\alpha}}{[\tilde{\kappa}\,\tilde{\zeta}]}\,,
\end{align}
where $\zeta_{\alpha},\tilde{\zeta}_{\dot\alpha}$ are constant/reference spinors. The above representatives for polarization tensors are chosen such that they obey the normalization \cite{Krasnov:2020bqr}
\be
\eps^{+}_{\alpha(2s-1)\,\dot\alpha}\eps_{-}^{\alpha(2s-1)\,\dot\alpha}=-1\,.
\ee
%\textcolor{blue}{ok, that should be fine. I think we dont need $\delta \A_{\alpha(2s-1)\,\dot\alpha}=\partial_{\alpha\dot\alpha}\xi_{\alpha(2s-2)}$, which is very fishy }
%\textcolor{blue}{??? not for the complex fields!! dont have so many conditions. But we may choose the modes which satisfy this condition! Probably this is the way to proceed} {\bf Only massless modes have to satisfy this condition.}
With the choice of polarization tensors in \eqref{YMhel}, it can be checked that 
\begin{subequations}
\begin{align}
 \partial_{\alpha}{}^{\dot\gamma}\A^{+}_{\alpha(2s-1)\,\dot\gamma}&=0\label{ph1}\,,\\
 \partial^{\beta\dot\alpha}\partial_{(\beta}{}^{\dot\gamma}\A^{-}_{\alpha(2s-1))\,\dot\gamma}&=0\label{nh1}\,.
\end{align}
\end{subequations}
The propagator between positive and negative helicity fields in the Lorenz gauge \eqref{LorG} is
\begin{align}\label{propagator}
    \langle \A^{+}_{\alpha(2s-1)\,\dot\alpha}(p)\A_{-}^{\beta(2s'-1)\,\dot\beta}(p')\rangle = \delta^4(p+p')\tilde\delta(s-s')\frac{\delta_{(\alpha_1}{}^{(\beta_1}\ldots \delta_{\alpha_{2s-1})}{}^{\beta_{2s'-1})}\delta_{\dot\alpha}{}^{\dot\beta}}{p^2}\,,
\end{align}
where $\tilde\delta$ is a Kronecker delta:
\begin{equation}
    \tilde\delta(x)=\begin{cases} 0\,,\qquad x\neq 0\,,\\
    1\,,\qquad x=0\,.
    \end{cases}
\end{equation}
Note that since the standard linearized gauge transformation $\delta \A_{\alpha(2s-1)\,\dot\alpha}=\p_{\alpha\dot\alpha}\xi_{\alpha(2s-2)}$ is not a symmetry of HS-IKKT theory a priori,\footnote{Recall that all higher-spin fields in HS-IKKT are ``would-be massive'' fields with more degrees of freedom than the massless ones.} there is no restriction on the positive helicity of a massless field compared to the cases studied in \cite{Adamo:2022lah,Tran:2022amg}.
To this end, we recall that since $\A_{\alpha(2s-1)\,\dot\alpha}$ has more un-dotted than dotted spinorial indices, they belong to what is so-called chiral representation (see footnote \ref{footnotechiral}).

%the positive \textcolor{blue}{redundant: "positive helicity" can only be $+1$?  I'd write "the helicity can only be positive $+1$ ..."} {\bf this is only true if we have $\delta \A_{\alpha(2s-1)\,\dot\alpha}=\p_{\alpha\dot\alpha}\xi_{\alpha(2s-2)}$} helicity of the external legs can only be $+1$ due to gauge invariance while external legs with negative helicity receive no constraint for the chiral field representation.

%%%%%%%%%%%%%%%%%%%%%%%%%%%%%%%%%%%%%%
\subsection{Tree-level amplitudes} 
Since we are working with complex-valued fields and complex kinematics, the on-shell tree-level 3-point amplitudes are \emph{not} vanishing a priori. They will act as seeds to construct higher-point tree-level $S$-matrices. 
%%%%%%%%%%%%%%%%%%%%%%%%%%%%%%%%%%%
\paragraph{3-point amplitudes.} There are eight possible helicity configurations: 
\begin{align*}
    (+,+,+)\,,\ (-,+,+)\,,\ (+,-,+)\,,\ (+,+,-)\,,\ (-,-,+)\,,\ (-,+,-)\,,\ (+,-,-)\,,\ (-,-,-)\,,
\end{align*}
 at 3-points, where we recall that the positions of fields in the cubic vertices are important. Since we work with complex kinematics where $\tilde{\kappa}^{\dot\alpha}$ is not the complex conjugate of $\kappa^{\alpha}$, the momentum conservation implies:\footnote{It is important to recall that $k_i$ is dimensionless since we work with dimensionless spinors.}
\begin{align}\label{conservedmomentum}
    \sum_{i=1}^3k_i=\sum_{i=1}^3\kappa_i^{\alpha}\tilde{\kappa}_i^{\dot\alpha}=0\qquad \Leftrightarrow\qquad \langle i\,j\rangle=0\quad \text{or}\quad [i\,j]=0\,,\quad \forall \ i,j=1,2,3\,.
\end{align}
As a result, there can be non-vanishing 3-point amplitudes with complex kinematics whose forms must be written explicitly only in terms of angled or square brackets (cf., \cite{Benincasa:2007xk,Benincasa:2011kn,Benincasa:2011pg}). This analytic continuation between real momentum and complex kinematics is significant for constructing higher-multiplicity scattering amplitudes from 3-point building blocks \cite{Britto:2005fq}.

\medskip

The tree-level 3-point amplitudes of the Yang-Mills sector are given by substituting the polarization tensors \eqref{YMhel} to the cubic interaction:
\be\label{cubic}
\tilde{\delta}(2-(s_2+s_3-s_1))\int_{\M}\d^{4}\tx\,\left(\partial_{\alpha}{}^{\dot\gamma} \A_{\alpha(2s_1-1)\,\dot\gamma}\,\p_{\alpha\dot\sigma}\A^{\alpha(2s_2-1)\,\dot\beta}\p_{\alpha}{}^{\dot\sigma}\A^{\alpha(2s_3-1)}{}_{\dot\beta}\right)\,.
\ee
It is a simple computation to show that both $\cM_{3}(1^+,2^+,3^+)$ and $\cM_{3}(1^-,2^-,3^-)$ vanish. Therefore, we can concentrate on the six others helicity configurations. 

Notice that $\cM_{3}(1^+,2^{h_2},3^{h_3})$ vanishes on-shell, which leaves us with only three possible non-vanishing contributions at cubic order. Namely,
\begin{align}
    (-,+,+)\,,\quad (-,-,+)\,,\quad (-,+,-)\,.
\end{align}
These are the vertices that feature minimal couplings -- the couplings with lowest number of derivatives given a triplet of external spins $(s_1,s_2,s_3)$.

\medskip

Upon substituting \eqref{YMhel} to the cubic vertex \eqref{cubic}, we obtain the $(-,+,+)$ or $\overline{\mbox{MHV}}_3$ scattering amplitude as:
\be\label{MHV-bar0}
\cM_{3}(1_{s_1}^-,2_{s_2}^+,3_{s_3}^+)=\tilde{\delta}(s_1-s_2-s_3+2)\,\frac{[2\,3]^2\,\la\zeta_2\,1\ra^{2s_2-2}\,\la\zeta_3\,1\ra^{2s_3-2}}{\la\zeta_2\,2\ra^{2s_2-2}\,\la\zeta_3\,3\ra^{2s_3-2}}\,,
\ee
where the overall momentum conserving delta function has been suppressed, and we ignored the overall factor of $\im$. The following useful relation:
\be\label{momcon1}
\la\zeta_2\,1\ra\,[1\,3]+\la\zeta_2\,2\ra\,[2\,3]=0\,, \qquad \la\zeta_3\,1\ra\,[1\,2]+\la\zeta_3\,3\ra\,[3\,2]=0\,,
\ee
can be obtained on the support of momentum conservation. Then taking advantage of \eqref{momcon1}, we arrive at the following result for the $\overline{\mbox{MHV}}_3$ amplitude:
\be\label{MHV-bar}
\cM_{3}(1_{s_1}^-,2_{s_2}^+,3_{s_3}^+)=\tilde{\delta}(s_1-s_2-s_3+2)\,\frac{[2\,3]^{2s_2+2s_3-2}}{[1\,2]^{2s_3-2}\,[3\,1]^{2s_2-2}}\,.
\ee
This is in agreement with the 3-pt amplitudes of the self-dual higher-spin gravity~\cite{Krasnov:2021nsq,Steinacker:2022jjv}. Next, we find the following MHV${}_3$ amplitudes:
\be\label{MHV}
\cM_{3}(1_{s_1}^{-},2_{s_2}^{-},3_{s_3}^{+})=\tilde\delta(s_1-s_2-s_3+2)\,\frac{1}{2}\,\frac{\langle 1\,2\rangle^{2(s_2+s_3)-2}[3\,1]}{\langle 2\,3\rangle^{2s_3-2}\langle 3\,1\rangle }+(1\leftrightarrow 2)\,,
\ee
where we have symmetrized the positions of two negative helicity external fields. The result of $\cM_3(1^-_{s_1},2^+_{s_2},3^-_{s_3})$ is similar where we simply swap $(2\leftrightarrow 3)$.

Observe that while all constant spinors have dropped out of the final MHV${}_3$ amplitudes, due to the appearance of both angled and square brackets in \eqref{MHV}, it is obvious that the above MHV${}_3$ amplitude vanishes even with complex kinematic. In fact, we can directly verify this statement by writing
\begin{align}
    \frac{[3\,1]}{\langle 3\,1\rangle}\sim\frac{(k_1+k_3)^2}{\langle 3\,1\rangle^2}=0
\end{align}
by virtue of \eqref{conservedmomentum} and momentum conservation. In the next subsection, by projecting the MHV${}_3$ amplitudes to light-cone gauge, we argue that they are spurious and can be removed by a local field redefinition. This is the general feature of any (higher-spin) gauge theories described by the chiral representation whose interactions have two or higher number of derivatives \cite{Tran:2022amg}. It would be interesting to investigate the gauge-matter and matter-matter sectors in the HS-IKKT matrix model to see if this pattern persists. We leave this investigation for future work.

%%%%%%%%%%%%%%%%%%%%%%%%%%%%%%%%%%%%%%%%
%%%%%%%%%%%%%%%%%%%%%%%%%%%%%%%%%%%

\paragraph{4-point amplitudes.} Besides contributions from the exchanges, we also have potential contributions from the contact interaction:
\be\label{quartic}
\tilde{\delta}(s_1+s_2-s_3-s_4)\,\int_{\M}\d^{4}\tx\,\Big(\p_{\alpha\dot\gamma_1}\A^{\alpha(2s_1-1)\,\dot\sigma}\p_{\alpha}{}^{\dot\gamma_1}\A^{\alpha(2s_2-1)}{}_{\dot\sigma}\Big)\Big(\p^{\alpha}{}_{\dot\gamma_2}\A_{\alpha(2s_3-1)}{}^{\dot\tau}\p^{\alpha\dot\gamma_2}\A_{\alpha(2s_4-1)\,\dot\tau}\Big)\,
\ee
when considering 4-point scattering amplitudes. Below we consider different cases of the 4-point scattering amplitudes between massless modes in the Yang-Mills sector of HS-IKKT theory. 

\medskip

 \underline{Case 1:} Consider the 4-point amplitudes with helicity configuration $(+,+,+,+)$. The exchange channels of this amplitude are obtained by gluing $(+,+,+)$ and $(-,+,+)$ vertices together. Since $\cM_3(1_{s_1}^+,2_{s_2}^+,2_{s_1}^+)=0$, one can show that all $\cM_{4}(1_{s_1}^+,2_{s_2}^+,3_{s_3}^+,4_{s_4}^+)$-related exchange channels in the Yang-Mills sector of HS-IKKT theory are zero. By making suitable choices for the reference spinors, the contributions from the contact terms also vanish. As a result, the 4-point amplitude $\cM_{4}(1^+,2^+,3^+,4^+)=0$.

\medskip

\underline{Case 2:} Next, consider the 4-point $\cM_{4}(1^-,2^+,3^+,4^+)$ amplitude:
\begin{equation*}
   \cM_{4}(1_{s_1}^-,2_{s_2}^+,3_{s_3}^+,4_{s_4}^+)= \cA_{4}^s+\cA_{4}^t+\cA_{4}^u+\cA_{4}^{\mathrm{cont}}\,,
\end{equation*}
where $\cA_4^{\bullet}=\{\cA_4^s\,,\cA^t_4\,,\cA^u_4\,,\cA^{\mathrm{cont}}\}$ is the set of contributions coming from the $s,t,u$-channel exchanges and the contact interaction. Let us keep the spins arbitrary for now, and denote the spin of the exchange as $\omega$. It is a simple computation to show that:
\begin{align}
    \cA_4^s=(-)^{\Lambda_4}\tilde\delta(4-s_2-s_3-s_4+s_1)\frac{[12]^{-s_1+s_2+\omega}[34]^{s_3+s_4-\omega}}{(k_1+k_2)^2}f(\zeta_2,\zeta_3,\zeta_4)\,,
\end{align}
where $\Lambda_4=-s_1+s_2+s_3+s_4$, and
\begin{align}
    f(\zeta_2,\zeta_3,\zeta_4)=\Big(\frac{\langle 
    \zeta_2\,1\rangle}{\langle \zeta_2\,2\rangle}\Big)^{s_2}\Big(\frac{\langle 
    \zeta_4\,1\rangle}{\langle 
    \zeta_4\,2\rangle}\Big)^{s_1}\Big(\frac{\langle 
    \zeta_3\,4\rangle}{\langle 
    \zeta_3\,3\rangle}\Big)^{s_3}\Big(\frac{\langle 
    \zeta_4\,3\rangle}{\langle 
    \zeta_4\,4\rangle}\Big)^{s_4}\Big(\frac{\langle\zeta_3\,1\rangle\langle \zeta_4\,2\rangle}{\langle\zeta_3\,3\rangle\langle \zeta_4\,4\rangle}\Big)^{\omega}\,
\end{align}
is a rational function $f$ whose homogeneity in reference spinors of the positive helicity external particles is zero.

Using residue gauge freedom to set $\zeta_{2}^{\alpha}=\zeta_{3}^{\alpha}=\zeta_{4}^{\alpha}=\kappa_{1}^{\alpha}$, it follows that $f(\zeta_2,\zeta_3,\zeta_4)=0$. Thus,  $\cA_4^s=0$. Similarly, we also obtain $\cA^t_4=0$ and $\cA^u_4=0$. Lastly, the contribution of contact interaction reads
\be\label{4ptcontact}
\tilde{\delta}(s_1+s_2-s_3-s_4)\,\frac{[1\,2][\tilde\zeta\,2][3\,4]^2\langle 1\,\zeta_3\rangle^{2s_3-2s_2}\langle 1\,\zeta_4\rangle^{2s_4-2}\langle \zeta_2\,\zeta_3\rangle^{2s_2-2}}{[1\,\tilde\zeta_1]\langle 2\,\zeta_2\rangle^{2s_2-2}\langle 3\,\zeta_3\rangle^{2s_3-2}\langle 4\,\zeta_4\rangle^{2s_4-2}} -\,(2\leftrightarrow3)\,.
\ee
Since we have chosen $\zeta_i=\kappa_1$ (for $i\neq 1$), the contributions from the contact terms also vanish in this case. Therefore, the final result is $\cM_{4}(1_{s_1}^-,2_{s_2}^+,3_{s_3}^+,4_{s_4}^+)=0$. The vanishing of this amplitude reflects the deep connection between chiral field representations and self-dual theories such as self-dual Yang-Mills, self-dual gravity, or self-dual/chiral higher-spin theories~\cite{Metsaev:1991mt,Metsaev:1991nb,Ponomarev:2016lrm,Ponomarev:2017nrr,Metsaev:2019aig,Metsaev:2019dqt,Tsulaia:2022csz}. Namely, tree-level amplitudes compose of $(-,+,+)$ vertices with all but one positive (or negative) helicity vanish for any number of external legs greater than or equal to three. 

\medskip

\underline{Case 3:} Next, we consider the 4-point MHV amplitude $\cM_4(1^-_{s_1},2^-_{s_2},3^+_{s_3},4^+_{s_4})$. We shall fix
\be\label{pgf}
    \begin{split}
    \zeta_{3}^{\alpha}=\zeta_{4}^{\alpha}=\kappa_1^{\alpha}\,,\\
    \tilde\zeta_1^{\dot\alpha}=\tilde\zeta_2^{\dot\alpha}=\tilde\kappa_4^{\dot\alpha}\,,
    \end{split}
\ee
to simplify the computation. The $s$-channel contribution is given by:\footnote{Once again, we ignore overall factor.}
\be\label{MHVschan}
\cA_{4}^{s}=\frac{\tilde{\delta}(s_1-s_2)\tilde\delta(4-s_3-s_4)}{2}\,\frac{\la1\,2\ra^{2s_1-2}\,\la\zeta_3\,1\ra^2\,\la\zeta_3\,2\ra^2\,[3\,4]^2}{\la3\,\zeta_3\ra^{2s_3-2}\,\la4\,\zeta_3\ra^{2s_4-2}}\,+\:(1\leftrightarrow2)\,.
\ee
Here the spin constraints fix the two negative helicity particles to have identical spin, while the spins of the two positive helicity particles have to sum up to four. Since $\zeta_3^{\alpha}=\kappa_1^{\alpha}$, the $s$-channel vanishes. Similar computations also lead to the vanishing of the $t$- and $u$- channels, as well as the contributions coming from the contact terms. As a result, the 4-point MHV amplitude is
\be\label{MHV4pt}
\cM_4(1^-_{s_1},2^-_{s_2},3^+_{s_3},4^+_{s_4})=0\,.
\ee
We find similar results in the case where the negative helicity particles are not consecutive, e.g. $\cM_4(1^-_{s_1},2^+_{s_2},3^-_{s_3},4^+_{s_4})=0$.

\paragraph{n-point amplitudes.} As a consequence of the above considerations, we conclude that all $n$-point tree-level amplitudes of the Yang-Mills sector in HS-IKKT model vanish, since they cannot be constructed from lower point amplitudes.

%%%%%%%%%%%%%%%%%%%%%%%%%%%%%%%%%%%%%%%%%%

\subsection{HS-IKKT vertices in the light-cone gauge}

The fact that \eqref{MHV4pt} is vanishing for generic higher-spin fields with $s>1$ is, in fact, not a surprise. As observed in \cite{Tran:2022amg}, what determines the existence of non-trivial higher-spin tree-level scattering is not spin but rather the number of transverse derivatives in the cubic interactions of type $\cV_3^{-\pm +}$. In particular, if the number of transverse derivatives is greater than or equal to two, non-trivial tree-level scattering amplitudes would be very unlikely to exist. To support this statement, let us project the cubic interactions \eqref{cubic} to the light-cone gauge in momentum space using the dictionary in \cite{Chalmers:1998jb,Bengtsson:2016jfk}. In particular, the map between spinors and momenta in the light-cone gauge are:
\begin{align}
    i]=2^{1/4}\binom{\bar{\kbold}_i\,\beta_i^{-1/2}}{-\beta_i^{1/2}}\,,\qquad  i\rangle =2^{1/4}\binom{\kbold_i\,\beta_i^{-1/2}}{-\beta_i^{1/2}}\,,
\end{align}
where $\kbold_i^+\equiv \beta_i$ and 
\begin{align}
    k_i^{\alpha\dot\alpha}=\begin{pmatrix} \beta_i & \bar{\kbold}_i\\
    \kbold_i & \kbold_i^-
    \end{pmatrix}\,.
\end{align}
Note that $\bar{\kbold}_i$ and $\kbold_i$ are referred to as \emph{transverse} derivatives. Using the above, we can express the square and angle brackets as
\begin{align}
    [i\,j]=\sqrt{\frac{2}{\beta_i\beta_j}}\PPb_{ij}\,,\qquad \langle i\,j\rangle=\sqrt{\frac{2}{\beta_i\beta_j}}\PP_{ij}\,
\end{align}
for $\PPb_{ij}=\bar \kbold_i\beta_j-\bar{\kbold}_j\beta_i$ and $\PP_{ij}= \kbold_i\beta_j- \kbold_j\beta_i$. By virtue of momentum conservation, one can show that
\begin{align} \label{PP1}
   \PPb_{12}=\PPb_{23}=\PPb_{31}= \PPb=\frac13\left[ (\beta_1-\beta_2)\bar{\kbold}_3+(\beta_2-\beta_3)\bar{\kbold}_1+(\beta_3-\beta_1)\bar{\kbold}_2\right]\,
\end{align}
at the level of 3-point amplitudes.\footnote{Note that this effect can also be achieved in the spinor formalism if we use the parametrization \eqref{lambdaparametrization}.} Thus, in terms of these new variables, the cubic vertices \eqref{cubic} reduce to
\begin{align}
    V_3^{\text{HS-IKKT}}=(x\,\PPb+y\,\PP^{2s_2-1})\PPb\,,
\end{align}
schematically, and $x,y$ are coefficients in terms of $\beta_i$ variables. According to light-cone recipe (see e.g. \cite{Metsaev:1991mt,Metsaev:1991nb,Metsaev:2005ar,Ponomarev:2016lrm}), one can always make a local field redefinition at cubic order if there is a combination of type $\PP\PPb$ to remove unnecessary data from the interactions. This goes hand in hand with the fact that there must be only one type of bracket in the final form of cubic amplitudes (cf., \eqref{conservedmomentum}): either angled or square bracket, but not both of them at the same time. As a result, the cubic interaction of HS-IKKT theory reduces further to
\begin{align}
    V_3^{\text{HS-IKKT}}\rightsquigarrow x\,\PPb^2\simeq V_3^{\text{self-dual HSGR}}
\end{align}
Thus, HS-IKKT theory in the flat limit is `secretly' a supersymmetric chiral higher-spin theory with two-derivative interactions (see e.g. \cite{Metsaev:2019aig,Metsaev:2019dqt,Tsulaia:2022csz}).\footnote{
We choose the word \emph{chiral} instead of self-dual here because the HS-IKKT theory has scalar fields in the spectrum.}

%%%%%%%%%%%%%%%%%%%%%%%%%%%%%%%%%

%%%%%%%%%%%%%%%%%%%%%%%%%%%%%%%%%
%%%%%%%%%%%%%%%%%%%%%%%%%%%%%%%%%
\section{Discussion}\label{sec:7}

In this work, we established the connections between the almost-commutative 4-hyperboloid and non-compact semi-classical twistor space $\P^{1,2}_N$ in the spinor formalism. Using this as a background in the IKKT model, we studied the resulting higher-spin gauge theory for two spaces with distinct signatures: $(i)$ a Euclidean 4-hyperboloid; $(ii)$ and a FLRW-like cosmological spacetime. We also furnished a simple vectorial description of HS-IKKT in the flat limit of these two cases, which simplifies  some of the technical steps in counting and organizing the degrees of freedom of higher-spin modes in HS-IKKT in \cite{Sperling:2018xrm} and \cite{Steinacker:2019awe}.

Armed with this setup,
we then shown that all tree-level $n$-point on-shell amplitudes (for $n\geq 4$) with an appropriate analytic continuation of the massless sector within the Yang-Mills part of the Euclidean HS-IKKT theory vanish in the flat limit. This result is expected, since HS-IKKT is a `parity-violating' higher-spin theory featuring two-derivative interactions when we express it in terms of chiral representation.\footnote{This representation gives the lowest number of derivatives in the interactions, and
unveils an 
intriguing relation between standard No-go theorems \cite{Weinberg:1964ew,Coleman:1967ad}
 in flat space and the number of derivatives in the interaction.} Indeed, as observed in \cite{Tran:2022amg}, massless higher-spin theories constructed from the chiral representation in the flat space with the number of derivatives in the cubic vertices higher than one must have trivial tree-level scattering amplitudes. Thus, the massless sector of HS-IKKT is also in argreement with the result of \cite{Tran:2022amg}.
% \textcolor{blue}{A simple intuitive way to see this is that the scale of the derivative interactions is set by the background curvature, which vanishes in the flat limit (? can we see this explicitly? why not NC scale?)}
 %\textcolor{blue}{why do we need highers spin here? It is a statement about higher derivative theories more generally, no?}
%\textcolor{blue}{It is also consistent with the fact that noncommutative $U(1)$ Yang-Mills theory reduces to a non-interacting Maxwell-like theory at the semi-classical level, i.e. without taking into account quantum effects}.
This can be explained by projecting the cubic vertices of the Yang-Mills sector in the flat limit to the light-cone gauge, we observe that the MHV${}_3$ amplitudes vanish so no other higher-point amplitudes can be formed using just the non-vanishing $\overline{\text{MHV}}_3$ amplitudes. 

This allows to identify the massless YM sector with self-dual higher-spin gravity with 2-derivative interactions considered in \cite{Krasnov:2021nsq}. In this sense, the massless sector of HS-IKKT theory in the flat limit falls into the class of (quasi-)chiral theories. We conclude that $S=1$ at tree-level, at least for the modes under consideration. However, the triviality of the $S$-matrix at tree level does not imply that the theory is trivial or not interesting; rather, it should be viewed as a consistency check. 
It would be then interesting to study the scattering amplitudes of HS-IKKT or any other higher-spin theories at loop-level, see the discussions in e.g. \cite{Skvortsov:2018jea,Adamo:2022lah,Monteiro:2022xwq}. 

The structure of the higher-spin gauge theory under consideration is quite interesting.
In contrast to more conventional attempts to formulate higher-spin gauge theory using free differential algebra (see e.g. \cite{Sharapov:2022awp,Sharapov:2022wpz,Sharapov:2022nps,Didenko:2022qga}) and the chiral formulation (see e.g. \cite{Haehnel:2016mlb,Adamo:2016ple,Tran:2021ukl,Tran:2022tft,Herfray:2022prf,Zinoviev:2021cmi,Basile:2022mif}), the framework of (HS-)IKKT matrix model 
leads naturally to a local action, albeit at the expense of manifest local Lorentz invariance  and a larger number of propagating degrees of freedom compared to massless or massive higher-spin gravities in 4-dimensional spacetime. This is due to the fact that the higher-spin fields are not divergence-free as discussed\footnote{Recall that in Lorentzian signature we have $2(2s+1)$ dof. and in Euclidean signature we have $3(2s+1)$ dof. for the would-be massive higher-spin fields of HS-IKKT theory.} in Section \ref{sec:shift-vector}. The lack of local Lorentz invariance is also reflected in the preferred frame, which encodes not only a metric but also a dilaton and an axion.
The Levi-Civita connection is accordingly replaced by the 
%It is well-known that a non-commutative spacetime possesses torsion   \textcolor{blue}{This is an overstatement. Will improve}, and this is no exception to (HS-)IKKT theory. As a consequence, there is no local Lorentz invariance to start with, and the usual Levi-Civita connection is modified by an extra term involving the contorsion. 
Weitzenb\"ock connection, whose torsion encodes   the Riemannian curvature; see e.g. \cite{Steinacker:2020xph,Fredenhagen:2021bnw,Battista:2022vvl}. Nevertheless, the propagation of all modes is governed by a universal effective metric, and the preferred frame naturally leads to a  
Cartan--type framework to 
describe gravitational couplings in (HS-)IKKT model. While the bare action is of Yang-Mills type, the Einstein-Hilbert action does arise at one loop, under certain assumptions for the background \cite{Steinacker:2023myp,Steinacker:2021yxt}. Non-Abelian gauge theory would then arise on a stack of such background branes, coupled to the effective metric. It is hence clear that the theory does contain interesting physics, even if the local scattering of its abelian sector vanishes at tree level.

From the physics point of view, it would be important to elaborate in more detail the spinorial formulation on the FLRW background with Minkowski signature. This leads to an unusual type of spinors adapted to the space-like isometries rather than the local Lorentz invariance, which we have only briefly touched upon. Due to the common origin from $\PT$, a close relation with the present Euclidean computation is expected. Moreover, the rather complicated organization of modes in \cite{Steinacker:2019awe,Sperling:2019xar} is expected to simplify in the spinorial formalism, which needs to be studied in more detail. This may allow to compute amplitudes directly in Minkowski signature, while avoiding complexification of the Euclidean case. This problem will be addressed in future work.

%\textcolor{blue}{I dont understand the following}
%Therefore, it is plausible  to expect that the HS-IKKT theory may admit a free differential algebra description the moment we introduce higher-spin fields to mitigate the effect of the Lorentz violation. 

%\medskip
%
%\textcolor{blue}{I dont understand how the following is related to the present paper. }

%It also allows us to reproduce all cubic interactions between massless fields previously only known in the light-cone gauge \cite{Bengtsson:1983pd,Bengtsson:1986kh,Metsaev:1991mt,Metsaev:1991nb,Metsaev:2005ar}. This chiral formulation for local higher-spin theories \cite{Krasnov:2021nsq} also enable us to reach a new sector of quasi-chiral higher-spin theories by considering a deformation away from the self-dual/chiral sector as in the case of Yang-Mills theory~\cite{Chalmers:1996rq}. 

%In addition to the massless local higher-spin models mentioned above, the chiral formulation can also be used to construct theories of partial massless and massive higher-spin particles, see e.g. . 

%Celestial story \cite{Ren:2022sws,Monteiro:2022lwm,Monteiro:2022xwq}

%Higher-spin holography\cite{Ponomarev:2022ryp,Ponomarev:2022qkx}

%%%%%%%%%%%%%%%%%%%%%%%%%%%%%%%%%%%%%%%%
\acknowledgments
We appreciates useful discussions with Zhenya Skvortsov. TT is grateful to Tim Adamo for useful discussion during the collaboration of \cite{Adamo:2022lah}. This research was partially completed at Corfu Summer Institute 2022 and the Humboldt Kolleg on ``Noncommutative and generalized geometry in string theory, gauge theory and related physical models''. The work of TT is partially supported by the Fonds de la Recherche Scientifique under Grants No. F.4503.20 (HighSpinSymm), No. F.4544.21 (Higher- SpinGraWave), and the funding from the European Research Council (ERC) under Grant No. 101002551. The work of HS is supported by the Austrian Science Fund (FWF) grant P32086.
%%%%%%%%%%%%%%%%%%%%%%%%%%%%%%%%%%%%

\appendix

%%%%%%%%%%%%%%%%%%%%%%%%%%%
\section{On the shift symmetry in the flat limit}
\label{app:1}

We verify that the $\vartheta$ transformation  \eqref{gaugetransform-theta} is \emph{not} a symmetry of HS-IKKT model in the flat limit. Recall that 
\begin{align}\label{decompositionappA}
    \sa_{\alpha\dot\alpha}=\sum_{s=0}^{\infty}=\lambda^{\beta(s)}\hat{\lambda}^{\beta(s)}\cA_{\beta(2s)|\alpha\dot\alpha}=\sum_{s=0}^{\infty}\lambda^{\beta(s)}\hat{\lambda}^{\beta(s)}\Big[A_{(\beta(2s)\alpha)\,\dot\alpha}+\eps_{\alpha\beta}\sA_{\beta(2s-1)\,\dot\alpha}\Big]\,,
\end{align}
and
\begin{align}
    \delta_{\vartheta} \cA_{\beta(2s)|\alpha\dot\alpha}=\epsilon_{\alpha\beta}\,\vartheta_{\beta(2s-1)\,\dot\alpha}\,.
\end{align}
Using the above, the linearized action in the flat limit transforms as
\begin{equation}
    \begin{split}
        \delta_{\vartheta}S_2=&2\sum_{s_i}\int \mho \lambda^{\beta(s_1)}\hat\lambda^{\beta(s_1)}\lambda^{\tau(s_2)}\hat\lambda^{\tau(s_2)}\{\ty^{\alpha}{}_{\dot\gamma},\cA_{\beta(2s_1)}{}^{\alpha\,\dot\gamma}\}\{\ty_{\alpha\dot\sigma},\eps_{\alpha\tau}\vartheta_{\tau(2s_2-1)}{}^{\dot\sigma}\}\,,
    \end{split}
\end{equation}
where we note that the term $\{\ty^{\alpha}{}_{\dot\gamma},\ty^{\alpha\dot\gamma}\}\{\sa_{\alpha\dot\sigma},\sa_{\alpha}{}^{\dot\sigma}\}$ vanishes in the flat limit using spinorial description of HS-IKKT theory (see Appendix \ref{app:2}) and therefore can be neglected henceforth. The above can be reduced further to
\begin{equation}\label{varthetastep1}
    \begin{split}
        \delta_{\vartheta}S_2=&\sum_{s_i}\int \mho \lambda^{\beta(s_1)}\hat\lambda^{\beta(s_1)}\lambda^{\tau(s_2)}\hat\lambda^{\tau(s_2)}\{\ty_{\tau\dot\gamma},\cA_{\beta(2s_1)}{}^{\alpha\,\dot\gamma}\}\{\ty_{\alpha\dot\sigma},\vartheta_{\tau(2s_2-1)}{}^{\dot\sigma}\}\\
        &+\sum_{s_i}\int \mho \lambda^{\beta(s_1)}\hat\lambda^{\beta(s_1)}\lambda^{\tau(s_2)}\hat\lambda^{\tau(s_2)}\{\ty^{\alpha}{}_{\dot\gamma},\cA_{\beta(2s_1)\tau}{}^{\dot\gamma}\}\{\ty_{\alpha\dot\sigma},\vartheta_{\tau(2s_2-1)}{}^{\dot\sigma}\}\,.%\\
    %&+2\sum_{s_i}\int\mho \lambda^{\rho(s_3)}\hat\lambda^{\rho(s_3)}\lambda^{\zeta(s_4)}\hat\lambda^{\zeta(s_4)}\{\ty_{\rho\dot\gamma},\ty^{\alpha\dot\gamma}\}\{\vartheta_{\rho(2s_3-1)\,\dot\sigma},\cA_{\zeta(2s_4)\alpha}{}^{\dot\sigma}\}\,.
    \end{split}
\end{equation}
The first term denoted as $\circled{I}$ in \eqref{varthetastep1} can be written as
\begin{equation}
    \begin{split}
        \circled{I}=&-\sum_{s_i}\int \mho \lambda^{\beta(s_1)}\hat\lambda^{\beta(s_1)}\lambda^{\tau(s_2)}\hat\lambda^{\tau(s_2)}\,\vartheta_{\tau(2s_2-1)}{}^{\dot\sigma}\{\ty_{\alpha\dot\sigma},\{\ty_{\tau\dot\gamma},\cA_{\beta(2s_1)}{}^{\alpha\,\dot\gamma}\}\}\,.
    \end{split}
\end{equation}
after integrating by part. Since 
\begin{align}
    \{\ty_{\tau\dot\gamma},\cA_{\beta(2s_1)}{}^{\alpha\dot\gamma}\}&\simeq  (\lambda_{\tau}\hat\lambda^{\delta}+\lambda^{\delta}\hat\lambda_{\tau})\p_{\delta\dot\gamma}\cA_{\beta(2s_1)}{}^{\alpha\dot\gamma}\,,
\end{align}
we get
\begin{equation}\label{random1}
    \begin{split}
        \circled{I}&\simeq\int\langle \lambda\,\hat\lambda\rangle \lambda^{\beta(s_1)}\hat\lambda^{\beta(s_1)}\lambda^{\tau(s_2)}\hat\lambda^{\tau(s_2-1)}\hat\lambda^{\delta}[\textcolor{red!70!black!100}{\lambda_{\alpha}\hat\lambda^{\zeta}}+\textcolor{green!60!black!100}{\hat\lambda_{\alpha}\lambda^{\zeta}}]\,\vartheta_{\tau(2s_2-1)}{}^{\dot\sigma}\p_{\zeta\dot\sigma}\p_{\delta\dot\gamma}\cA_{\beta(2s_1)}{}^{\alpha\,\dot\gamma}\\
        &-\int\langle \lambda\,\hat\lambda\rangle \lambda^{\beta(s_1)}\hat\lambda^{\beta(s_1)}\lambda^{\tau(s_2-1)}\hat\lambda^{\tau(s_2)}\lambda^{\delta}[\textcolor{red!70!black!100}{\lambda_{\alpha}\hat\lambda^{\zeta}}+\textcolor{green!60!black!100}{\hat\lambda_{\alpha}\lambda^{\zeta}}]\,\vartheta_{\tau(2s_2-1)}{}^{\dot\sigma}\p_{\zeta\dot\sigma}\p_{\delta\dot\gamma}\cA_{\beta(2s_1)}{}^{\alpha\,\dot\gamma}
    \end{split}
\end{equation}
Observe that by grouping the red and green terms in \eqref{random1} together and using the identity $\lambda^{[\alpha}\hat\lambda^{\beta]}=-\frac{1}{2}\eps^{\alpha\beta}\langle\lambda\,\hat\lambda\rangle$, we can simplify the above to
\begin{equation}
    \begin{split}
        \circled{I}\simeq&-\frac{1}{2}\int \textcolor{red!70!black!100}{\lambda_{\alpha}\hat\lambda^{\zeta}}\langle \lambda\,\hat\lambda\rangle^2  \lambda^{\beta(s_1)}\hat\lambda^{\beta(s_1)}\lambda^{\tau(s_2-1)}\hat\lambda^{\tau(s_2-1)}\,\vartheta_{\tau(2s_2-2)\delta}{}^{\dot\sigma}\p_{\zeta\dot\sigma}\p^{\delta}{}_{\dot\gamma}\cA_{\beta(2s_1)}{}^{\alpha\,\dot\gamma}\\
        &+\frac{1}{2}\int \textcolor{green!60!black!100}{\hat\lambda_{\alpha}\lambda^{\zeta}} \langle\lambda\,\hat\lambda\rangle^2 \lambda^{\beta(s_1)}\hat\lambda^{\beta(s_1)}\lambda^{\tau(s_2-1)}\hat\lambda^{\tau(s_2-1)}\,\vartheta_{\tau(2s_2-2)\delta}{}^{\dot\sigma}\p_{\zeta\dot\sigma}\p^{\delta}{}_{\dot\gamma}\cA_{\beta(2s_1)}{}^{\alpha\,\dot\gamma}\,,
    \end{split}
\end{equation}
which can be reduced one more time to 
\begin{equation}
    \begin{split}
        \circled{I}\simeq\frac{1}{4}\int\langle \lambda\,\hat\lambda\rangle^3  \lambda^{\beta(s_1)}\hat\lambda^{\beta(s_1)}\lambda^{\tau(s_2-1)}\hat\lambda^{\tau(s_2-1)}\,\vartheta_{\tau(2s_2-2)\delta}{}^{\dot\sigma}\p_{\zeta\dot\sigma}\p^{\delta}{}_{\dot\gamma}\cA_{\beta(2s_1)}{}^{\zeta\,\dot\gamma}\,.
    \end{split}
\end{equation}

Decomposing $\cA_{\beta(2s)|\alpha\dot\alpha}=A_{\beta(2s)\alpha\,\dot\alpha}+\eps_{\alpha\beta}\sA_{\beta(2s-1)\,\dot\alpha}$, we get
\begin{align}
    \p_{\zeta\dot\sigma}\p^{\delta}{}_{\dot\gamma}\cA_{\beta(2s_1)}{}^{\zeta\,\dot\gamma}= \p_{\zeta\dot\sigma}\p^{\delta}{}_{\dot\gamma}A_{\beta(2s_1)}{}^{\zeta\,\dot\gamma}-\p_{\beta\dot\sigma}\p^{\delta}{}_{\dot\gamma}\sA_{\beta(2s_1-1)}{}^{\dot\gamma}\,.
\end{align}
The contribution associated with $A_{\beta(2s_1)}{}^{\zeta\,\dot\gamma}$ vanishes since there are a pair of $\lambda^{\beta}\hat\lambda^{\beta}$ that contract with $A_{\beta(2s_1)}{}^{\zeta\,\dot\gamma}$. Therefore, after integrating out fiber coordinates we obtain a trace in $A_{\beta(2s_1)}{}^{\zeta\,\dot\gamma}$ which is zero due to tracelessness. This means that
\begin{align}
    \circled{I}=-\frac{1}{4}\int\langle \lambda\,\hat\lambda\rangle^3  \lambda^{\beta(s_1)}\hat\lambda^{\beta(s_1)}\lambda^{\tau(s_2-1)}\hat\lambda^{\tau(s_2-1)}\,\vartheta_{\tau(2s_2-2)\delta}{}^{\dot\sigma}\p_{\beta\dot\sigma}\p^{\delta}{}_{\dot\gamma}\sA_{\beta(2s_1-1)}{}^{\dot\gamma}\,.
\end{align}
Notice that $\circled{I}\neq 0$. So the only way $\vartheta$ can be a symmetry of HS-IKKT theory is that $\circled{I}$ combines with other contributions in \eqref{varthetastep1} to give zero. We show below that it is \emph{not} the case. 

Consider the second line of \eqref{varthetastep1} for which we denote as $\circled{II}$. Using \eqref{effectivevielbeins1}, we get
\begin{equation}
    \begin{split}
        \circled{II}&\simeq\int \mho \lambda^{\beta(s_1)}\hat\lambda^{\beta(s_1)}\lambda^{\tau(s_2)}\hat\lambda^{\tau(s_2)}\big[\lambda^{\alpha}\hat\lambda^{\delta}+\lambda^{\delta}\hat\lambda^{\alpha}\big]\big[\lambda_{\alpha}\hat\lambda^{\zeta}+\lambda^{\zeta}\hat\lambda_{\alpha}\big]\times\\
        &\times \p_{\delta\dot\gamma}\cA_{\beta(2s_1)\tau}{}^{\dot\gamma}\p_{\zeta\dot\sigma}\vartheta_{\tau(2s_2-1)}{}^{\dot\sigma}\\&\simeq\sum_{s_i}\int \mho \lambda^{\beta(s_1)}\hat\lambda^{\beta(s_1)}\lambda^{\tau(s_2)}\hat\lambda^{\tau(s_2)}\langle\lambda\,\hat\lambda\rangle\big(\hat\lambda^{\delta}\lambda^{\zeta}-\lambda^{\delta}\hat\lambda^{\zeta}\big)\p_{\delta\dot\gamma}\cA_{\beta(2s_1)\tau}{}^{\dot\gamma}\p_{\zeta\dot\sigma}\vartheta_{\tau(2s_2-1)}{}^{\dot\sigma}\,,
    \end{split}
\end{equation}
which can be simplifed further to
\begin{equation}
    \begin{split}
       \circled{II} &\simeq\frac{1}{2}\sum_{s_i}\int \mho \lambda^{\beta(s_1)}\hat\lambda^{\beta(s_1)}\lambda^{\tau(s_2)}\hat\lambda^{\tau(s_2)}\langle\lambda\,\hat\lambda\rangle^2\eps^{\delta\zeta}\p_{\delta\dot\gamma}\cA_{\beta(2s_1)\tau}{}^{\dot\gamma}\p_{\zeta\dot\sigma}\vartheta_{\tau(2s_2-1)}{}^{\dot\sigma}\\
        &\simeq-\frac{1}{2}\sum_{s_i}\int \mho \lambda^{\beta(s_1)}\hat\lambda^{\beta(s_1)}\lambda^{\tau(s_2)}\hat\lambda^{\tau(s_2)}\langle\lambda\,\hat\lambda\rangle^2\cA_{\beta(2s_1)\tau\,\dot\gamma}\Box\vartheta_{\tau(2s_2-1)}{}^{\dot\gamma}
    \end{split}
\end{equation}
where we have made an integration by part and used the identity $\p_{\delta\dot\gamma}\p^{\delta}{}_{\dot\sigma}=-\Box\eps_{\dot\gamma\dot\sigma}$. 

Since one of the spinors $\lambda^{\tau}$ or $\hat\lambda^{\tau}$ is contracted with $\cA_{\alpha(2s_1)\tau\,\dot\gamma}$, it results in a trace of this coefficient when we integrate out fiber coordinates. However, based on the decomposition \eqref{decompositionappA}, it is clear that we will obtain $A^{\alpha}{}_{\alpha\beta(2s-2)\,\dot\alpha}$ which is traceless. The remaining contribution from the second modes $\sA$ gives
\begin{align}
   \circled{II}\simeq \int\mho \langle\lambda\,\hat\lambda\rangle^2 \lambda^{\beta(s_1)}\hat\lambda^{\beta(s_1)}\lambda^{\tau(s_2)}\hat\lambda^{\tau(s_2)}\eps_{\beta\tau}\sA_{\beta(2s_1-1)\,\dot\gamma}\Box\vartheta_{\tau(2s_2-1)}{}^{\dot\gamma}\,,
\end{align}
which clearly does not vanish by itself. In addition, $\circled{I}$ and $\circled{II}$  will not cancel each others. As such $\vartheta$ is not a symmetry of the HS-IKKT model in the flat limit.\footnote{Note that $\vartheta$ is nevertheless a symmetry of the self-dual sector of HS-IKKT in the flat limit \cite{Steinacker:2022jjv}.} As a result, there are more propagating degrees of freedom in HS-IKKT compared to massless higher-spin fields in (A)dS$_{4,5}$. Intriguingly, in the flat limit, the would-be massive degrees of freedom modes can be mapped into a helicity basis. We can further select all the massless modes by imposing the Lorenz gauge condition $\p^{\gamma\dot\gamma}\A_{\alpha(2s-2)\gamma\,\dot\gamma}=0$ as discussed in the main text.

%%%%%%%%%%%%%%%%%%%%%%%%%%%
\section{Averaging over the fibers}\label{app:2}
The explicit spacetime action 
%\textcolor{blue}{well this is euclidean ...}
of the Yang-Mills sector in the Euclidean case can be obtained by avaraging over the $\P^1$ fiber  using \eqref{eq:bridge}. Consider the first term in \eqref{kineticYM} where
\begin{align}
   I_1=2\int_{\P^1}\tK\,\{\ty^{\alpha}{}_{ \dot\gamma},\sa^{\alpha\dot\gamma}\}\{\ty_{\alpha\dot\sigma},\sa_{\alpha}{}^{\dot\sigma}\}\,,\qquad \tK=\frac{\langle \lambda\, \d\lambda\rangle \wedge \langle \hat\lambda\, \d\hat\lambda\rangle}{\langle \lambda\, \hat\lambda\rangle^2}\,.
\end{align}
In the flat limit, the main contribution from the Poisson brackets yield
\begin{align}
   I_1=2\sum_{s,s'} \int_{\P^1}\tK\, \cE^{\alpha}{}_{\dot\gamma,\beta\dot\beta}\pl^{\beta\dot\beta}\sa^{\alpha\,\dot\gamma}\cE_{\alpha\dot\sigma,\delta\dot\delta}\pl^{\delta\dot\delta}\sa_{\alpha}{}^{\dot\sigma}\,.
\end{align}
where $\cE^{\alpha\dot\alpha\beta\dot\beta}\simeq 2\lambda^{(\alpha}\hat\lambda^{\beta)}\epsilon^{\dot\alpha\dot\beta}$. The contraction between two effective spinor vielbein gives

%The condition of having equal number of $\lambda,\hat\lambda$ then forces $s=s'$. \textcolor{blue}{? I dont see $s$ here} 
\begin{align}
    \cE^{\alpha}{}_{\dot\gamma,\beta\dot\beta}\cE_{\alpha\dot\sigma,\delta\dot\delta}=-2\langle \lambda\,\hat\lambda\rangle \lambda_{[\beta}\hat\lambda_{\delta]}\epsilon_{\dot\gamma\dot\beta}\epsilon_{\dot\sigma\dot\delta}=\epsilon_{\beta\delta}\epsilon_{\dot\gamma\dot\beta}\epsilon_{\dot\sigma\dot\delta}\,\langle \lambda\,\hat\lambda\rangle^2\,, 
\end{align}
where $2\lambda_{[\alpha}\hat\lambda_{\beta]}=-\epsilon_{\alpha\beta}\langle \lambda\,\hat\lambda\rangle$. We obtain
\begin{equation}\label{cancelwithI3}
    \begin{split}
    I_1=&+2\int_{\P^1}\tK\,\langle \lambda\,\hat\lambda\rangle^2\p_{\circ\dot\gamma}\sa^{\alpha\,\dot\gamma}\,\p^{\circ}{}_{\dot\sigma}\sa_{\alpha}{}^{\dot\sigma}-8\int_{\P^1}\tK\, \lambda_{(\alpha_2}\hat\lambda_{\circ)}\lambda_{(\alpha_1}\hat\lambda_{\bullet)}\p^{\circ}{}_{\dot\gamma}\sa^{\alpha_1\dot\gamma}\p^{\bullet}{}_{\dot\sigma}\sa^{\alpha_2\dot\sigma}\,.
    \end{split}
\end{equation}
Observe that the second term above cancels the gauging fixing term 
\begin{equation}
    \begin{split}
    \int 2\{\ty_{\alpha\dot\alpha},\sa^{\alpha\dot\alpha}\}^2=8\int_{\P^1}\lambda_{(\alpha_2}\hat\lambda_{\circ)}\lambda_{(\alpha_1}\hat\lambda_{\bullet)}\p^{\circ}{}_{\dot\gamma}\sa^{\alpha_1\dot\gamma}\p^{\bullet}{}_{\dot\sigma}\sa^{\alpha_2\dot\sigma}\,.
    \end{split}
\end{equation} 
Thus, 
\begin{align}
    I_1+\int 2\{\ty_{\alpha\dot\alpha},\sa^{\alpha\dot\alpha}\}^2 = 2\int_{\P^1}\tK\,\langle \lambda\,\hat\lambda\rangle^2\p_{\circ\dot\gamma}\sa^{\alpha\,\dot\gamma}\,\p^{\circ}{}_{\dot\sigma}\sa_{\alpha}{}^{\dot\sigma}\,.
\end{align}
Next, the second contribution in \eqref{kineticYM} reads\be
    \begin{split}
    I_2&=\int_{\P^1}\tK\,\{\ty^{\alpha}{}_{\dot\gamma},\ty^{\alpha\dot\gamma}\}\{\sa_{\alpha\dot\sigma},\sa_{\alpha}{}^{\dot\sigma}\}=-2\int_{\P^1}\tK\{\ty^{\alpha\dot\gamma},\sa_{\alpha}{}^{\dot\sigma}\}\{\ty^{\alpha}{}_{\dot\gamma},\sa_{\alpha\dot\sigma}\}
    \end{split}
\ee
where we have made an integration by parts and used Jacobi identity. In the flat limit, $I_2$ reduces to
\begin{align}
    I_2=\frac{1}{4}\int_{\P^1}\tK \langle \lambda\,\hat\lambda\rangle^2\eps^{\alpha_1\alpha_2}\Big(\sa_{\alpha_1}{}^{\dot\sigma}\Box \sa_{\alpha_2\dot\sigma}-\sa_{\alpha_2\dot\sigma}\Box\sa_{\alpha_1}{}^{\dot\sigma}\Big)=0\,.
\end{align}
Therefore, the contribution coming from the term $\{\ty^{\alpha}{}_{\dot\gamma},\ty^{\alpha\dot\gamma}\}\{\sa_{\alpha\dot\sigma},\sa_{\alpha}{}^{\dot\sigma}\}$ vanishes in the flat limit. We conclude that the kinetic action in the flat limit reduces to
\begin{align}\label{K2generation}
    S_2=\int \d^4 \tx\int \tK\, \sa^{\alpha
    \,\dot\alpha}\Box \sa_{\alpha\,\dot\alpha}\,.
\end{align}
Expanding $\sa_{\alpha\dot\alpha}=\lambda^{\beta(s)}\hat\lambda^{\beta(s)}\Big[A_{\beta(2s)\alpha\,\dot\alpha}+\eps_{\alpha\beta}\sA_{\beta(2s-1)\,\dot\alpha}\Big]$, we  can elaborate the following explicit cases:
%\textcolor{blue}{what is level? $s$?} {\bf yes} 
\begin{itemize}
    \item[$\bullet$] \underline{Spin 1.} The kinetic action \eqref{K2generation} consists of the following contributions coming from spin-1 fields:
    \begin{align}
        S_2^{(1)}=\int A^{\alpha\dot\alpha}\Box A_{\alpha\dot\alpha}-\sA^{\beta\dot\alpha}\Box A^{\beta}{}_{\dot\alpha}\lambda_{\beta}\hat\lambda_{\beta}+A_{\beta}{}^{\dot\alpha}\Box \sA_{\beta\dot\alpha}\lambda^{\beta}\hat\lambda^{\beta}+\langle\lambda\,\hat\lambda\rangle^2\sA^{\alpha\dot\alpha}\Box\sA_{\alpha\dot\alpha}\,.
    \end{align}
    Making an appropriate field redefinition to fit with the form of the integral \eqref{eq:bridge}, we get
    \begin{align}
        S_2^{(1)}=-\int A^{\alpha\dot\alpha}\Box A_{\alpha\dot\alpha}+\sA^{\alpha\dot\alpha}\Box\sA_{\alpha\dot\alpha}\,.
    \end{align}
    Thus, the two modes decouple from each others at the quadratic level, and hence are two independent modes. Intriguingly, we can make the following change of variables
    \begin{align}
        \A_{\pm}^{\alpha\dot\alpha}=A^{\alpha\dot\alpha}\pm \im\,\sA^{\alpha\dot\alpha}\,,
    \end{align}
    to bring the kinetic term at level-1 to a more standard form:
    \begin{align}
        S_2^{(1)}=-\int \A_{-}^{\alpha\dot\alpha}\Box\A^+_{\alpha\dot\alpha}\,.
    \end{align}
  %  This is how one can diagonalize these two independent modes in spinorial formalism.
    
    \item[$\bullet$] \underline{Spin 2.} There are the following contributions at level 2:
    \begin{equation}
        \begin{split}
            S_2^{(2)}=&\int A^{\beta(2)\alpha\,\dot\alpha}\Box A_{\zeta(2)\alpha\,\dot\alpha}\lambda_{\beta}\hat\lambda_{\beta}\lambda^{\zeta}\hat\lambda^{\zeta}-\sA^{\beta(3)\dot\alpha}\Box A^{\beta}{}_{\zeta(2)\,\dot\alpha}\lambda_{\beta(2)}\hat\lambda_{\beta(2)}\lambda^{\zeta}\hat\lambda^{\zeta}\\
            +&\int A_{\beta}{}^{\zeta(2)\,\dot\alpha}\Box \sA_{\beta(3)\,\dot\alpha}\lambda^{\beta(2)}\hat\lambda^{\beta(2)}\lambda_{\zeta}\hat\lambda_{\zeta}+\langle \lambda\,\hat\lambda\rangle^2\lambda_{\beta}\hat\lambda_{\beta}\sA^{\beta(2)\alpha\,\dot\alpha}\Box \sA_{\zeta(2)\alpha\,\dot\alpha}\lambda^{\zeta}\hat\lambda^{\zeta}\,,
        \end{split}
    \end{equation}
    which gives
    \begin{align}
        S_2^{(2)}=-\int A^{\alpha(3)\dot\alpha}\Box A_{\alpha(3)\,\dot\alpha}+\sA^{\alpha(3)\,\dot\alpha}\Box \sA_{\alpha(3)\,\dot\alpha}\,.
    \end{align}
\end{itemize}
Inductively, we get the full kinetic term as
\begin{equation}
    \begin{split}
    S_2&=-\sum_{s\geq 1}\int \d^4 \tx\Big(A^{\alpha(2s-1)\,\dot\alpha}\Box A_{\alpha(2s-1)\,\dot\alpha}+\sA^{\alpha(2s-1)\,\dot\alpha}\Box \sA_{\alpha(2s-1)\,\dot\alpha}\Big)\\
    &=-\sum_{s\geq 1}\int \d^4\tx \,\A_-^{\alpha(2s-1)\,\dot\alpha}\Box\A^+_{\alpha(2s-1)\,\dot\alpha}\,.
    \end{split}
\end{equation}
where
\begin{align}
    \A^{\alpha(2s-1)\,\dot\alpha}_{\pm}:=A^{\alpha(2s-1)\,\dot\alpha}\pm\im\,\sA^{\alpha(2s-1)\,\dot\alpha}\,.
\end{align}
The same change of variables can also be made at cubic and quartic interactions. Therefore, we may interpret $\A_{\pm}$ as higher-spin fields with positive/negative ``helicity''. 
%This field redefinition is suggestive since $A$ and $\sA$ interact with each others at cubic and quartic orders.

We can now  understand better the role of the 
``would-be massive'' higher-spin fields. These extra dof arise because the above fields $A^{\alpha(2s-1)\,\dot\alpha}$ etc. are not divergence-free. Hence they contain extra dof arising as pure divergence mode, which should behave like ordinary pure gauge modes in the flat limit, and are thus expected to decouple. 
This is consistent with the lack of a scale parameter in the flat limit.

%In this sense, the ``would-be massive'' higher-spin fields in HS-IKKT theory become effectively `massless' and all the would-be massive modes will almost decouple from the system once flat limit is taken (due to the lack of a mass-like parameter). Note that this field redefinition is suggestive since $A$ and $\sA$ interact with each others at cubic and quartic orders.

Next, the cubic vertices read:
\be
    \begin{split}
    I_3&=2\int_{\P^1}\tK\, \{\ty^{\alpha}{}_{\dot\gamma},\sa^{\alpha\dot\gamma}\}\{\sa_{\alpha\dot\sigma},\sa_{\alpha}{}^{\dot\sigma}\}=2\int_{\P^1}\tK\,\cE^{\alpha}{}_{\dot\gamma,\beta\dot\beta}\p^{\beta\dot\beta}\sa^{\alpha\dot\gamma}\,\cE^{\circ\dot\circ,\bullet\dot\bullet}\p_{\circ\dot\circ}\sa_{\alpha\dot\sigma}\,\p_{\bullet\dot\bullet}\sa_{\alpha}{}^{\dot\sigma}\,\\
    &=-2\int_{\P^1}\tK\big(\lambda^{\alpha}\hat\lambda_{\beta}+\lambda_{\beta}\hat\lambda^{\alpha}\big)\p^{\beta}{}_{\dot\gamma}\sa^{\alpha\dot\gamma}\big(\lambda^{\circ}\hat\lambda^{\bullet}+\lambda^{\bullet}\hat\lambda^{\circ}\big)\p_{\circ\dot\circ}\sa_{\alpha\dot\sigma}\p_{\bullet}{}^{\dot\circ}\sa_{\alpha}{}^{\dot\sigma}\,.
    \end{split}
\ee
This means that 
\begin{align}
    I_3=-2\int(\lambda^{\alpha}\hat\lambda_{\beta}\lambda^{\circ}\hat\lambda^{\bullet}+\lambda^{\alpha}\hat\lambda_{\beta}\lambda^{\bullet}\hat\lambda^{\circ}+\lambda_{\beta}\hat\lambda^{\alpha}\lambda^{\circ}\hat\lambda^{\bullet}+\lambda_{\beta}\hat\lambda^{\alpha}\lambda^{\bullet}\hat\lambda^{\circ})\p^{\beta}{}_{\dot\gamma}\sa^{\alpha\dot\gamma}\p_{\circ\dot\circ}\sa_{\alpha\dot\sigma}\p_{\bullet}{}^{\dot\circ}\sa_{\alpha}{}^{\dot\sigma}\,.
\end{align}
From here, we can reposition indices to obtain $\epsilon$ symbols using \eqref{eq:bridge}. While there is no shortcut to contract indices, we observe that there are many contributions cancel each others, as well as contributions vanish on-shell (cf., \eqref{YMhel}). For instance, contributions such as
\begin{align}
    A\p^{\alpha\dot\alpha}B\p_{\alpha\dot\alpha}C\,,
\end{align}
can be discarded since it vanishes on-shell on support of momentum conservation in the flat limit. Furthermore, contributions that produce both types of angled and square brackets in terms of physical (but complex) spinors at 3-points will also vanish (see discussion in Section \ref{sec:6}). The vertices that produce these contributions will be `irrelevant' when computing scattering amplitudes, and thus can be neglected.

Notice that due to gravitational interactions coming from the Poisson structure \eqref{eq:Poissonspinors}, the lowest possible spins entering the cubic vertices should be $(1,1,2)$. In expanding higher-spin modes, we obtain
\begin{align}
    I_3=4\sum_{s_i}\int_{\P^1} \digamma^{\alpha,\beta|\circ,\bullet|\zeta(2s_2)|\rho(2s_3)}_{\tau(2s_1)}\p_{\beta}{}^{\dot\beta}\cA^{\tau(2s_1)\alpha}{}_{\dot\beta}\p_{\circ\dot\diamond}\cA_{\zeta(2s_2)\alpha\,\dot\sigma}\p_{\bullet}{}^{\dot\diamond }\cA_{\rho(2s_3)\alpha}{}^{\dot\sigma}\,,
\end{align}
where we have introduced the notation
\begin{align}
   \digamma^{\alpha,\beta|\circ,\bullet|\zeta(2s_2)|\rho(2s_3)}_{\tau(2s_1)}= \lambda^{(\alpha}\hat{\lambda}^{\beta)}\lambda^{(\circ}\hat{\lambda}^{\bullet)}\lambda^{\zeta(s_2)}\hat\lambda^{\zeta(s_2)}\lambda^{\rho(s_3)}\hat\lambda^{\rho(s_3)}\lambda_{\tau(s_1)}\hat\lambda_{\tau(s_1)}
\end{align}
etc., for convenience. The results can be summarized into the form
\begin{align*}
    \p A \{A,A\} +\p A\{A,\sA\} +\p A\{\sA,A\} + \p \sA \{A,A\}\\
    +\p \sA\{\sA,A\}+\p\sA\{A,\sA\}+\p A\{\sA,\sA\}+\p\sA\{\sA,\sA\}
\end{align*}
where $\{\,,\}$ indicate the contributions coming from the Poisson brackets. Here, the positions of fields are important since we are working with almost-commutative field theory. In terms of $\A_{\pm}$ we have all possible configuration of ``helicities'' at cubic order. Very roughly,
\begin{align}
    I_3\sim V_{+++}+V_{-++}+V_{+-+}+V_{++-}+V_{--+}+V_{-+-}+V_{+--}+V_{---}\,.
\end{align}
Without specifying the $\pm$ subscript below $\A$, we can write the cubic action as
\be    
   S_3 \simeq 4\sum_{s_2+s_3=s_1+2}\int d^4\tx \,\p_{\alpha\dot\alpha} \A_{\alpha(2s_1-1)}{}^{\dot\alpha}\p_{\alpha\dot\gamma}\A^{\alpha(2s_2-1)\,\dot\sigma}\p_{\alpha}{}^{\dot\gamma}\A^{\alpha(2s_3-1)}{}_{\dot\sigma}  +\widetilde{S}_3   \,,
\ee
where $\widetilde{S}_3$ denotes the irrelevant part of the action for massless sector, i.e. contributions that vanish on-shell. Here, the un-dotted indices in the
partial derivatives are contracted to those of the gauge potentials in all possible way. This way of writing is possible due to our symmetrization convention, i.e. the sum of all coefficients coming from contracting of $\eps$ tensors in \eqref{eq:bridge} with fields and derivatives is one. While the non-commutative twistor approach produces more contributions than the standard approach in twistor literature, we observe that the non-vanishing contributions in non-commutative twistor approach coincide with the ones using the standard approach of twistor theory in the flat limit.

\medskip

Finally, let us look at the quartic term:
\begin{equation}
    \begin{split}
    I_4&=\frac 12 \int_{\P^1}\tK\,\{\sa^{\alpha}{}_{\dot\gamma},\sa^{\alpha\dot\gamma}\}\{\sa_{\alpha\dot\sigma},\sa_{\alpha}{}^{\dot\sigma}\}=\frac 12\int_{\P^1}\tK\,\cE^{\circ\dot\circ,\bullet\dot\bullet}\p_{\circ\dot\circ}\sa^{\alpha}{}_{\dot\gamma}\p_{\bullet\dot\bullet}\sa^{\alpha\dot\gamma}\,\cE^{\diamond\dot\diamond,\blacklozenge\dot\blacklozenge}\p_{\diamond\dot\diamond}\sa_{\alpha\dot\sigma}\p_{\blacklozenge\dot\blacklozenge}\sa_{\alpha}{}^{\dot\sigma}\\
    &=2\sum_{s_i}\int_{\P^1}\tK\, \digamma^{\circ,\bullet|\diamond,\blacklozenge|\zeta(2s_3)|\tau(2s_4)}_{\rho(2s_1)|\delta(2s_2)}\p_{\circ\dot\circ}\cA^{\rho(2s_1)\alpha}{}_{\dot\gamma}\p_{\bullet}{}^{\dot\circ}\cA^{\delta(2s_2)\alpha\dot\gamma}\p_{\diamond\dot\diamond}\cA_{\zeta(2s_3)\,\dot\sigma}\p_{\blacklozenge}{}^{\dot\diamond}\cA_{\tau(2s_4)\alpha}{}^{\dot\sigma}\,.
    \end{split}
\end{equation}
\normalsize
Thanks to the plane-wave solutions \eqref{YMhel} and the specific gauge choice we made in Section \ref{sec:6}, all contributions resulting from the integral over $\P^1$ vanish on-shell. For this reason, we can either ignore the quartic, or write it as
\begin{align}
    S_4\simeq 2\int d^4\tx\, \Big(\p_{\alpha\dot\gamma_1}\A^{\alpha(2s_1-1)\,\dot\sigma}\p_{\alpha}{}^{\dot\gamma_1}\A^{\alpha(2s_2-1)}{}_{\dot\sigma}\Big)\Big(\p^{\alpha}{}_{\dot\gamma}\A_{\alpha(2s_3-1)}{}^{\dot\tau}\p^{\alpha\dot\gamma_2}\A_{\alpha(2s_4-1)\,\dot\tau}\Big)+\widetilde{S}_4\,.
\end{align}
Unlike the construction of higher-spin Yang-Mills \cite{Adamo:2022lah}, all possible configuration of helicities are allowed due to the helicity-like solutions $\A_{\pm}$.

%%%%%%%%%%%%%%%%%%%%%%%%%%%
\paragraph{Vertices from the self-dual sector.} To obtain the above result, one can also start from the self-dual sector. This provides a simpler way to contract indices in the interactions. But we first need to know how the higher-spin modes of the $\bs_{\alpha\alpha}$ field talk to each others. Recall that the action of the YM sector in first order formalism has the form cf. Section \ref{sec:5}

\begin{align}
    S^{\text{YM}}=\int_{\PS}\mho\Big(\bs_{\alpha\alpha}\ff^{\alpha\alpha}-\frac{1}{2}\bs_{\alpha\alpha}\bs^{\alpha\alpha}\Big) 
\end{align}
Now, we let $\bs_{\alpha\alpha}$ to have the following higher-spin expansion (see the discussion around \eqref{reality-cA} for the reality condition of $\hs$-valued $\ff^{\alpha\alpha}$ and $\bs_{\alpha\alpha}$):
\begin{align}
    \bs_{\alpha\alpha}=\Big(B_{\beta(2s)\alpha\alpha}-\im \eps_{\beta\alpha}\sB_{\beta(2s-1)\alpha}\Big)\lambda^{\beta(s)}\hat\lambda^{\beta(s)}\,.
\end{align}
It can be checked that there is no mixed term between $B$ and $\sB$. For instance, consider spin-1 case 
\small
\begin{align}\label{B-spin1}
    \int\bs_{\alpha\alpha}\bs^{\alpha\alpha}\big|_{s=1}&= \int B_{\alpha\alpha}B^{\alpha\alpha}+\im B_{\alpha\alpha}\eps^{\beta\alpha}\sB^{\beta\alpha}\lambda_{\beta}\hat\lambda_{\beta}-\im B_{\alpha\alpha}\eps^{\beta\alpha}\sB^{\beta\alpha}\lambda_{\beta}\hat\lambda_{\beta}+\lambda^{\beta}\hat\lambda^{\beta}\eps_{\beta\alpha}\eps^{\gamma\alpha}\sB_{\beta\alpha}\sB^{\gamma\alpha}\lambda_{\gamma}\hat\lambda_{\gamma}\nonumber\\
    &=\int B_{\alpha\alpha}B^{\alpha\alpha}+\lambda^{\beta}\hat\lambda^{\beta}\eps_{\beta\alpha}\eps^{\gamma\alpha}\sB_{\beta\alpha}\sB^{\gamma\alpha}\lambda_{\gamma}\hat\lambda_{\gamma}\,.
\end{align}
\normalsize 
%The second term on the rhs. can be written as
%\begin{align}
%   \int -B_{\alpha\alpha}\eps^{\beta_1\alpha}\sB_{\beta_2}{}^{\alpha}\lambda_{\beta_1}\hat\lambda^{\beta_2}-B_{\alpha\alpha}\eps_{\beta_1}{}^{\alpha}\sB^{\beta_2\alpha}\lambda_{\beta_2}\hat\lambda^{\beta_1}=0
%\end{align}
%which gives zero after integrating out fiber coordinates. 
We can shorten the last term in \eqref{B-spin1} to
\begin{align}
   \int -\frac{1}{4}\langle \lambda\,\hat\lambda\rangle^2 \sB_{\alpha\alpha}\sB^{\alpha\alpha}+\frac{1}{2}\lambda^{\beta}\hat\lambda^{\beta}\sB_{\beta}{}^{\gamma}\sB_{\beta}{}^{\gamma}\lambda_{\gamma}\hat\lambda_{\gamma}\sim \int \d^4 x\,\sB_{\alpha\alpha}\sB^{\alpha\alpha}\,.
\end{align}
Therefore, the $B$ and $\sB$ modes do not couple to each other (other higher-spin cases are analogous). Observe that we can use $\sB$ as Lagrangian multipliers for the fuzzy Lorenz gauge condition \eqref{gaugefixing} 
\begin{align}
    \int \mho \,\sB_{\beta(2s)}\{\ty_{\alpha\dot\alpha},\cA^{\beta(2s)|\alpha\dot\alpha}\}
\end{align}
so that only the first modes $B_{\alpha(2s)}$ propagate and $\sB$ becomes non-dynamical. Therefore, it is suggestive to treat $B_{\alpha(2s)}$ as negative helicity modes. To proceed, we will look at the kinetic terms
\begin{align}
    \sum_{s_1,s_2}\int \mho \lambda^{\beta(s_1)}\hat\lambda^{\beta(s_1)}B_{\beta(2s_1)\alpha\alpha}\{\ty^{\alpha}{}_{\dot\alpha},\cA^{\gamma(2s_2)|\alpha\dot\alpha}\}\lambda_{\gamma(s_2)}\hat\lambda_{\gamma(s_2)}\,.
\end{align}
By decomposing $\cA^{\beta(2s)|\alpha\dot\alpha}=A^{\beta(2s)\alpha\,\dot\alpha}+\im \eps^{\alpha\beta}\sA^{\beta(2s-1)\,\dot\alpha}$, we can check that
%\begin{align}
%    L_{2,1}&=\int \mho \lambda^{\beta(s)}\hat\lambda^{\beta(s)}\lambda_{\gamma(s)}\hat\lambda_{\gamma(s)}B_{\beta(2s)\alpha\alpha}\Big(\{\ty^{\alpha}{}_{\dot\alpha},A^{\gamma(2s)\alpha\,\dot\alpha}\}+\im \eps^{\alpha\gamma}\{\ty^{\alpha}{}_{\dot\alpha},\sA^{\gamma(2s+1)\,\dot\alpha}\Big)\,,\nonumber\\
%    &\sim \int \d^4x  B_{\alpha(2s+2)}\p^{\alpha}{}_{\dot\alpha}\A_+^{\alpha(2s+1)\,\dot\alpha}\,.
%\end{align}
%Here, we recall that $\A_+^{\alpha(2s-1)\,\dot\alpha}=A^{\alpha(2s-1)\,\dot\alpha}+\im \sA^{\alpha(2s-1)\,\dot\alpha}$ are positive higher-spin fields defined in \eqref{Apm-def-euclid}. Similarly, 
%\begin{align}
%    L_{2,2}&=-\im \int \mho \lambda^{\beta(s)}\hat\lambda^{\beta(s)}\lambda_{\gamma(s)}\hat\lambda_{\gamma(s)}\sB_{\beta(2s)\alpha}\eps_{\alpha\beta}\{\ty^{\alpha}{}_{\dot\alpha},\cA^{\gamma(2s)|\alpha\dot\alpha}\}\,,\nonumber\\
%    &=-\frac{\im}{2} \int \mho \lambda^{\beta(s)}\hat\lambda^{\beta(s)}\lambda_{\gamma(s)}\hat\lambda_{\gamma(s)}\sB_{\beta(2s)\alpha}\Big(\{\ty_{\beta\dot\alpha},\cA^{\gamma(2s)|\alpha\dot\alpha}\}+\{\ty^{\alpha}{}_{\dot\alpha},\cA^{\gamma(2s)|\ \ \dot\alpha}_{\qquad \beta}\Big)\,,\nonumber\\
%    &\sim -\im \int \d^4x  \sB_{\alpha(2s+2)}\p^{\alpha}{}_{\dot\alpha}\A_+^{\alpha(2s+1)\,\dot\alpha}
%\end{align}
%By combining $L_{2,1}$ and $L_{2,2}$ we have the standard kinetic terms 
\begin{align}\label{L2Euclidean}
    L_2&=\int \mho \lambda^{\beta(s)}\hat\lambda^{\beta(s)}\lambda_{\gamma(s)}\hat\lambda_{\gamma(s)}B_{\beta(2s)\alpha\alpha}\Big(\{\ty^{\alpha}{}_{\dot\alpha},A^{\gamma(2s)\alpha\,\dot\alpha}\}+\im \eps^{\alpha\gamma}\{\ty^{\alpha}{}_{\dot\alpha},\sA^{\gamma(2s+1)\,\dot\alpha}\}\Big)\,\nonumber\\
    &=\int \d^4x B_{\alpha(2s)}\p^{\alpha}{}_{\dot\alpha}\A_+^{\alpha(2s-1)\,\dot\alpha}\,.
\end{align}
Thus, as always, the kinetic terms decrible the coupling between negative and positive helicity fields. From here, we can express $A$ and $\sA$ in terms of $\A_{\pm}$ as
\begin{align}
    A^{\alpha(2s-1)\,\dot\alpha}=\frac{\A_+^{\alpha(2s-1)\,\dot\alpha}+\A_-^{\alpha(2s-1)\,\dot\alpha}}{2}\,,\quad \sA^{\alpha(2s-1)\,\dot\alpha}=\frac{\A_+^{\alpha(2s-1)\,\dot\alpha}-\A_-^{\alpha(2s-1)\,\dot\alpha}}{2\im}
\end{align}
to study the cubic interactions $\int \bs\{\cA,\cA\}$.

In this first-order formalism, the interaction terms have much less structure compared to the second-order case. In particular, after integrating out fiber coordinates, we end up with three main structures
\begin{subequations}\label{cubic-mainstructures}
    \begin{align}
    B^-_{\alpha(2s_1)}\p^{\alpha}{}_{\dot\gamma}\A_{\pm}^{\alpha(2s_2-1)\,\dot\alpha}\p^{\alpha\dot\gamma}\A_{\pm}^{\alpha(2s_3-1)}{}_{\dot\alpha}\,,\\
    B^-_{\alpha(2s_1)}\p_{\alpha\dot\gamma}\A_{\pm}^{\alpha(2s_2-1)\,\dot\alpha}\p^{\alpha\dot\gamma}\A_{\pm}^{\alpha(2s_3-1)}{}_{\dot\alpha}\,,\\
    B^-_{\alpha(2s_1)}\p_{\alpha\dot\gamma}\A_{\pm}^{\alpha(2s_2-1)\,\dot\alpha}\p_{\alpha}{}^{\dot\gamma}\A_{\pm}^{\alpha(2s_3-1)}{}_{\dot\alpha}\,,\label{survive-cubic}
\end{align}
\end{subequations}
where, as always, all un-dotted indices of the lower level are understood to be contracted in all possible way with the indices of the upper level. Notice that the difference between the above structures are merely the positions of the undotted indices in the partial derivatives. 

Besides the polarization tensors \eqref{YMhel} for the $\A_{\pm}$ fields, we set 
\begin{align}
    \eps_{\alpha(2s)}^-=\lambda_{\alpha(2s)}
\end{align}
to the the polarization tensor associated with the $B_{\alpha(2s)}$ negative helicity fields \cite{Krasnov:2021nsq}. Note that there is no auxiliary/reference spinors $\eps^-_{\alpha(2s)}$. Then, upon plugging in the plane-wave solutions, it can be checked that  \eqref{survive-cubic} is the only structure that survives. It is worth mentioning that there are no combination such as 
\begin{align}\label{LCcombinationtrivial}
    B_{\alpha(2s_1)}\{\A^{\alpha(2s_2-1-m)\beta(m)\,\dot\alpha},\A^{\alpha(2s_3-1-m)}{}_{\beta(m)\,\dot\alpha}\}
\end{align}
where un-dotted indices of the $\A$ fields contracted with each others since it results in trivial amplitudes. In fact, this can also be understood from the light-cone point of view where we set $\A^{\alpha(2s-2)0\,\dot 0}=0$ and choose $\A^{1(2s-1)\,\dot 0}$ to be the components carry physical degrees of the $\A$ fields. Then, for any combination such as \eqref{LCcombinationtrivial}, it will vanish on the nose in the light-cone gauge.

Further scrutiny shows that non-trivial cubic amplitudes will come from
\begin{align}
    B^-_{\alpha(2s_1)}\p_{\alpha\dot\gamma}\A_{+}^{\alpha(2s_2-1)\,\dot\alpha}\p_{\alpha}{}^{\dot\gamma}\A_{+}^{\alpha(2s_3-1)}{}_{\dot\alpha}\,.
\end{align}
These are the vertices that give us the $\overline{\text{MHV}}_3$ amplitudes \eqref{MHV-bar} in the main text. As a result, the $\bs_{\alpha\alpha}\ff^{\alpha\alpha}$ can be written as
\begin{align}\label{spacetimeBF}
   \int \mho\, \bs_{\alpha\alpha}\ff^{\alpha\alpha}\sim \int \d^4 x B^-_{\alpha(2s)}F^{\alpha(2s)}
\end{align}
where for convenience we have defined
\begin{align}
    F_+^{\alpha(2s)}:=\p^{\alpha\dot\alpha}\A_+^{\alpha(2s-1)}{}_{\dot\alpha}+\frac{1}{2}\sum_{m+n=s+2}\p_{\alpha\dot\gamma}\A_+^{\alpha(2s_2-1)\,\dot\sigma}\p_{\alpha}{}^{\dot\gamma}\A_+^{\alpha(2s_3-1)}{}_{\dot\sigma}\,.
\end{align}

Of course, the above BF action \eqref{spacetimeBF} in flat space is only the self-dual sector which features the $\overline{\text{MHV}}_3$ amplitudes. To go back to the second-order formalism, we can integrate out the $B^-_{\alpha(2s)}$ which results in 
\begin{align}
   S_{\text{YM}}=\int F_{\alpha(2s)}F^{\alpha(2s)}
\end{align}
As this stage, one can give the higher-spin fields $\A^{\alpha(2s-1)\,\dot\alpha}$ all possible helicity when we consider cubic and quartic interactions. The reason is that the negative helicity modes $B^-_{\alpha(2s)}$ need to be replaced by $\A_-^{\alpha(2s-1)\,\dot\alpha}$. Lastly, so as not to forget about other structures, we can write the final action as
\begin{align}\label{gravHSYMaffine}
    S^{\text{YM}}=\sum_s\int \d^4\tx \,F_{\alpha(2s)}F^{\alpha(2s)}+\widetilde{S}_3+\widetilde{S}_4\,.
\end{align}
where $\widetilde S_i$ for $i=3,4$ denote terms that are irrelevant when studying scattering amplitudes.

%%%%%%%%%%%%%%%%%%%%%%%%%%%

%%%%%%%%%%%%%%%%%%%%%%%%%%%%%%%%%

%%%%%%%%%%%%%%%%%%%%%%%%%%%%%%%%%%%%%%%%

\setstretch{0.8}
\footnotesize
\bibliography{twistor}
\bibliographystyle{JHEP}

\end{document}